\documentclass[prd, aps, twocolumn, showpacs, tightenlines, nofootinbib, eqsecnum, amsmath, amsfonts, floatfix, superscriptaddress]{revtex4-1}

\usepackage[utf8]{inputenc}
\usepackage{mathrsfs}
\usepackage{euscript}
\usepackage{epsfig}
\usepackage{graphics}
\usepackage{graphicx}
\usepackage{amsmath}
\usepackage{amssymb}
\usepackage{bm}
\usepackage[usenames,dvipsnames,svgnames,table]{xcolor}
\usepackage{xspace}
\usepackage{wasysym}
\usepackage{times}
\usepackage{appendix}
\usepackage{lipsum}
\usepackage[nolist,nohyperlinks]{acronym}
\usepackage{float}
\usepackage[colorlinks=true,linkcolor=dodgerblue,citecolor=dodgerblue,urlcolor=dodgerblue]{hyperref}
\usepackage{natbib, ifthen}
\usepackage{hyperref}
\usepackage{booktabs}

\usepackage{longtable}

\definecolor{ZurichBlue}{rgb}{.255,.41,.884} 		
\definecolor{ZurichRed}{rgb}{0.9, 0.1, 0} 			
\definecolor{ZurichGreen}{rgb}{.196,.504,.396} 		
\definecolor{ZurichYellow}{rgb}{1,.648,0} 			
\definecolor{dodgerblue}{rgb}{0.12, 0.56, 1.0}
\definecolor{azure}{rgb}{0.0, 0.5, 1.0}
\definecolor{alizarincrimson}{rgb}{0.82, 0.1, 0.26}
\definecolor{mediumpurple}{rgb}{0.58, 0.44, 0.86}
\definecolor{lasallegreen}{rgb}{0.03, 0.47, 0.19}
\definecolor{my_gray}{rgb}{0,0,0}

\DeclareMathAlphabet{\pazocal}{OMS}{zplm}{m}{n}

\DeclareSymbolFontAlphabet{\mathrsfs}{rsfs}
\DeclareMathAlphabet\mathbfcal{OMS}{cmsy}{b}{n}

\newcommand{\CITEME}[1]{\textcolor{alizarincrimson}{CITEME!}}
\newcommand{\citeme}[1]{\textcolor{alizarincrimson}{citation required}}

\usepackage{listings,xcolor}

\newcommand{\ba}{\begin{align}}
\newcommand{\ea}{\end{align}}

\newcommand{\n}{\newline}

\def\lm{{\ell m}}   

\def\ii{{\rm i}}

\def\pph{p_{\varphi}}
\def\prs{p_{r_{\ast}}}

\renewcommand{\Re}{\operatorname{Re}}

\def\TEOBResumS{\texttt{TEOBResumS}}
\def\EOBResumM{\texttt{{EOBResumMultipoles$^+$}}}
\def\TEOBiResumSM{\texttt{TEOBiResumMultipoles}}

\newcommand{\be}{\begin{equation}}
\newcommand{\ee}{\end{equation}}

\begin{document}

\title{A Multipolar Effective One Body Model for Non-Spinning Black Hole Binaries}

\author{Alessandro \surname{Nagar}}
\affiliation{Centro Fermi - Museo Storico della Fisica e Centro Studi e Ricerche ``Enrico Fermi'', 00184 Roma, Italy}
\affiliation{INFN Sezione di Torino, Via P.~Giuria 1, 10125 Torino, Italy}
\affiliation{Institut des Hautes Etudes Scientifiques, 91440 Bures-sur-Yvette, France}
\author{Geraint \surname{Pratten}}
\affiliation{Universitat de les Illes Balears, Crta. Valldemossa km 7.5, E-07122, Palma, Spain}
\affiliation{School of Physics and Astronomy and Institute for Gravitational Wave Astronomy, University of Birmingham, Edgbaston, Birmingham, B15 9TT, United Kingdom}
\author{Gunnar \surname{Riemenschneider}}
\affiliation{INFN Sezione di Torino, Via P.~Giuria 1, 10125 Torino, Italy}
\affiliation{Dipartimento di fisica, Universit\`a di Torino, Via P.~Giuria~1, 10125 Torino, Italy}
\author{Rossella \surname{Gamba}}
\affiliation{INFN Sezione di Torino, Via P.~Giuria 1, 10125 Torino, Italy}
\affiliation{Dipartimento di fisica, Universit\`a di Torino, Via P.~Giuria~1, 10125 Torino, Italy}

\begin{abstract}
  We introduce \TEOBiResumSM{}, a nonspinning inspiral-merger-ringdown waveform model built within the
  effective one body (EOB) framework that includes gravitational waveform modes beyond the dominant quadrupole
  $(\ell,|m|) = (2,2)$. The model incorporates: (i) an improved Pad\'e resummation of the factorized
  waveform amplitudes $\rho_\lm^{\rm orb}$ entering the EOB-resummed waveform where the 3PN, 
  mass-ratio dependent, terms are hybridized with test-mass limit terms up to 6PN relative order
  for most of the multipoles up to $\ell=6$ included; (ii) an improved determination of the effective 
  5PN function $a_6^c(\nu)$ entering the EOB interaction potential done using the most recent,
  error-controlled, nonspinning numerical relativity (NR) waveforms from the 
  Simulating eXtreme Spacetimes (SXS) collaboration;
  and (iii) a NR-informed phenomenological description of the multipolar ringdown. 
  Such representation stems from 19 NR waveforms with mass ratios up to $m_1/m_2=18$ 
  as well as  test-mass waveform data, although it does not incorporate mode-mixing effects.
  The NR-completed higher modes through merger and ringdown  considered here are:  
  $(\ell,|m|) = \lbrace (2,1), (3,3), (3,2),(3,1),(4,4), (4,3),(4,2), (4,1),(5,5)\rbrace$. 
  For simplicity, the other subdominant modes, up to $\ell=8$, are approximated by 
  the corresponding, purely analytical, factorized and resummed EOB waveform. To attempt an estimate
  of (some of) the underlying analytic uncertainties of the model, we also contrast the effect 
  of the  6PN-hybrid Pad\'e-resummed  $\rho_\lm$'s with the standard $3^{+2}$~PN, Taylor-expanded,
  ones used in previous EOB works. The maximum unfaithfulness $\bar{F}$ against the SXS
  waveforms including all NR-completed modes up to $\ell=m=5$  is always $\lesssim 2$\%
  for binaries with total mass $M$ as $50 M_{\odot} \leq M \lesssim 200 M_{\odot}$.
  The Pad\'e-resummed multipolar EOB model for nonspinning binaries discussed here
  defines the foundations of a multipolar EOB waveform model for spin-aligned
  binaries that will be introduced in a companion paper.
\end{abstract}

\date{\today}


\maketitle

\section{Introduction}
The recent observation made by LIGO \cite{TheLIGOScientific:2014jea} and Virgo \cite{TheVirgo:2014hva} 
of gravitational wave signals (GWs) from eleven coalescing compact binaries 
marked the beginning of the era of gravitational wave astronomy.
Of these detections, ten have been associated to binary black holes
(BBH) \cite{Abbott:2016blz,Abbott:2016nmj,Abbott:2017vtc,Abbott:2017gyy,Abbott:2017oio,LIGOScientific:2018mvr}
as well as the detection of a coalescing binary neutron star (BNS)~\cite{TheLIGOScientific:2017qsa}.

One of the standard tools in modelled GW data analysis is matched filtering,
implicitly demanding high-fidelity, low-bias waveform models. Search pipelines can detect
GW signals from binary black holes by cross-correlation of the data against
theoretical waveform templates for the expected signal.
Bayesian parameter estimation allows us to infer the source properties
of the binary by comparing the data against our analytical waveform families.
Null tests of General Relativity (GR) are often predicated on the comparison
of the data to the faithful representations of GR given by our waveform models.
It is therefore vital that our waveform models are accurate and incorporate
as much physics as possible. A deficiency in many waveform families to date
has been that they only model the leading quadrupole $(\ell=2,m=2)$ mode
of gravitational radiation. For weak signal to noise ratio (SNR) observations 
or for binaries where intrinsic asymmetries are suppressed, i.e. comparable mass ratios
and near equal spin configurations, this may be sufficient. 
This simplification was sufficiently accurate for detecting the binary black hole sources observed during the 
first two LIGO - Virgo observing runs (O1 and O2),
with no compelling evidence for higher modes seen in the 
parameter estimation ~\cite{Abbott:2016wiq,LIGOScientific:2018mvr}. 

However, with the sensitivity of Advanced LIGO and Virgo ever increasing,
 systematic errors will start to dominate statistical errors leading to
potentially large biases in our parameter estimation~\cite{OShaughnessy:2014shr,Varma:2016dnf}
and could degrade the performance of our search pipelines~\cite{Capano:2013raa,Bustillo:2016gid,Harry:2017weg}. 
Specifically, this could be the case for binaries that have high inclination angles or where
there are stronger intrinsic asymmetries, such as one BH being more massive
than the other or large unequal spin effects. Similarly, at large inclinations,
the modeling of gravitational wave modes beyond the dominant mode becomes
increasingly important as higher modes are geometrically suppressed in the face-on/off limit.

A key result in GR is the no-hair theorem, in which the quasi-normal-modes (QNMs)
of an isolated BH in GR may only depend on the BH's mass $M_{\rm BH}$, angular momentum
$J_{\rm BH}$ or charge $Q$. In vacuum, testing the no-hair theorem requires us to identify 
at least two QNMs in the ringdown, necessitating accurate higher-mode waveforms for binary black holes. 
Recent studies on such black hole spectroscopy have demonstrated the feasibility of performing such
tests of general relativity~\cite{Dreyer:2003bv,Berti:2005ys,Kamaretsos:2011um,Gossan:2011ha,Kamaretsos:2012bs,Meidam:2014jpa,London:2014cma,Berti:2016lat,Cardoso:2016ryw,Yang:2017zxs,Brito:2018rfr,Carullo:2018sfu}. 

Within the effective-one-body (EOB) approach, the first model including higher-order
modes in the nonspinning case was presented in Ref.~\cite{Pan:2011gk}, now known as {\tt EOBNRv2HM},
followed by Ref.~\cite{Damour:2012ky}, that employed different EOB dynamics.
More recently Ref.~\cite{Cotesta:2018fcv} improved the {\tt EOBNRv2HM} model, 
generalizing it to the case of spin-aligned BBHs. This model is called {\tt SEOBNRv4HM}
and it is currently the only available EOB model with higher harmonics. In the {\tt IMRPhenom} family of 
waveform models, \cite{Mehta:2017jpq} presented a non-spinning model calibrated against NR and \cite{London:2017bcn}
used approximate analytical relations to construct an uncalibrated higher mode model for spin-aligned BBHs. 

In this paper we follow up on the work of Ref.~\cite{Damour:2012ky}.
To do so, we improve the nonspinning sector of \TEOBResumS~\cite{Nagar:2018zoe}
by completing the EOB-resummed multipolar waveform through merger and ringdown
with all subdominant multipoles up to $\ell=m=5$ included.
This result crucially relies on an improved representation of the merger-ringdown
waveform that builds upon the post-merger analytical description of
Refs.~\cite{Damour:2014yha,Nagar:2016iwa} that is informed by suitable fits
of a few ($19$) Numerical Relativity (NR) nonspinning waveforms. Comparable-mass NR data are also
complemented by the test-mass waveform data of~\cite{Harms:2014dqa}.
Our important conceptual finding is that, once that the NR information
is represented analytically, the procedure to match the post-merger
waveform to the inspiral EOB waveform is precisely the same for
all multipoles, without any special ad-hoc tuning and/or calibration
for each mode. Moreover, we also improve the dynamical sector of the model,
by augmenting the multipolar waveform amplitudes with  test-particle information
and/or Pad\'e resumming them. This in turn needs a new determination of the
effective coefficient $a_6^c$ that enters the EOB interaction potential at
5th post-Newtonian (PN) order and is informed by NR simulations.
The nonspinning dynamics presented here will then be used for the generalization
of the multipolar waveform model to the spin-aligned case, that will be
presented in a separate publication.

A complementary approach was presented in~\cite{Borhanian:2019kxt}, which studies the structure of the multipolar waveform throughout the transition from the 
perturbative inspiral regime to the highly non-perturbative merger using a combination of post-Newtonian theory and numerical relativity data. There it is 
shown that the dependencies of the dominant modes amplitudes on the symmetric mass ratio and binary spin, to leading order in post-Newtonian theory, 
agree remarkably well with those from numerical relativity even in a regime in which this approximation should have broken down. 

This paper is organized as follows: In Sec.~\ref{sec:HM_phenom} we recall the
phenomenology due to the presence of higher waveform modes. Section~\ref{sec:EOB}
discusses in detail all the building blocks of the nonspinning EOB model, with
special emphasis on the analytical structure of the waveform multipoles;
Section~\ref{sec:nrinfo} focuses on how the EOB dynamics is informed by NR
simulations through the determination of $a_6^c$, highlighting the differences
with respect to previous work; Section~\ref{sec:postmerger} illustrates in
high detail the construction of the merger and postmerger multipolar waveform
model, reporting all the NR-informed fits needed, while the phasing accuracy
of the model is assessed in Sec.~\ref{sec:acc}, either via standard time-domain
phasing comparisons or via unfaithfulnesses analyses. Our concluding remarks
are reported in Sec.~\ref{sec:conc}.
Unless stated otherwise, we use geometrized units with $G=c=1$ and the
following notation: $M\equiv m_1+m_2$, $\mu\equiv m_1 m_2/M$, $\nu\equiv\mu/M$ with the
convention that the mass ratio is $q\equiv m_1/m_2\geq 1$. 


\begin{figure}[t]
\begin{center}
\includegraphics[width=0.95\columnwidth]{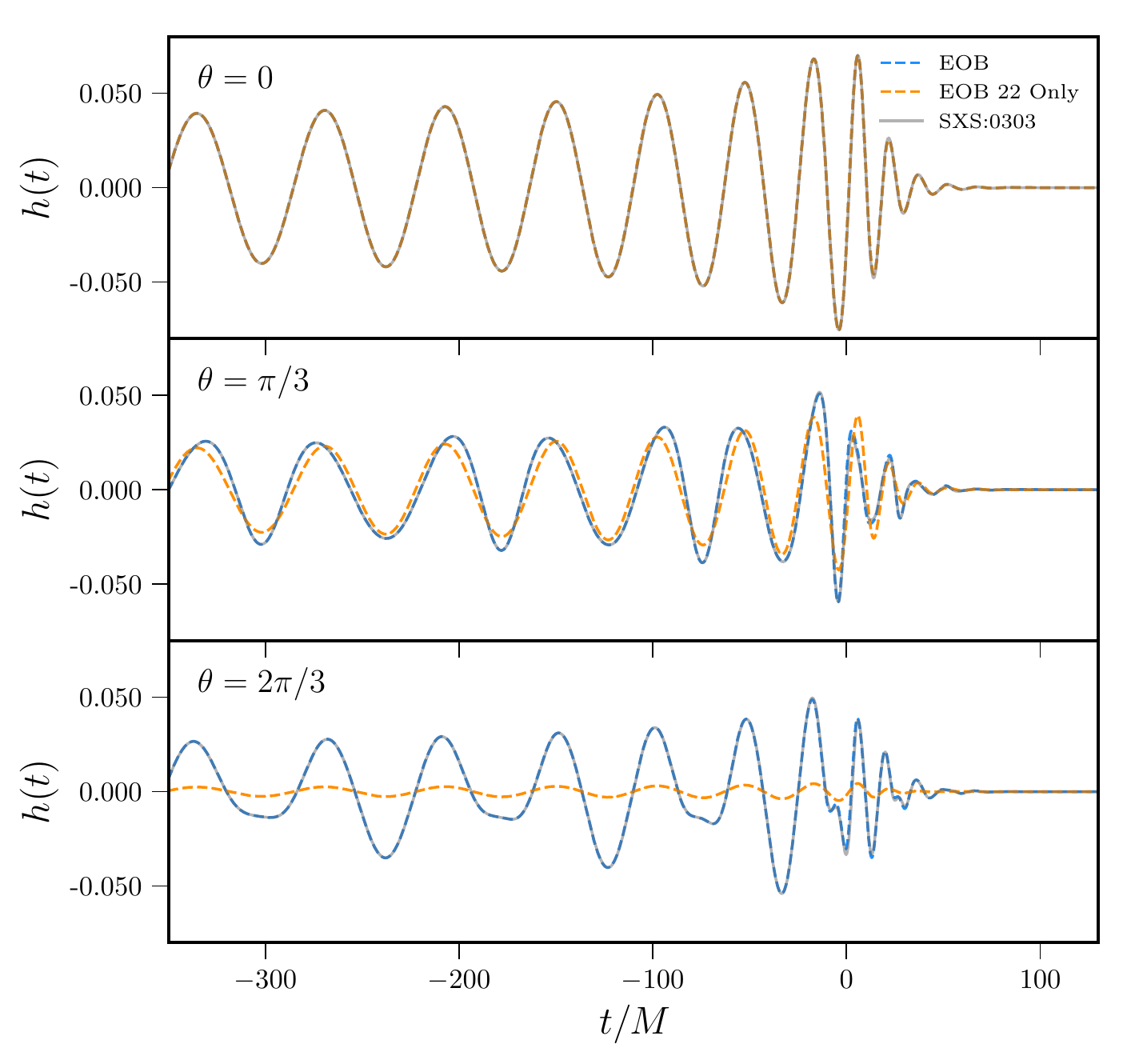}
\caption{Illustrating the effect of higher multipoles.
Comparison of $h (t)$ between the EOB model and SXS:0303, a non-spinning binary of mass ratio $q=10$.
}
\label{fig:EOBvsNRHM}
\end{center}
\end{figure}

\section{Waveform phenomenology due to higher-order modes}
\label{sec:HM_phenom}
The time-domain complex gravitational wave strain can be
written in terms of the spin weighted spherical harmonics as 
\begin{align}
  &h_{+} ( \theta , \varphi ; t) - {\rm i} h_{\times} ( \theta , \varphi ; t) \nonumber\\
  &\qquad\quad = \dfrac{GM}{{\cal D}_Lc^2}\displaystyle\sum_{\ell = 2}^{\infty} \displaystyle\sum_{m = -\ell}^{\ell} \,  \, h_{\ell m}  (t) _{-2}Y_{\ell m} (\theta, \varphi) \ ,
\end{align}
\n
where $\lbrace \theta,\varphi \rbrace$ are the spherical angles in a
coordinate system with the z-axis aligned along the orbital angular momentum
and ${D_L}$ the luminosity distance from the source. The spherical harmonic modes
$h_{\ell m} (t)$ are functions of the intrinsic parameters of the system with
the angular dependence of the emitted gravitational radiation
being governed by the spin weighted spherical harmonic basis functions.
The gravitational wave strain measured by a detector will therefore be highly dependent
on the sky location and inclination of the binary with respect to the observer,
as the spherical harmonics effectively act as a geometric factor contributing to
the relative strength of the higher modes\footnote{Note that the $(\ell,m) \neq (2,2)$
  modes are denoted by numerous names in the literature: subdominant modes, higher-order modes
  or higher modes.}. For face-on/face-off binaries ($\theta \sim 0$),
the dominant contribution to the signal is from the quadrupolar
($\ell = 2$, $m=\pm2$) mode, as shown in the top panel
of Figure~\ref{fig:EOBvsNRHM}. However, for more generic orientations,
the higher modes can be of a comparable order of magnitude to the $(2,2)$ mode
with the relative contribution of the higher modes depending on the underlying
symmetries of the binary. This can be seen in the bottom two panels of
Fig.~\ref{fig:EOBvsNRHM}. We can cleary see that the pure quadrupolar
waveforms fails to capture the morphology of the signal for generic
orientations. An additional simplification can be made for non-precessing
binaries as there will also be a reflection symmetry about the orbital
plane such that $h_{\ell m} = (-1)^{\ell} \, h_{\ell -m}^{\ast}$.
This means that we can restrict our discussion to modes with $m > 0$
and can use the above relation in order to reconstruct the negative $m$ modes.


\section{Nonspinning effective one body model}
\label{sec:EOB}
The dynamics of the multipolar EOB model we discuss here
stems from a slightly modified version of the nonspinning sector
of \TEOBResumS{}~\cite{Nagar:2018zoe}. The analytical modifications
mainly regard the radiation reaction sector, where we incorporate 
more analytical information than previously used in~\cite{Nagar:2018zoe}.
In addition, due to the availability of NR waveforms from the SXS
collaboration with a smaller overall error budget,
it was possible to improve the model by re-informing the effective 5PN parameter $a_6^c$.
For self-consistency, we detail all the various building blocks of
the model, highlighting the core changes with respect to~\cite{Nagar:2018zoe}.

\subsection{Hamiltonian}
\label{sec:Ham}
The conservative dynamics of two bodies of masses $m_1$ and $m_2$ 
is described by a Hamiltonian $H_{\rm{EOB}} (Q^i,P_i)$
in terms of the relative motion of the binary, where $R=|Q^i|$ is the binary separation in
EOB coordinates. The Hamiltonian will depend on two radial functions $A(R)$ and $B(R)$,
where $A(R)$ is the EOB radial potential. In the above notation, the EOB Hamiltonian is given by
\begin{align}
\hat{H}_{\rm{EOB}} \equiv \frac{H_{\rm{EOB}}}{\mu} = \frac{1}{\nu} \sqrt{1 + 2 \nu \left( \hat{H}_{\rm{eff}} - 1\right) },
\end{align}
where the usual effective EOB Hamiltonian is 
\begin{align}
\hat{H}_{\rm{eff}} &= \sqrt{ p^2_{r_{\ast}} + A(r, \nu) \left( 1 + \frac{p_\varphi^2}{r^2} + \mathcal{Q}_4 \right) },
\end{align}
\n
with $\mathcal{Q}_4 = 2 \nu (4 - 3 \nu) p^4_{r_{\ast}}$ taken at 3PN accuracy~\cite{Damour:2000we}.
Here, $\lbrace r, p_{r_{\ast}}, \varphi, p_{\varphi} \rbrace$
are the usual dimensionless EOB phase space variables in polar coordinates~\cite{Damour:2014sva}
and we have replaced the conjugate momentum $p_r$ by the tortoise rescaled variable
$p_{r_{\ast}} = (A/B)^{1/2} \, p_r$, where $r_{\ast} = \int dr (A/B)^{1/2}$.
We restrict our attention to orbits confined to the equatorial plane,
$\theta = \pi/2$. The dimensionless EOB phase space variables are related
to the dimensionful variables $(R,P_R,\varphi,P_{\varphi})$ by
\begin{align}
r = \frac{R}{GM}, \; p_{r_{\ast}} = \frac{P_{R_{\ast}}}{\mu}, \; p_{\varphi} = \frac{P_{\varphi}}{\mu G M}, \; t = \frac{T}{GM}. 
\end{align}
\n
Hamilton's equations naturally follow from the above expressions and can be explicitly written as
\begin{align}
\frac{d \varphi}{d t} &= \Omega = \frac{\partial \hat{H}_{\rm EOB}}{\partial \pph} , \\
\frac{d r}{d t} &= \left( \frac{A}{B} \right)^{1/2} \, \frac{\partial \hat{H}_{\rm EOB}}{\partial \prs} , \\
\frac{d \pph}{d t} &= \hat{\mathcal{F}}_{\varphi}, \\
\frac{d \prs}{d t} &= -\left( \frac{A}{B} \right)^{1/2}  \frac{\partial \hat{H}_{\rm EOB}}{\partial r}.
\end{align}
\n
where $\hat{\mathcal{F}}_{\varphi} = \mathcal{F}_{\varphi} / \mu$ is a
radiation reaction term and $\mathcal{F}_{r_{\ast}}$ is explicitly put to
zero\footnote{This can be considered a gauge choice. This is also chosen to be like this
  due to the current absence of a robust strategy for resumming such a radial
  contribution~\cite{Bini:2012ji} to improve its behavior close to merger.
  See also Ref.~\cite{Nagar:2015xqa} for the effect of such nonresummed
  ${\cal F}_r$ on the binary energetics.}. It incorporates all multipoles,
up to $\ell=8$, in special resummed form~\cite{Damour:2008gu} as detailed
in Sec.~\ref{sec:wave} below. Note that although the effect of gravitational wave absorption through the black hole horizons
is small in the non-spinning case~\cite{Bernuzzi:2012ku}, it is explicitly included in the model
following Refs.~\cite{Nagar:2011aa,Bernuzzi:2012ku}.

The EOB radial potential $A(r)$ is taken with the full 4PN-accurate analytical
information augmented by the 5PN logarithmic term~\cite{Damour:2009sm,Blanchet:2010zd,Barausse:2011dq,Bini:2013zaa,Damour:2014yha},
as
\begin{multline}
A^{\rm{PN}}_{\rm{orb}} (u) = 1 - 2 u + 2 \nu u^3 + \nu a_4 u^4 + \nu \bigg[ a^c_5 (\nu) \\
+a^{\rm{log}}_5 \ln u \bigg] u^5 + \nu \bigg[ a^c_6 (\nu) + a^{\rm{log}}_6 \ln u \bigg] u^6,
\end{multline}
where $u\equiv 1/r$. The 4PN and 5PN logarithmic coefficients read
\begin{align}
  a^{\log}_5 &= \frac{64}{5},\\
  a^{\log}_6 (\nu) &= - \frac{7004}{105} - \frac{144}{5} \nu ,
\end{align}
\n
while the 4PN coefficient, $a^c_5 (\nu)$,
is~\cite{Bini:2013zaa}:
\begin{align}
a^c_5 (\nu) &= a^{c0}_5 + \nu a^{c1}_5 , \\
a^{c0}_5 &= - \frac{4237}{60} + \frac{2275}{512} \pi^2 + \frac{256}{5} \ln 2 + \frac{128}{5} \gamma_E , \\
a^{c1}_5 &= - \frac{221}{6} + \frac{41}{32} \pi^2 ,
\end{align}
\n
where $\gamma_E$ is the Euler constant. The 5PN coefficient $a_6^c(\nu)$
is analytically known just at linear order\footnote{In fact, the linear in $\nu$ part of the $A$ function
is analytically known up to 22PN order~\cite{Kavanagh:2015lva}.}
in $\nu$~\cite{Barausse:2011dq}; the other coefficient $a_6^c(\nu)$ 
is here seen as an effective PN parameter that is determined, as usual,
by phasing comparison with NR simulations. Before doing so, 
the PN-expanded radial potential $A(u)$ is Pad\'e resummed as
\begin{align}
A(u ; \nu ; a^c_6) &= P^{1}_{5} \left[ A^{\rm{PN}}_{\rm{orb}} (u) \right] ,
\end{align}
\n
The other potential entering the Hamiltonian, $B(r)$, is instead taken at 3PN
accuracy and is incorporated, as usual, by means of the function
$D \equiv A \, B$ that is resummed by its $P^0_{3}$ version that reads
\begin{align}
D(u) &= \frac{1}{1 + 6 \nu u^2 + 2(26 -3 \nu) \nu u^3} .
\end{align}

\subsection{Resummed waveform and radiation reaction: \\two different multipolar EOB models}
\label{sec:wave}
The structure of the EOB waveform in the nonspinning case is
nowadays standard~\cite{Damour:2008gu}. The strain multipoles
$h_{\ell m}$ are written in factorized and resummed form as
\be
h_{\lm}=h_{\lm}^{(N,\epsilon)}\hat{S}^{(\epsilon)}_{\rm eff} \hat{h}_{\lm}^{\rm tail}(y)\left[\rho^{\rm orb}_\lm(x)\right]^\ell\hat{h}_\lm^{\rm NQC},
\ee
where $\epsilon=(0,1)$ denotes the parity of $\ell+m$,
$\hat{h}_\lm^{(N,\epsilon)}$ is the Newtonian (or leading-order)
contribution to each mode, $\hat{S}_{\rm eff}^{(\epsilon)}$ the effective source
of the field, $\hat{h}_\lm^{\rm tail}(y)$ the tail factor~\cite{Damour:2007xr,Damour:2008gu},
$\rho^{\rm orb}_\lm$ the residual amplitude corrections and $\hat{h}_\lm^{\rm NQC}$
the next-to-quasi-circular (NQC) correction factor, that will be discussed
in Sec.~\ref{sec:nqc} below. The superscript ``orb'' stands for {\it orbital}
and we explicitly write it here to ensure that our notation is consistent with that used
in other work. The tail factor is written as
\be
\hat{h}_\lm^{\rm tail}(y)\equiv T_\lm(y)e^{\rm i\delta_\lm(y)},
\ee
where $y\equiv v_\Omega^2\equiv \Omega^{2/3}$ and the $\delta_\lm(y)$ are the
residual phase corrections that incorporate several test-particle limit terms
and are resummed using Pad\'e approximants following Ref.~\cite{Damour:2012ky}.

The general form of the Newtonian prefactor of the circularized waveform is
\begin{equation}
  \label{eq:hNewt}
h_\lm^{(N,\epsilon)}=\nu c_{\l+\epsilon}(\nu)n_\lm^{(\epsilon)}x^{(\ell+\epsilon)/2} Y^{(\ell-\epsilon,-m)}(\pi/2,\varphi),
\end{equation}
where $Y(\pi/2,\varphi)$ are the scalar spherical harmonics, $n_\lm^{(\epsilon)}$
are parity-dependent constants given in Eqs.~(5)-(6) of Ref.~\cite{Damour:2008gu},
while $c_{\l+\epsilon}(\nu)$ encode the leading-order $\nu$ dependence. For circularized
binaries, $x\equiv v_\Omega^2$ is the frequency parameter. However, as pointed out
long ago~\cite{Damour:2006tr,Damour:2007xr}, the $\ell=m=2$ waveform amplitude during
the plunge is better represented by relaxing the Newtonian Kepler's constraint and
using $x=v_\varphi^2\equiv (r_\omega \Omega)^2$ with $r_\omega\equiv r\psi^{1/3}$,
where $\psi$ is a suitably defined function such that $v_\varphi$ and $r_\omega$
satisfy Kepler's law $1=\Omega^2 r_\omega^3$ during the
adiabatic inspiral. In the remainder of this section, we shall assume that the circular
variable $x$ is always replaced by $v_\varphi^2$, since this is used in our EOB implementation
of the radiation reaction. We shall however introduce exceptions to this rule when discussing
the Newtonian multipolar prefactors in Sec.~\ref{sec:hNewt} below.

The $\rho_\lm^{\rm orb}$ functions implemented in v1.0 of
\TEOBResumS{}~\cite{Nagar:2018zoe,Nagar:2018plt,Akcay:2018yyh} are
taken following the original prescription of Refs.~\cite{Damour:2008gu,Damour:2009kr},
i.e. as truncated PN series at $3^{+2}$~PN order of the form
$\rho_\lm= 1 + c_1^\lm x + c_2^\lm x^2 + \dots$. Here, $3^{+2}$~PN accuracy
denotes that the 3PN-accurate, full $\nu$-dependent, waveform information
has been augmented by test-mass terms such that the PN-order of the PN-expanded
(nonresummed) flux is globally 5PN. In practice, this means that $\rho_{22}(x)$
is given by a fifth-order polynomial in $x$,
i.e. $\rho_{22}(x) = 1+ c_1^{22}(\nu) x + c_2^{22}(\nu) x^2 + c_3^{22}(\nu)x^3 + c_4^{22} x^4 + c_5^{22} x^5$,
where we have explicitly highlighted the $\nu$ dependent terms, while
the subdominant $\rho_\lm$'s are polynomials of progressively lower order
so as to be compatible with global 5PN accuracy, once multiplied by the corresponding
Newtonian prefactors.
In recent years, the analytical knowledge of the test-particle waveform and fluxes
has been pushed to much higher PN orders~\cite{Fujita:2014eta}, notably up to 22PN
in the nonspinning case~\cite{Fujita:2012cm}. It is therefore meaningful to revise and possibly improve
the current choices implemented in \TEOBResumS{}. This was explored by means
of a new factorization and resummation paradigm applied to {\it spinning} waveform
amplitudes introduced in~\cite{Nagar:2016ayt} and recently improved in~\cite{Messina:2018ghh}.
The basic idea behind this approach is to (i) factorize the
orbital (nonspinning) part from the spin-dependent one, so that
each $\rho_\lm(x)$ function is written as the product $\rho_{\lm}\equiv \rho_\lm^{\rm orb}\hat{\rho}_\lm^{\rm S}$,
where each function is of the form $1+\dots$, and (ii) to properly resum each factor.
For example, in Ref.~\cite{Messina:2018ghh} it was proposed to use
Pad\'e approximants for the orbital factor $\rho_{\lm}^{\rm orb}$,
while $\hat{\rho}^S_\lm$ was replaced by its inverse-Taylor representation, 
as suggested in~\cite{Nagar:2016ayt}. In particular, focusing on
the test-mass limit, Ref.~\cite{Messina:2018ghh} pointed out that keeping
each $\rho_\lm$ at 6PN order (i.e. as sixth-order polynomials) yields,
after the resummation strategy described above, an excellent agreement
between the analytical, resummed, residual amplitudes and the exact ones
up to the light-ring, {\it even for} a quasi-extremal Kerr black hole.
As a consequence, we here discuss two different treatments of the orbital $\rho_\lm^{\rm orb}(x)$
that will yield two separate EOB models.
\begin{itemize}
\item[(i)] {\it Fully resummed waveform model}. In this case, we follow the recipe of
  Ref.~\cite{Messina:2018ghh} by first hybridizing the 3PN, $\nu$-dependent,
  terms in the $\rho_\lm$'s with test-particle terms so that the functions
  are globally 6PN accurate. Then, each hybridized $\rho_\lm$ is Pad\'e resummed
  according to the choices of approximants listed in Table~I of~\cite{Messina:2018ghh}.
  This allows us to construct an EOB model whose multipolar waveform amplitudes, 
  and thus also the radiation reaction, are consistent with the choice of the resummed
  amplitudes in the large-mass ratio limit. Essentially the same choices will also be retained 
  in the spinning case, though the factorization paradigm is applied to only the
  $m=\text{odd}$ multipoles up to $\ell=5$. Keeping then the nomenclature of~\cite{Nagar:2016ayt,Messina:2018ghh}
  for consistency, we shall address this waveform model as {\tt TEOBiResumSMultipoles},
  where the {\tt i} prefix refers to the fact that, whenever applied, the spin-dependent
  factors in the waveform are resummed by taking their inverse Taylor expansions.

\item[(ii)] {\it Improved Taylor-expanded waveform model}. In this implementation, we use the
  usual Taylor-expanded expression of the $\rho_\lm^{\rm orb}$ at $3^+2$~PN order, 
  with the only exception being $\rho_{44}^{\rm orb}$, which also incorporates the
  (relative) 5PN test-particle coefficient. More precisely,
  this function reads
  \begin{align}
  \label{eq:rho44}
  \rho_{44}^{\rm orb}(x) &= 1 + c_1^{44}(\nu)x+c_2^{\rm 44}(\nu)x^2 \\
                      &+ c^{44}_3 x^3 + c^{44}_4 x^4 + c^{44}_5 x^5, \nonumber
  \end{align}
  where the coefficient
  \begin{align}
  c_5^{44}&\equiv -\dfrac{17154485653213713419357}{568432724020761600000}\nonumber\\
              &+ \dfrac{22324502267}{3815311500}{\rm eulerlog}_4(x)
  \end{align}
  is omitted in the current version of the \TEOBResumS{} model.
  We have, however, verified that including this term is useful to improve the agreement
  between the $(4,4)$ analytical and numerical waveform amplitudes up to merger.
  In addition to this, another key difference with respect to \TEOBResumS{} relates to 
  the approximation of the
  second time derivative of the radial separation (which enters the NQC factors)
  and a more demanding determination (with respect to NR uncertainties) of
  the effective 5PN parameters $a_6^c$. Though these choices will not be ported
  to the spinning case, they allow us to define a self-consistent, multipolar,
  EOB model for non-spinning binaries that we denote 
  {\tt EOBResumMultipoles$^+$}.

\end{itemize}
The differences in the analytical representation of the $\rho_\lm^{\rm orb}$'s reflect in two
different determinations of the function $a_6^c(\nu)$ by NR/EOB phasing comparisons.
We shall address this issue in Sec.~\ref{sec:nrinfo} below, by relying on SXS waveforms
with reduced phase uncertainty that were not available at the time of Ref.~\cite{Nagar:2015xqa}.

\subsection{Newtonian prefactors in the waveform}
\label{sec:hNewt}
As mentioned above, the standard procedure to improve the behavior
of the Newtonian prefactor of the circularized waveform during the
plunge is to replace $x\to v_\varphi^2$ multipole by
multipole. However, such simple choice makes the amplitude of some
multipoles too small towards merger with respect to the corresponding
NR one. This in turn makes the standard NQC factor (that will be detailed
below) unable to correctly modify the bare EOB multipole. In fact, one
experimentally finds that the NQC amplitude correction factor is particularly
efficient when the peak of the purely analytical EOB waveform amplitude is
larger than the NR one. To implement this condition efficiently, we
then act as follows. Instead of completely replacing $v_\Omega$ with $v_\varphi$,
we just replace some of the powers entering the Newtonian prefactor,
with a choice that depends on the multipole. The aim of such pragmatic
choice is essentially to mimic the effect of the missing noncircular
terms in the specific multipole, and help the action of the NR-informed
NQC factor. In practice we found that the following choices best recover
NR amplitudes:
\begin{align}
  h_{22}^{(N,0)} &= -8\sqrt{\dfrac{\pi}{5}} \nu\,v_\varphi^2 e^{-2\ii\varphi},\\
  h_{21}^{(N,1)} &= -\dfrac{8{\rm i}}{3}\sqrt{\dfrac{\pi}{5}} \nu\sqrt{1-4\nu}\,v_\varphi^3 e^{-\ii\varphi},\\
  h_{33}^{(N,0)}&=3\ii\sqrt{\dfrac{6\pi}{7}}\nu\sqrt{1-4\nu}\,v_\varphi v_\Omega^2 e^{-3\ii \varphi},\\
  h_{32}^{(N,1)}&=\dfrac{8}{3}\sqrt{\dfrac{\pi}{7}}\nu(1-3\nu)\,v_\varphi^2 v_\Omega^2 e^{-2\ii \varphi},\\
  h_{31}^{(N,0)}&=-\dfrac{\ii}{3}\sqrt{\dfrac{2\pi}{35}}\nu\sqrt{1-4\nu}v_\Omega^3 e^{-\ii\varphi},\\
  h_{44}^{(N,0)}&=-\dfrac{64}{9}\sqrt{\dfrac{\pi}{7}} \nu(1-3\nu)\,v_\varphi^2v_\Omega^2 e^{-4 \ii \varphi},\\
  h_{43}^{(N,1)}&-\dfrac{9\ii}{5}\sqrt{\dfrac{2\pi}{7}}\nu(2\nu-1)\sqrt{1-4\nu}\,v_\varphi^2v_\Omega^3e^{-3\ii\varphi},\\
  h_{42}^{(N,0)}&=\dfrac{8\sqrt{\pi}}{63}\nu(1-3\nu)\, v_\varphi v_\Omega^3e^{-2\ii \varphi},\\
  h_{41}^{(N,1)}&=\dfrac{\ii}{105}\sqrt{2\pi}\nu(2\nu-1)\sqrt{1-4\nu}\,v_\Omega^5e^{-\ii\varphi},\\
  h_{55}^{(N,0)}&=\dfrac{125\ii}{12}\sqrt{\dfrac{5\pi}{66}}\nu(2\nu-1)\sqrt{1-4\nu}v_\Omega^4v_\varphi e^{-5\ii \varphi}.
\end{align}
The other Newtonian prefactors in the EOB waveform are obtained replacing $x=v_\varphi^2$
in Eq.~\eqref{eq:hNewt}. Note however that, to reduce to the minimum the modifications
with respect to the standard version of \TEOBResumS{}, we only modified the Newtonian
prefactor that enters the waveform, while keeping untouched the corresponding quantity
in the flux. In addition, as in previous work, the NQC correction factor is applied
{\it only} to the $(2,2)$ flux and not to the other modes. This is done for simplicity,
although, for consistency, the flux should be modified consistently with the waveform.
Previous work~\cite{Damour:2012ky} explored the effect of incorporating also the
$(2,1)$ and $(3,3)$ NQC corrections in the radiation reaction. The result of this choice
was the need of determining new values of the parameters entering the EOB interaction
potential, i.e. new values of $a_6^c$. Given the effective character of this quantity,
we do not think, at this stage, that it is worth increasing the complexity of the model.
If the need arises (e.g. to increase the consistency between the NR and EOB fluxes
up to merger), it will be straightforward to modify the model so as to take this into account.

\subsection{Next-to-Quasi-Circular waveform corrections}
\label{sec:nqc}
Let us turn now to the discussion of the NQC correction factor to the multipolar
wavefoms $\hat{h}_\lm^{\rm NQC}$. For each $(\ell,m)$, it reads
\begin{align}
  \label{eq:hlmNQC}
\hat{h}_{\ell m}^{\mathrm{NQC}}= \left( 1+ a_1^{\ell m} n^\lm_1 + a_2^\lm n^\lm_2\right) e^{{\rm i}\left(b_1^\lm n^\lm_3+b_2^\lm n^\lm_4\right)},
\end{align}
where each multipole is characterized by 4 parameters $\left(a_1^\lm,a_2^\lm,b_1^\lm,b_2^\lm\right)$
and four functions $(n_1^\lm,n_2^\lm,n_3^\lm,n_4^\lm)$ that explicitly depend on the
radial momentum and on the radial acceleration. The choice of these functions
can, in principle, depend on the multipole. We implement a few modifications
with respect to previous EOB works. Let us see this in detail.

For the $\ell=m=2$ mode, the NQC functions read~\cite{Damour:2014sva}
\begin{align}
n^{22}_1 & = \left(\frac{p_{r_\star}}{r\Omega}\right)^2, \\
n^{22}_2 & = \frac{\underline{\ddot{r}}^{(0)}}{r\Omega^2},\\
n^{22}_3 & = \frac{p_{r_\star}}{r\Omega},\\
n^{22}_4 & = (r\Omega) p_{r_*}\ .
\end{align}
Here $\underline{\ddot{r}}^{(0)}$ is an approximation to $\ddot{r}^{(0)}$,
the second time-derivative of the radial separation along the conservative
dynamics, that is obtained by neglecting the contributions proportional
to the radiation reaction $\mathcal{F}_\varphi$~\cite{Damour:2014sva}. 
This quantity reads 
\be
\ddot{r}^{(0)}=\dot{p}_{r_*} \dfrac{\partial\dot{r}}{\partial{p_{r_*}}} + \dot{r}\partial_r\dot{r},
\ee
and the additional approximation we do is to neglect the second term
$\dot{r}\partial_r\dot{r}$, so to define
\be
\underline{\ddot{r}}^{(0)}\equiv \dot{p}_{r_*} \dfrac{\partial\dot{r}}{\partial{p_{r_*}}}.
\ee
The reason for doing so is that this new function has a milder growth
towards merger than the complete one and was found to be more robust
in the spinning case~\footnote{Note that this was originally implemented
  in this form in the spinning sector of \TEOBResumS{}, while $\ddot{r}^{(0)}$
  was kept in the nonspinning sector.}.

For the $\ell=2$, $m=1$ mode, one pragmatically finds that
a slightly different basis delivers a more controllable
behavior of the correcting factor, that reads
\begin{align}
  n_{1}^{21} & = n_1^{22},\\
  n_{2}^{21} & = n_1^{21} \Omega^{2/3},\\
  n_3^{21}  & = n_3^{22},\\
  n_4^{21}  & = n_3^{21}\Omega^{2/3}.
\end{align}
For all other modes with $\ell\geq 3$, one simply uses
\begin{align}
n_1^{\lm} &= n_1^{22},\\
n_2^{\lm} &= n_2^{22},\\
n_3^{\lm} &= n_3^{22},\\
n_4^\lm  & = n_3^{22}\Omega^{2/3}.
\end{align}
The determination of the parameters $(a_1^\lm,a_2^\lm,b_1^\lm,b_2^\lm)$
is achieved by matching the NQC-modified waveform multipole
to the corresponding NR one at a specified, $\nu$-dependent-time,
precisely by imposing there a $C^2$ contact condition between the
EOB and NR the amplitudes and frequencies~\cite{Damour:2012ky,Damour:2014sva}. 
The determination of the NQC parameters relies on the need of
connecting the EOB time axis with the NR time axis, in correspondence
to the point where the waveform information needed to determine the
NQC is extracted from the NR multipolar waveform.
The important point on the EOB time axis is defined by the
{\it peak of the orbital frequency} $\Omega$.
For the $\ell=m=2$ mode, the NQC determination time is chosen as 
\begin{align}
\label{eq:DtNQC}
t^{\rm EOB}_{\rm NQC}\equiv t^{\rm peak}_{\Omega}-\Delta t_{\rm NQC}
\end{align}
with $\Delta t_{\rm NQC}=1$, which is identified $2$ dimensionless
time units after the peak of the $(\ell,m)=(2,2)$ mode:
\begin{align}
t^{\rm EOB}_{\rm NQC} \leftrightarrow t^{\rm NR}_{\rm NQC}\equiv t^{\rm NR}_{A_{22}^{\rm max}} + 2.
\end{align}
This prescription was elaborated and tested in previous 
works~\cite{Damour:2012ky,Nagar:2015xqa,Nagar:2017jdw,Nagar:2018zoe}
and then we give it here without additional explanations. As a consequence, 
we proceed in the same  way for the other modes. More precisely, for 
each multipole the NQC extraction point is taken to be $2$ dimensionless 
time units after the peak of the corresponding multipole, that is
\be
\label{tNQC_extr}
t^{\rm NR}_{\rm NQC-\lm}\equiv t^{\rm NR}_{A_\lm^{\rm max}}+2.
\ee
Finally, on the EOB time-axis one has to locate the peak of each
multipole with respect to the peak of the $\ell=m=2$ mode, that is
\be
t^{\rm EOB}_{A_{22}^{\rm max}}\equiv t_{\rm NQC}^{\rm EOB}-2,
\ee
while the peaks of the other modes on the EOB time axes are located
at
\be
t^{\rm EOB}_{A_\lm^{\rm max}}\equiv t^{\rm EOB}_{A_{22}^{\rm max}}+\Delta t_\lm^{\rm NR},
\ee
where
\be
\Delta t_\lm^{\rm NR}\equiv t^{\rm peak}_{\lm}-t_{22}^{\rm peak}
\ee
are the delays of the peak of the subdominant modes with respect to
the peak of the $\ell=m=2$ mode, that can be accurately fitted from
NR simulation data, see Eqs.~\eqref{eq:Dt_peak_lm}
and~\eqref{eq:Dt_hat_peak_lm} below.
The NQC parameters are determined by imposing the following conditions between
the EOB multipoles and the corresponding NR values
\begin{align}
A^{\rm EOB}_{\ell m}\left(t^{\rm EOB}_{\rm NQC}+\Delta t_{\ell m}^{\rm NR} \right) & = A^{\rm NR}_{\ell m}\left(t^{\rm NR}_{\rm NQC-\lm}\right),\\
\dot{A}^{\rm EOB}_{\ell m}\left(t^{\rm EOB}_{\rm NQC,}+\Delta t_{\ell m}^{\rm NR} \right) & = \dot{A}^{\rm NR}_{\ell m}\left(t^{\rm NR}_{\rm NQC-\lm}\right),\\
\omega^{\rm EOB}_{\ell m}\left(t^{\rm EOB}_{\rm NQC}+\Delta t_{\ell m}^{\rm NR} \right) & = \omega^{\rm NR}_{\ell m}\left(t^{\rm NR}_{\rm NQC-\lm}\right),\\
\dot{\omega}^{\rm EOB}_{\ell m}\left(t^{\rm EOB}_{\rm NQC}+\Delta t_{\ell m}^{\rm NR} \right) & = \dot{\omega}^{\rm NR}_{\ell m}\left(t^{\rm NR}_{\rm NQC-\lm} \right).
\end{align}
This set of equations is solved for $\left\lbrace a_1^{\ell m},a_2^{\lm},b_1^{\lm},b_2^{\lm} \right\rbrace$. 
As the coefficients $a_j^{\ell m}$ only affect the modulus of the waveform, they will implicitly
affect the computation of radiation reaction force, modifying the EOB dynamics. In order to
enforce consistency of the radiation reaction terms, the NQC parameters are iteratively
determined until convergence at a given tolerance is reached. Typically this method
requires on the order of 3 iterations before an acceptable level of convergence is achieved.
Note that in this procedure the only arbitrariness is in having fixed $\Delta t_{\rm NQC}=1$ in
Eq.~\eqref{eq:DtNQC} above. This is inspired by the time-delay between the the peak of the
orbital frequency and the peak of the $\ell=m=2$ waveform, that is $\simeq 2.56$
(see Fig.~4 of Ref.~\cite{Damour:2012ky}) so that the same structure is approximately
preserved also in the comparable mass case.

\section{NR-informing the EOB dynamics: $a_6^c(\nu)$}
\label{sec:nrinfo}

We proceed now with a new determination of $a_6^c$. To do so,
we only used SXS waveforms that have the smallest nominal
uncertainty. Part of these dataset was not publicly available
at the time of Ref.~\cite{Nagar:2015xqa}. Particularly useful
to this aim are datasets SXS:BBH:0169, SXS:BBH:0259, SXS:BBH:0297 and
SXS:BBH:0302, that have a rather small 
numerical phase uncertainty at merger, estimated taking the difference
between the two highest resolutions,  see Table~\ref{tab:a6c}.
The final outcome of this analysis is that the current,
NR-informed, espression for $a_6^c(\nu)$ used in \TEOBResumS{}
\be
\label{eq:a6c_old}
a_6^c (\nu) = 3097.3 \nu^2 - 1330.6 \nu + 81.38,
\ee
that was obtained in Ref.~\cite{Nagar:2015xqa} and never changed since
then, will be replaced by two different analytical expressions, one for
each choice of waveform amplitude resummation. Each choice of $a_6^c(\nu)$
will define a different, multipolar, EOB model.
Of the several nonspinning waveform at our disposal, only the $7$ listed
in Table~\ref{tab:a6c} are needed to do so. The last column of the table
lists the nominal phase uncertainty at merger of each dataset. This number
corresponds to the phase difference between the highest and second highest
resolutions evaluated at the merger time of the highest resolution waveform.
\begin{table}
  \caption{\label{tab:a6c} Determination of $a_6^c$. Best values of $a_6^c$ to obtain a EOB/NR
    dephasing at merger of the order of the nominal numerical error, $\delta\phi^{\rm NR}_{\rm mrg}$,
    with the two choices of radiation reaction (Pad\'e resummed or improved Taylor-expanded).
    These values are then fitted in one case with Eq.~\eqref{eq:a6c_pade}
    and in the other by Eq.~\eqref{eq:a6c_taylor}. The differences between the two effective
    representations of $a_6^c(\nu)$ are illustrated in Fig.~\ref{fig:a6c}.}
  \begin{ruledtabular}
    \begin{tabular}{ccccc}
      ID & $q$  & $a^6_c$ [Pad\'e]& $a^6_c$ [Taylor] & $\delta\phi^{\rm NR}_{\rm mrg}$ [rad] \\
      \hline
      \hline
      SXS:BBH:0002 & 1.00   & $-42$ & $-47$    & $-0.063$ \\
      SXS:BBH:0007 & 1.50    & $-47$ & $-52$    & $-0.0186$ \\
      SXS:BBH:0169 & 2.00   &  $-59$ & $-63$   & $-0.0271$\\
      SXS:BBH:0259 & 2.50   & $-54$  & $-58$   & $-0.0080$ \\
      SXS:BBH:0030 & 3.00   & $-52$ & $-57$   & $-0.0870$ \\
      SXS:BBH:0297 & 6.50   & $-27$ & $-36$   & $-0.053$ \\
      SXS:BBH:0298 & 7.00   & $-26$ & $\dots$ & $-0.0775$ \\
      SXS:BBH:0302 & 9.50   & $-17$ & $-25$   & $+0.0206$

    \end{tabular}
  \end{ruledtabular}
  \end{table}

\begin{figure}[t]
\center
\includegraphics[width=0.4\textwidth]{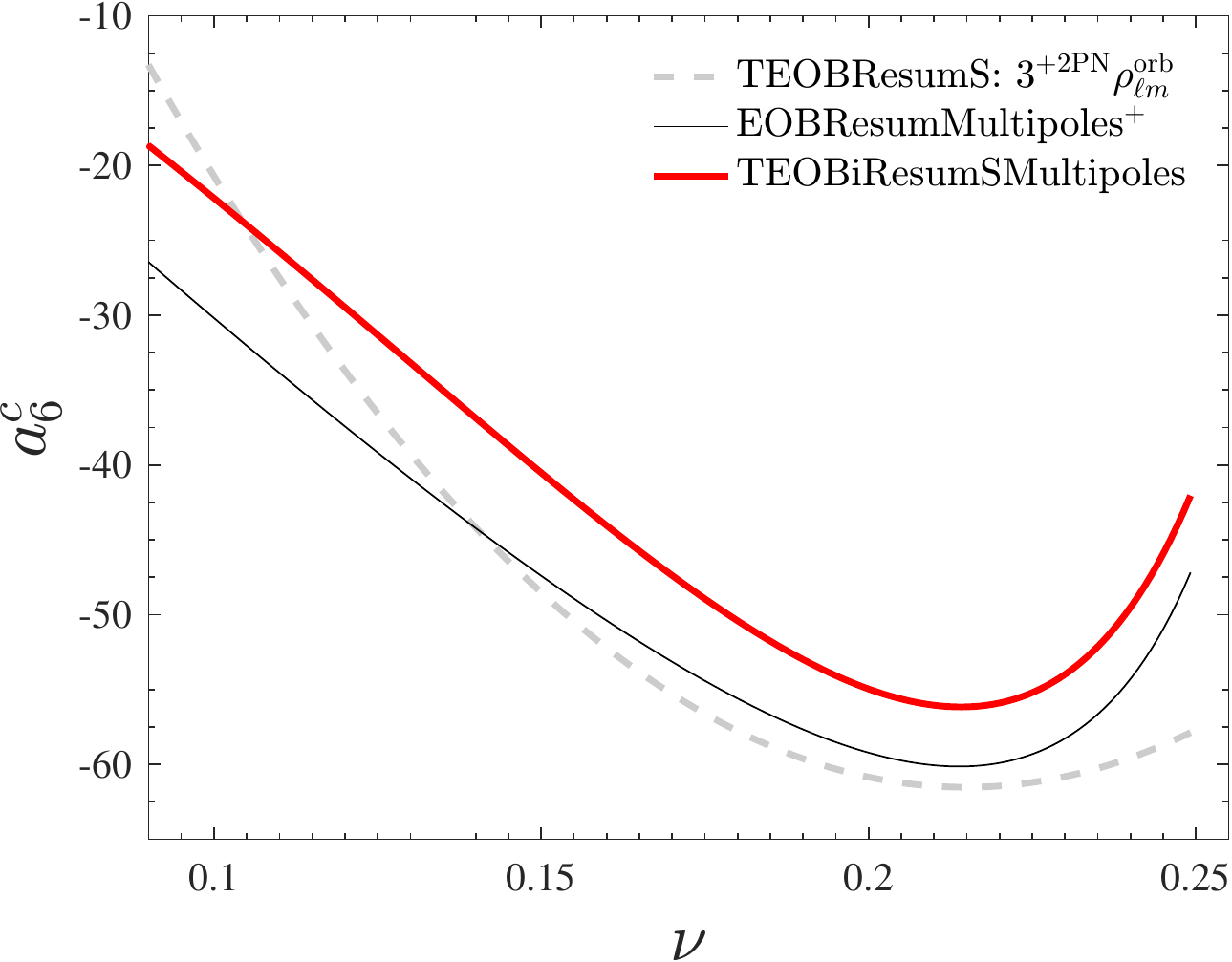}
\caption{\label{fig:a6c}Effective 5PN function $a_6^c(\nu)$ entering the EOB interaction potential.
  The previous \TEOBResumS{} function (dashed, grey line) determined
  in Ref.~\cite{Nagar:2015xqa} is contrasted with the two new determinations
  of $a_6^c$ corresponding to the \TEOBiResumSM{} or to the \EOBResumM{} choices.
  The difference is due to changes in the radiation reaction as well as
  to the reduced nominal error bars of the NR data used.}
\end{figure}

\subsubsection{\TEOBiResumSM{}}
The best values of $a_6^c$ determined by EOB/NR phasing comparison
are listed in the third column of Table~\ref{tab:a6c}.
To fit them properly we use the following rational function
\be
\label{eq:a6c_pade}
a_6^c =n_0\dfrac{1+n_1 \nu + n_2 \nu^2 + n_3 \nu^3}{1+d_1\nu},
\ee
where the parameters are determined to be
\begin{align}
  n_0 &= \;\;\;5.9951,\\
  n_1 &= -34.4844,\\
  n_2 &= -79.2997,\\
  n_3 &= \;\;\;713.4451,\\
  d_1 &= -3.167.
\end{align}

\subsubsection{\EOBResumM{}}
With the $3^{+2}$~PN accurate, PN-expanded, version of the $\rho_\lm$'s,
the best values of $a_6^c$ yielding an EOB/NR phase difference compatible
with the NR uncertainty are listed in the fourth column of Table~\ref{tab:a6c}.
These numbers can be accurately fitted with a rational function of the form
\be
\label{eq:a6c_taylor}
a_6^c = n_0\dfrac{1+n_1 \nu + n_2 \nu^2}{1+d_1\nu},
\ee
where
\begin{align}
  n_0 &= \;\;\;9.3583,\\
  n_1 &= -47.4566,\\
  n_2 &= \;\;\;162.4461,\\
  d_1 &= -3.425.
\end{align}
\begin{table*}
	\caption{\label{tab:NR_Waveforms} 
	Properties of the 19 non-spinning NR SXS waveforms used to inform the postpeak waveform fits
	that complete the EOB multipolar waveform. From left to right, the columns report: the
	identification number; the SXS classification name; the mass ratio $q\equiv m_1/m_2\geq 1$;
	the symmetric mass ratio $\nu$; the number of orbits 
	between $t=0$ and the time when a common event horizon is formed; the orbital
        eccentricity $\epsilon$ determined at relaxed measurement time;
        the $\ell=m=2$ phase-difference $\delta\phi^{\rm NR}_{\rm mrg}$ between the highest
        and second highest resolution accumulated between $t=600M$ and the peak amplitude
        of the highest resolution data; the highest resolution available ${\rm Lev_h}$
        and the second-highest resolution available ${\rm Lev_l}$.}
\begin{ruledtabular}
\begin{tabular}{r | c | c c | c c c | c c}
\# & name & $q$ & $\nu$ & $\#$ orbits & $\epsilon\left[10^{-3}\right]$ & $\delta\phi^{\rm NR}_{\rm mrg}\left[{\rm rad}\right]$ & ${\rm Lev_h}$ & ${\rm Lev_l}$\\\hline
$1$ & SXS:BBH:0180 & $1$\;\;\, & $0.2500$ & $28.1825$ & $0.0514$ & $-0.4247$ & Lev4 & Lev3 \\
$2$ & SXS:BBH:0007 & $1.5$ & $0.2400$ & $29.0922$ & $0.4200$ & $-0.0186$ & Lev6 & Lev5 \\
$3$ & SXS:BBH:0169 & $2$\;\;\, & $0.2222$ & $15.6805$ & $0.1200$ & $-0.0271$ & Lev5 & Lev4 \\
$4$ & SXS:BBH:0259 & $2.5$ & $0.2041$ & $28.5625$ & $0.0590$ & $-0.0080$ & Lev5 & Lev4 \\
$5$ & SXS:BBH:0030 & $3$\;\;\, & $0.1875$ & $18.2228$ & $2.0100$ & $-0.0870$ & Lev5 & Lev4 \\
$6$ & SXS:BBH:0167 & $4$\;\;\, & $0.1600$ & $15.5908$ & $0.0990$ & $-0.5165$ & Lev5 & Lev3 \\
$7$ & SXS:BBH:0295 & $4.5$ & $0.1488$  & $27.8067$ & $0.0520$ & $+0.2397$ & Lev5 & Lev4 \\
$8$ & SXS:BBH:0056 & $5$\;\;\, & $0.1389$  & $28.8102$ & $0.4900$ & $+0.4391$ & Lev5 & Lev4 \\
$9$ & SXS:BBH:0296 & $5.5$ & $0.1302$ &  $27.9335$ & $0.0520$ & $+0.4427$ & Lev5 & Lev4 \\
$10$ & SXS:BBH:0166 & $6$\;\;\, & $0.1224$  & $21.5589$ & $0.0440$ & \dots & Lev5 & \dots \\
$11$ & SXS:BBH:0297 & $6.5$ & $0.1156$  & $19.7111$ & $0.0640$ & $-0.0543$ & Lev5 & Lev4 \\
$12$ & SXS:BBH:0298 & $7$\;\;\, & $0.1094$  & $19.6757$ & $0.0610$ & $-0.0775$ & Lev5 & Lev4 \\
$13$ & SXS:BBH:0299 & $7.5$ & $0.1038$  & $20.0941$ & $0.0590$ & $-0.0498$ & Lev5 & Lev4 \\
$14$ & SXS:BBH:0063 & $8$\;\;\, & $0.0988$ & $25.8255$ & $0.2800$ & $+1.0094$ & Lev5 & Lev4 \\
$15$ & SXS:BBH:0300 & $8.5$ & $0.0942$ & $18.6953$ & $0.0570$ & $-0.0804$ & Lev5 & Lev4 \\
$16$ & SXS:BBH:0301 & $9$\;\;\, & $0.0900$ & $18.9274$ & $0.0550$ & $-0.1641$ & Lev5 & Lev4 \\
$17$ & SXS:BBH:0302 & $9.5$ & $0.0862$ & $19.1169$ & $0.0600$ & $+0.0206$ & Lev5 & Lev4 \\
$18$ & SXS:BBH:0185 & \;\,$9.99$ & $0.0827$ & $24.9067$ & $0.3055$ & $+0.3714$ & Lev3 & Lev2 \\
$19$ & SXS:BBH:0303 & $10$\;\;\;\;\; & $0.0826$ & $19.2666$ & $0.0510$ & $+0.2955$ & Lev5 & Lev4 \\
\end{tabular}
\end{ruledtabular}
\end{table*}
Note that in this case a function with only 4 parameters is sufficient
and we didn't use the $q=7$ data.
Figure~\ref{fig:a6c} illustrates the differences between the NR-informed
function $a_6^c(\nu)$ of Eq.~\ref{eq:a6c_old} from Ref.~\cite{Nagar:2015xqa},
implemented in \TEOBResumS{}, and the two new ones~\eqref{eq:a6c_pade}-\eqref{eq:a6c_taylor}.
Two things are noticeable: (i) the need to comply with the rather small
error bars in the NR data forces  $a_6^c(\nu)$ to be {\it essentially linear}
as $\nu\to 0$. Such qualitative behavior is consistent with the analytical
expectation, though the $\nu=0$ numerical values obtained from the fits are not.
In fact, in one case the limit is $\simeq 6$ and in the other is $\simeq 9$,
both very different from the value of $a_6^c(0)$ known analytically
from gravitational self-force (GSF) calculations~\cite{Barausse:2011dq,Bini:2013rfa},
$a_6^c(0)\simeq -134.072$ (see Eq.~(67a) of~\cite{Bini:2013rfa}).
This difference is not surprising seen that (i) our $a_6^c$ is an effective
parameter that enters a Pad\'e approximant and (ii) that it {\it does} depend
(as we showed) on the analytical choices made to construct the radiation
reaction and on the details of the NQC correction factors~\footnote{As a side remark, 
we note that we also NR-informed the model using $\ddot{r}^{(0)}$
instead of $\underline{\ddot{r}^{(0)}}$, which yields another (though qualitatively similar)
determination of $a_6^c$ that is around 15 for $\nu=0$.}.
Clearly, seen the effective nature of $a_6^c(\nu)$, the nonspinning
models that we are constructing are not expected to give
a faithful representation of the true physics for large mass-ratio binaries
(e.g. extreme-mass-ratio inspirals) because of the lack of the correct
linear-in-$\nu$ analytical information in the interaction potential.
Evidently, this is not a {\it conceptual} issue, since, in principle,
high-order analytical information could be incorporated in the $A$
function that could then be resummed accordingly.
A dedicated investigation is needed to assess whether such GSF-augmented
$A$ function would improve the agreement with NR data as is or it (still)
would need to be additionally informed by some other effective parameter,
e.g. like the 5PN coefficient in the $A$ potential proportional to $\nu^2$.
As a positive final note, we shall check below that our NR-informed
determinations of $a_6^c(\nu)$ are still sufficiently valid for
$\nu\simeq 0.0499$ ($q=18$), as they yield an excellent EOB/NR
waveform agreement with a (relatively short) $q=18$ NR waveform
obtained with the {\tt BAM} code.

\section{Multipolar ringdown waveform}
\label{sec:postmerger}

\begin{figure*}[t]
  \center
  \includegraphics[width=0.22\textwidth]{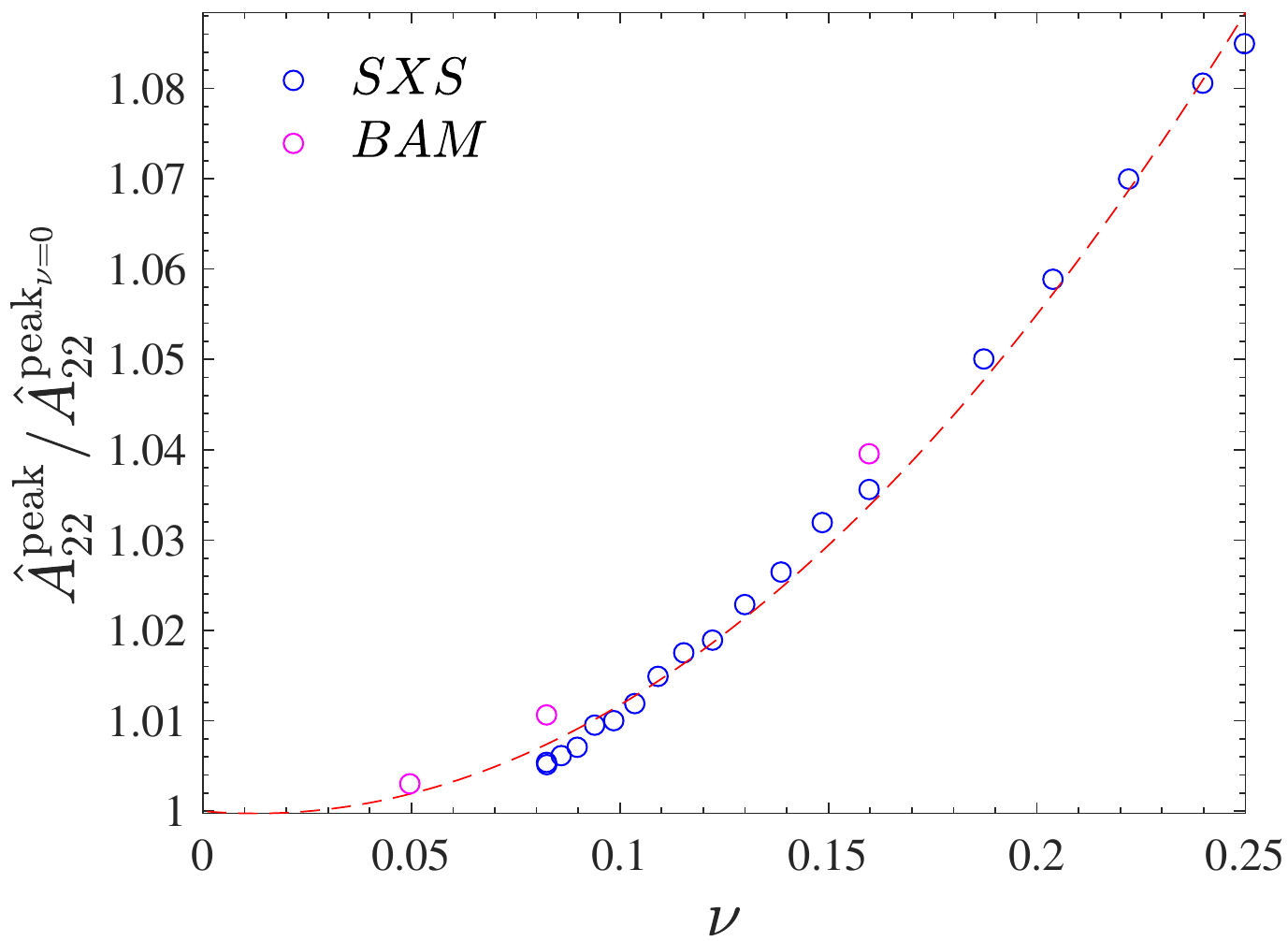}
  \hspace{1mm}
  \includegraphics[width=0.22\textwidth]{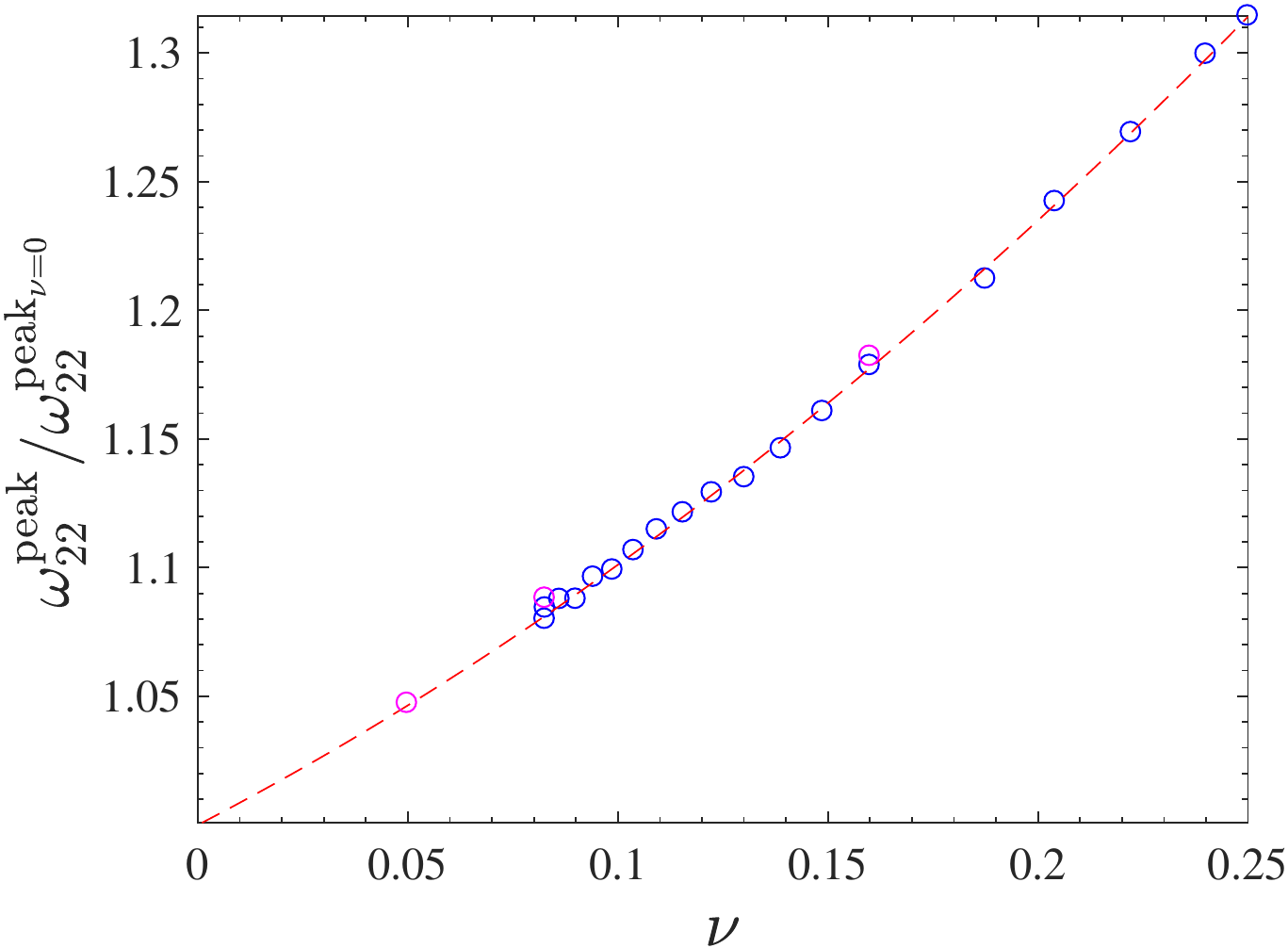}
  \hspace{1mm}
  \includegraphics[width=0.22\textwidth]{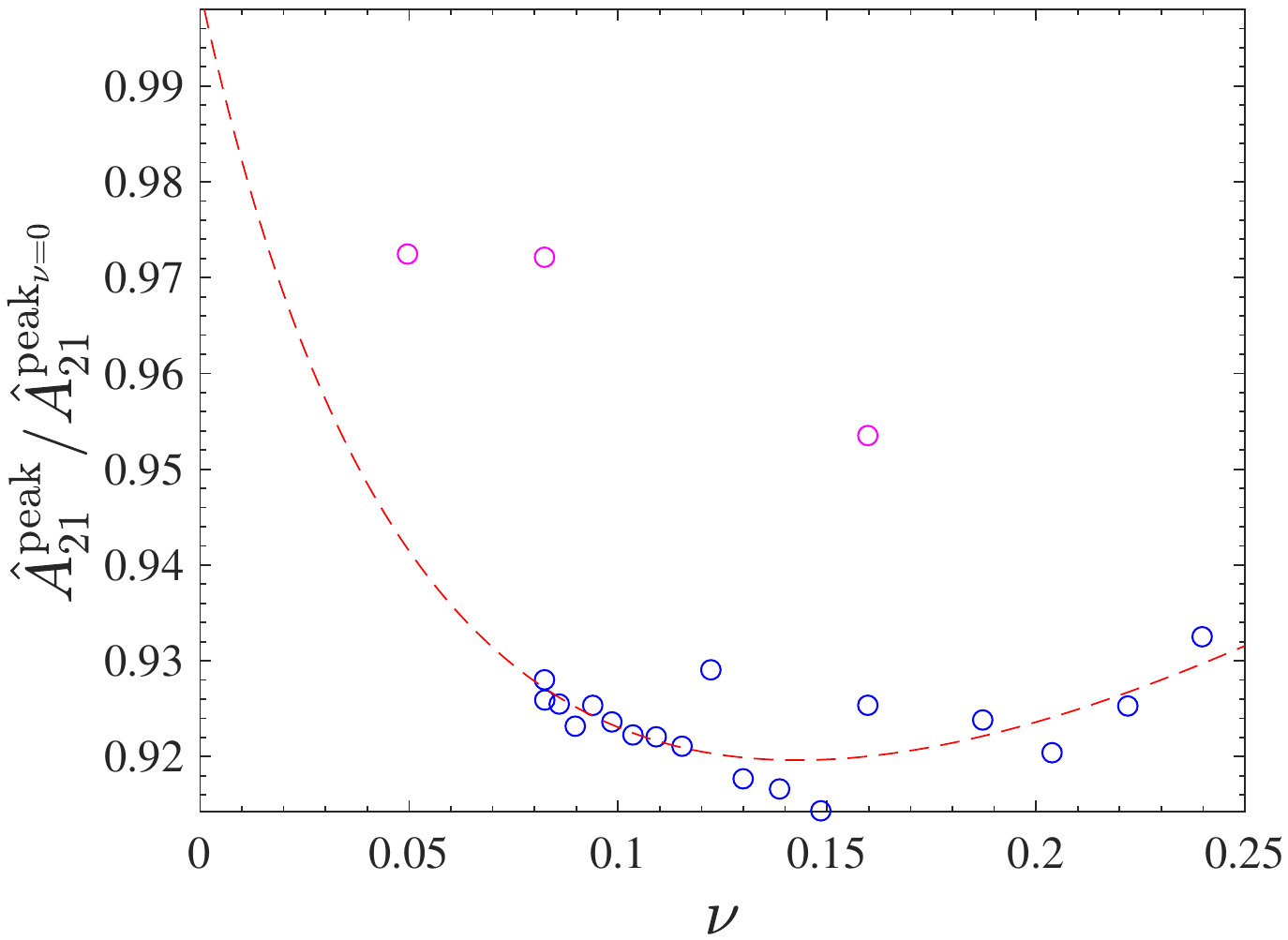}
  \hspace{1mm}
  \includegraphics[width=0.22\textwidth]{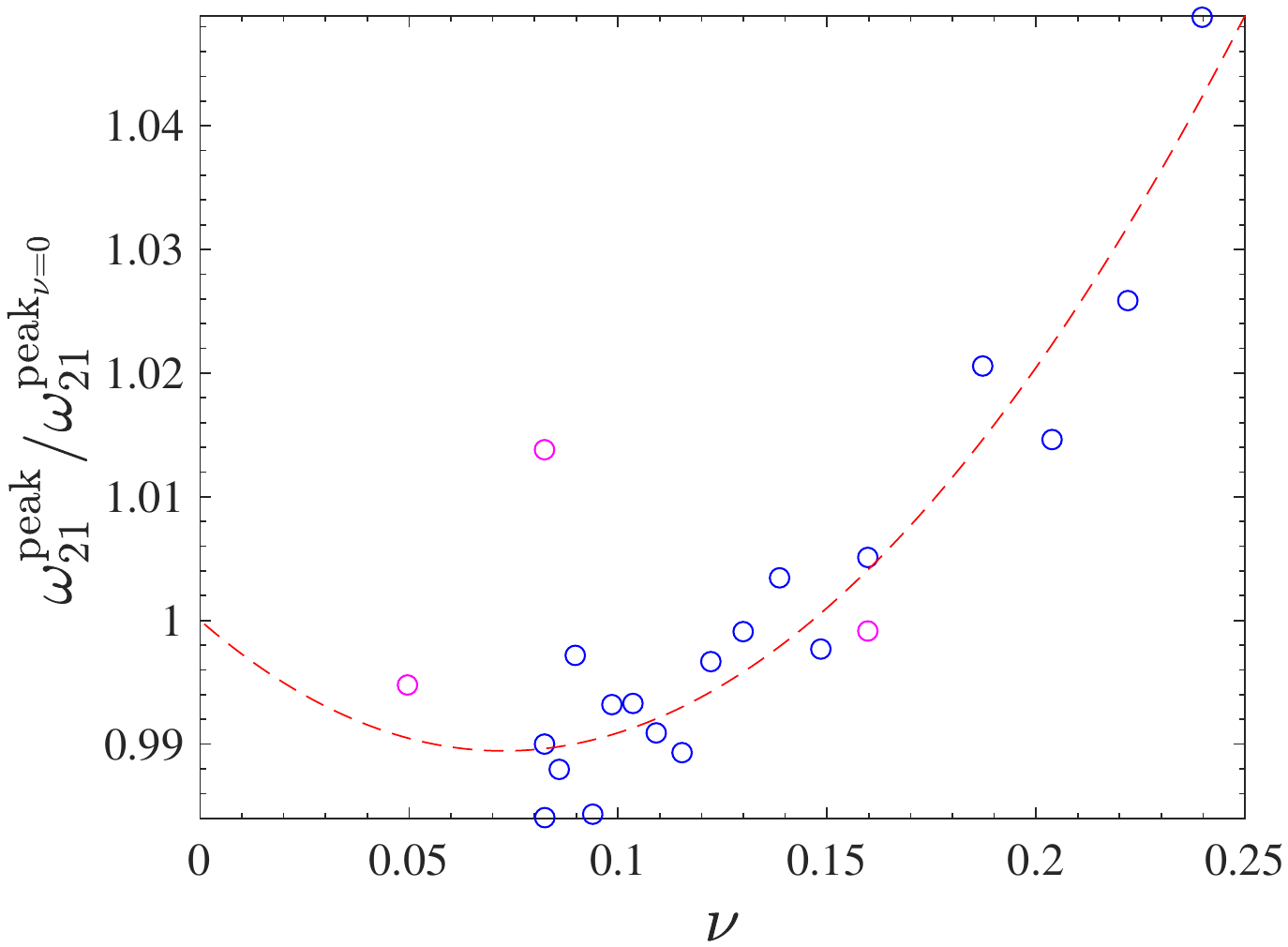}\\
  \includegraphics[width=0.22\textwidth]{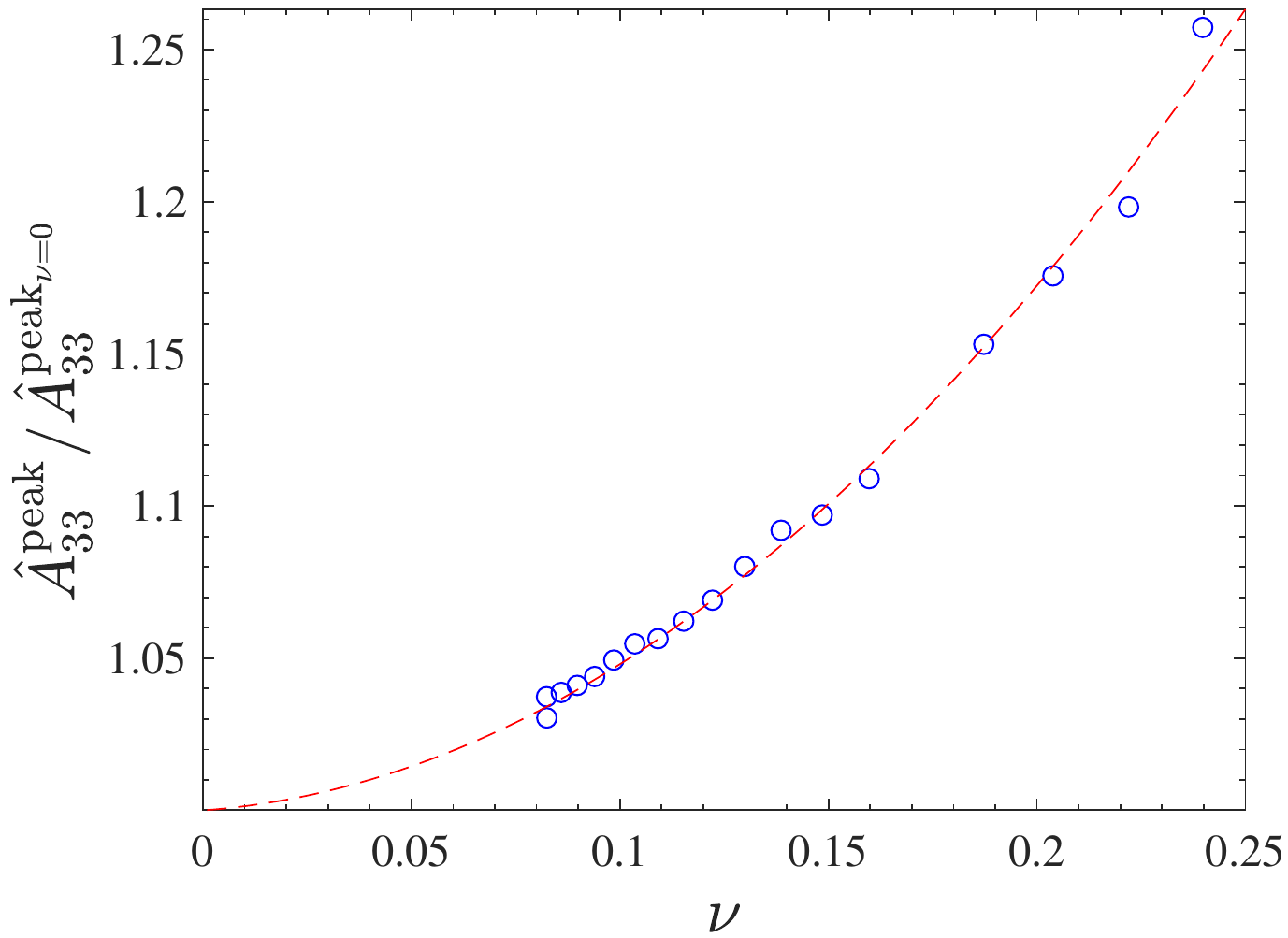}
  \hspace{1mm}
  \includegraphics[width=0.22\textwidth]{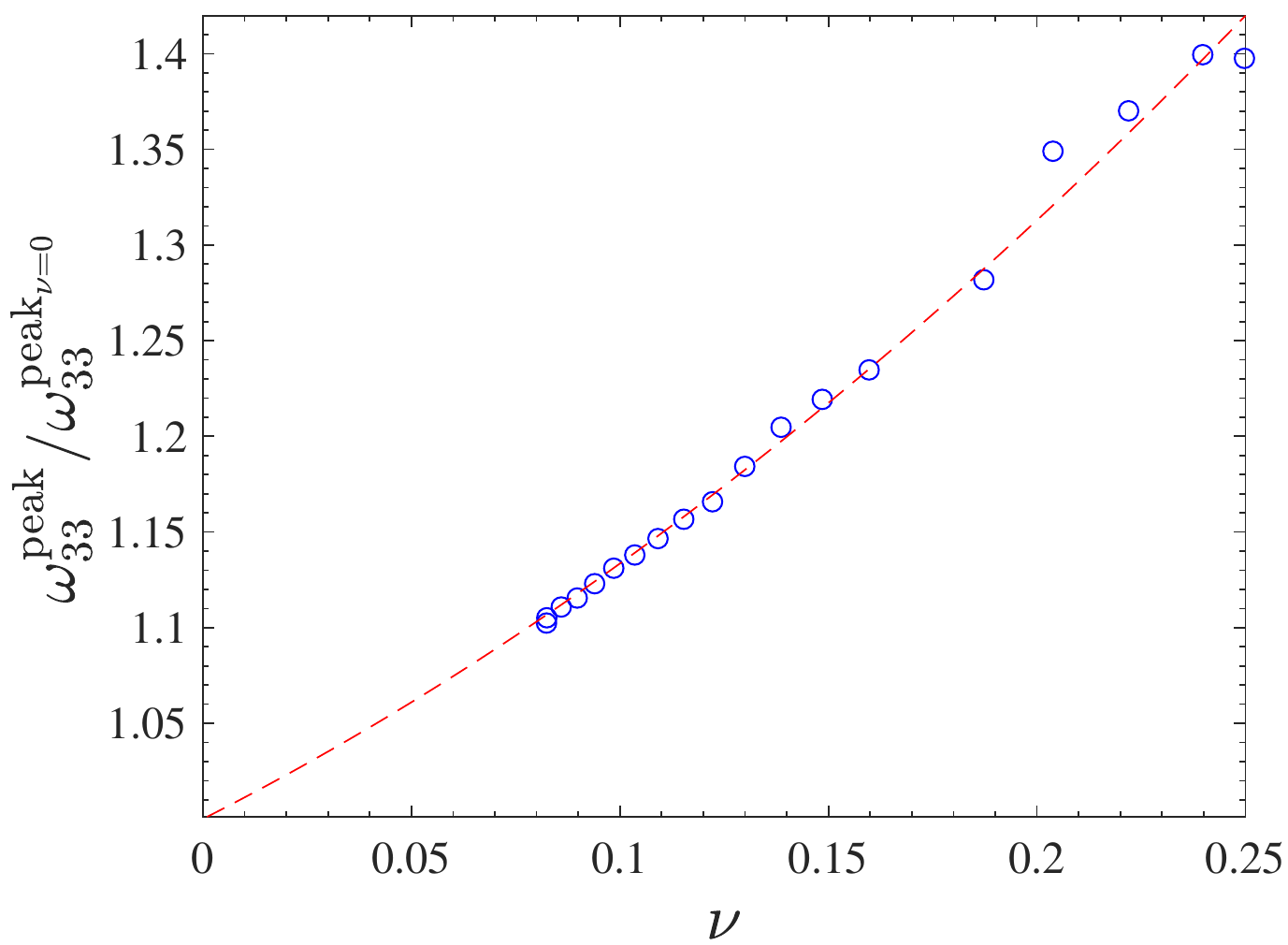}
  \hspace{1mm}
  \includegraphics[width=0.22\textwidth]{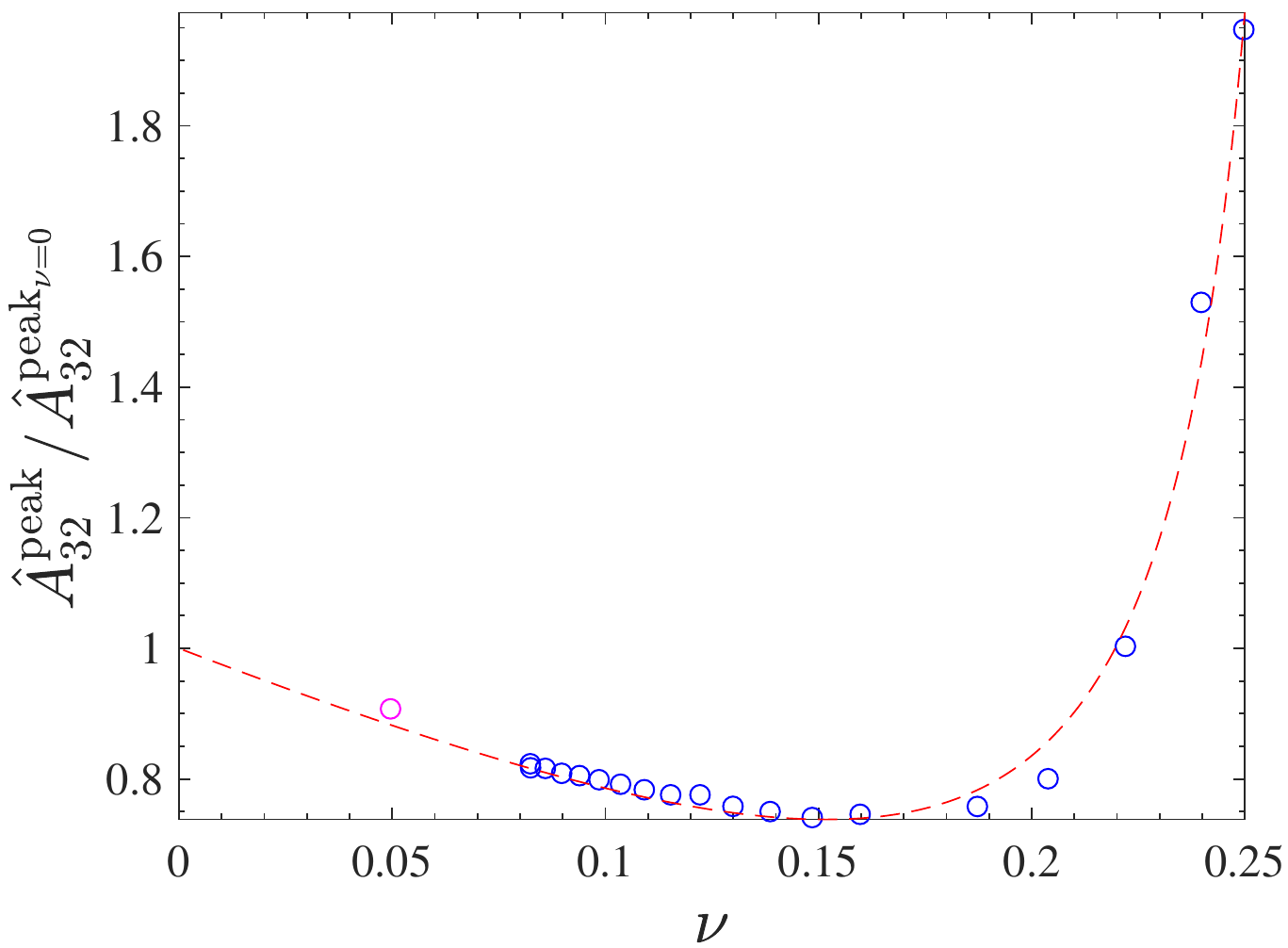}
  \hspace{1mm}
  \includegraphics[width=0.22\textwidth]{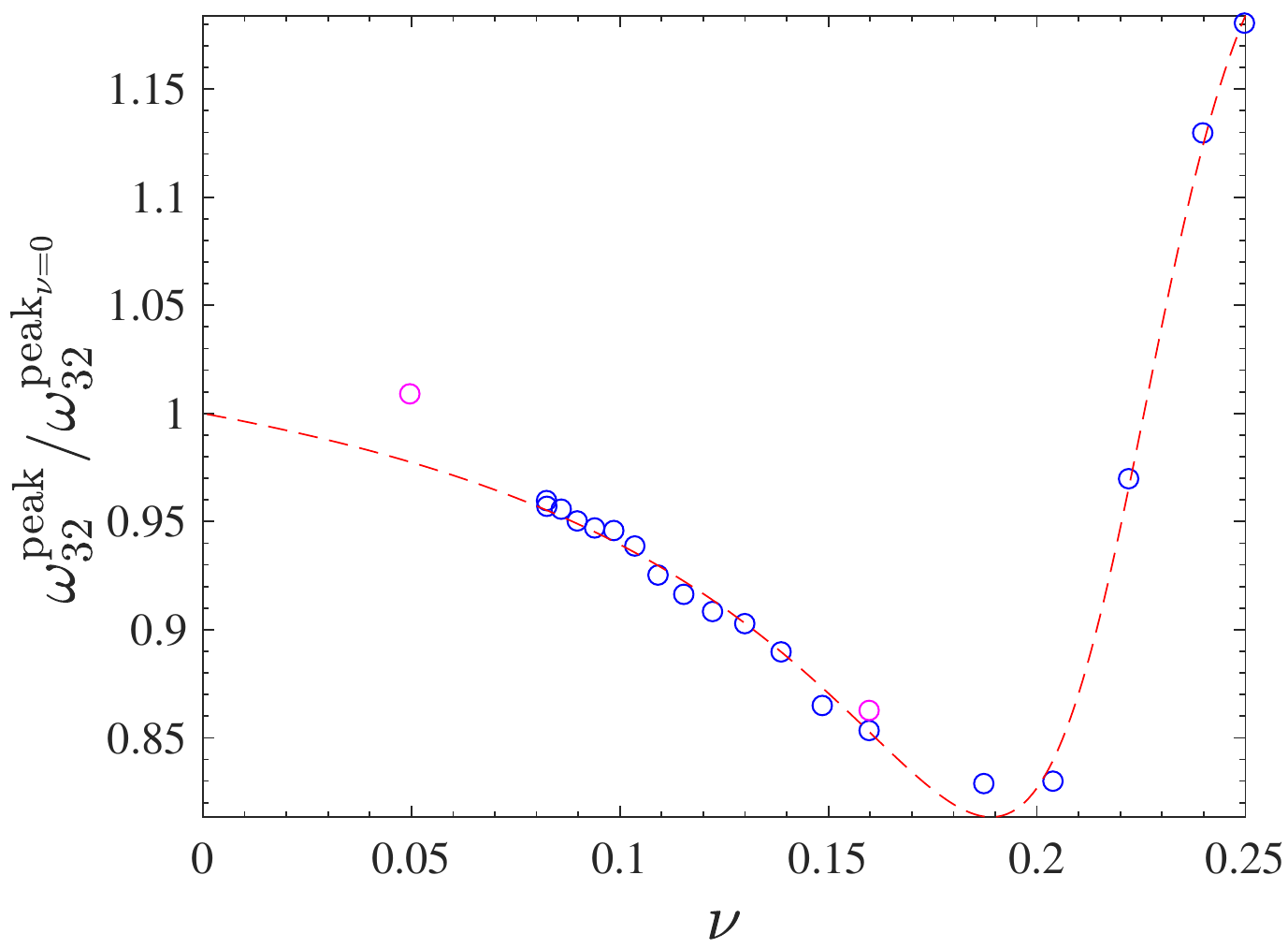}\\
  \includegraphics[width=0.22\textwidth]{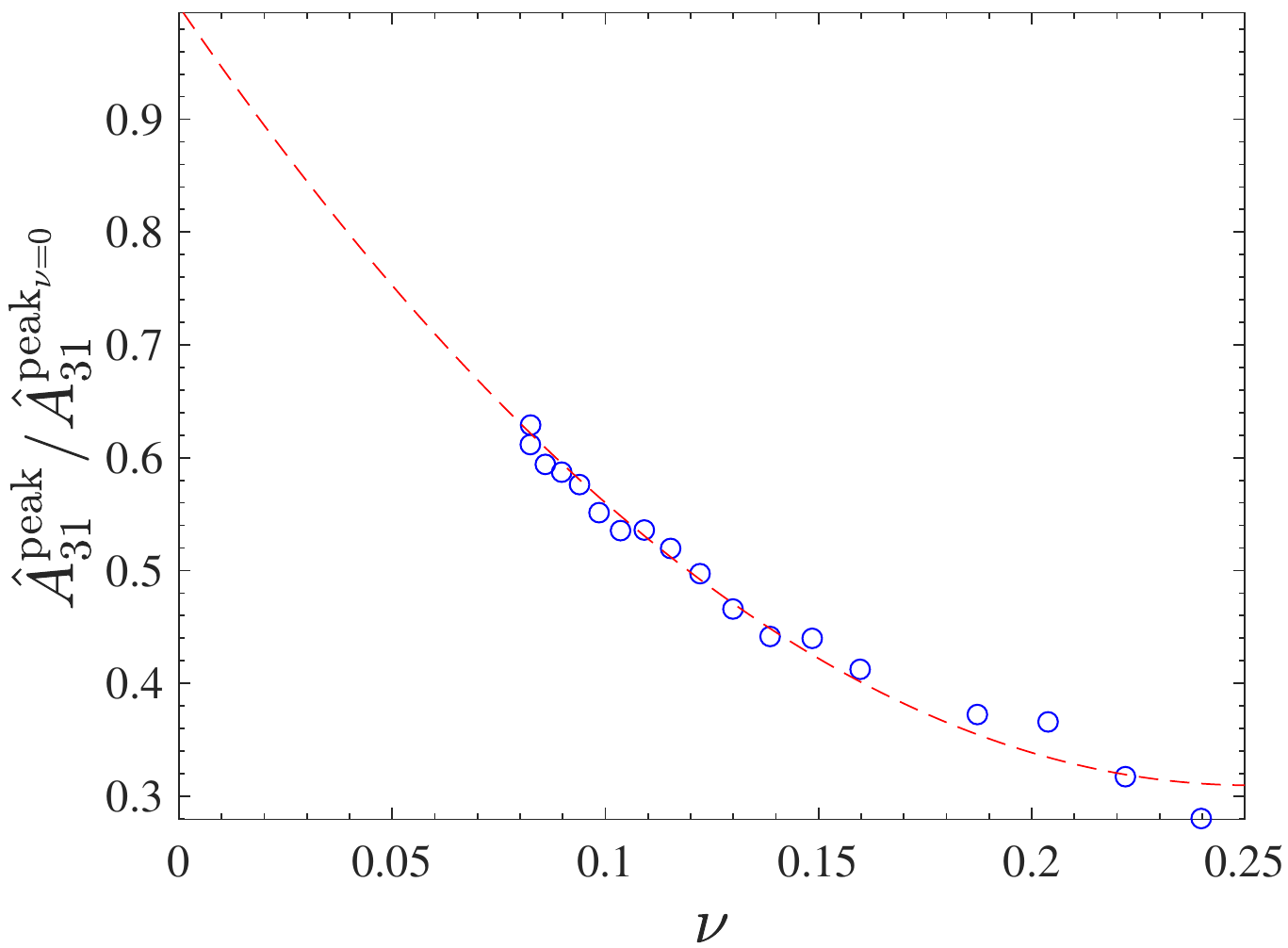}
  \hspace{1mm}
  \includegraphics[width=0.22\textwidth]{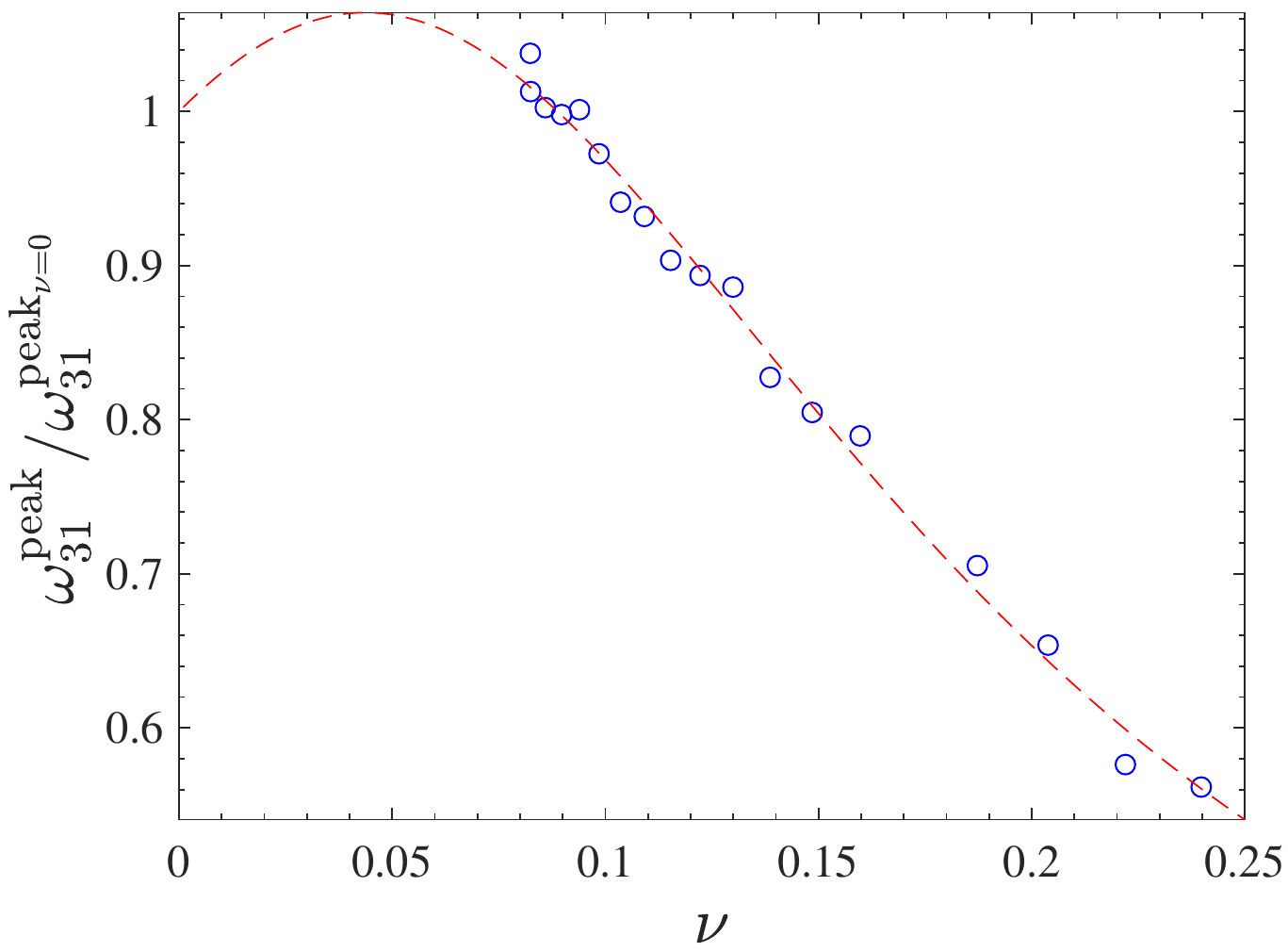}
  \hspace{1mm}
  \includegraphics[width=0.22\textwidth]{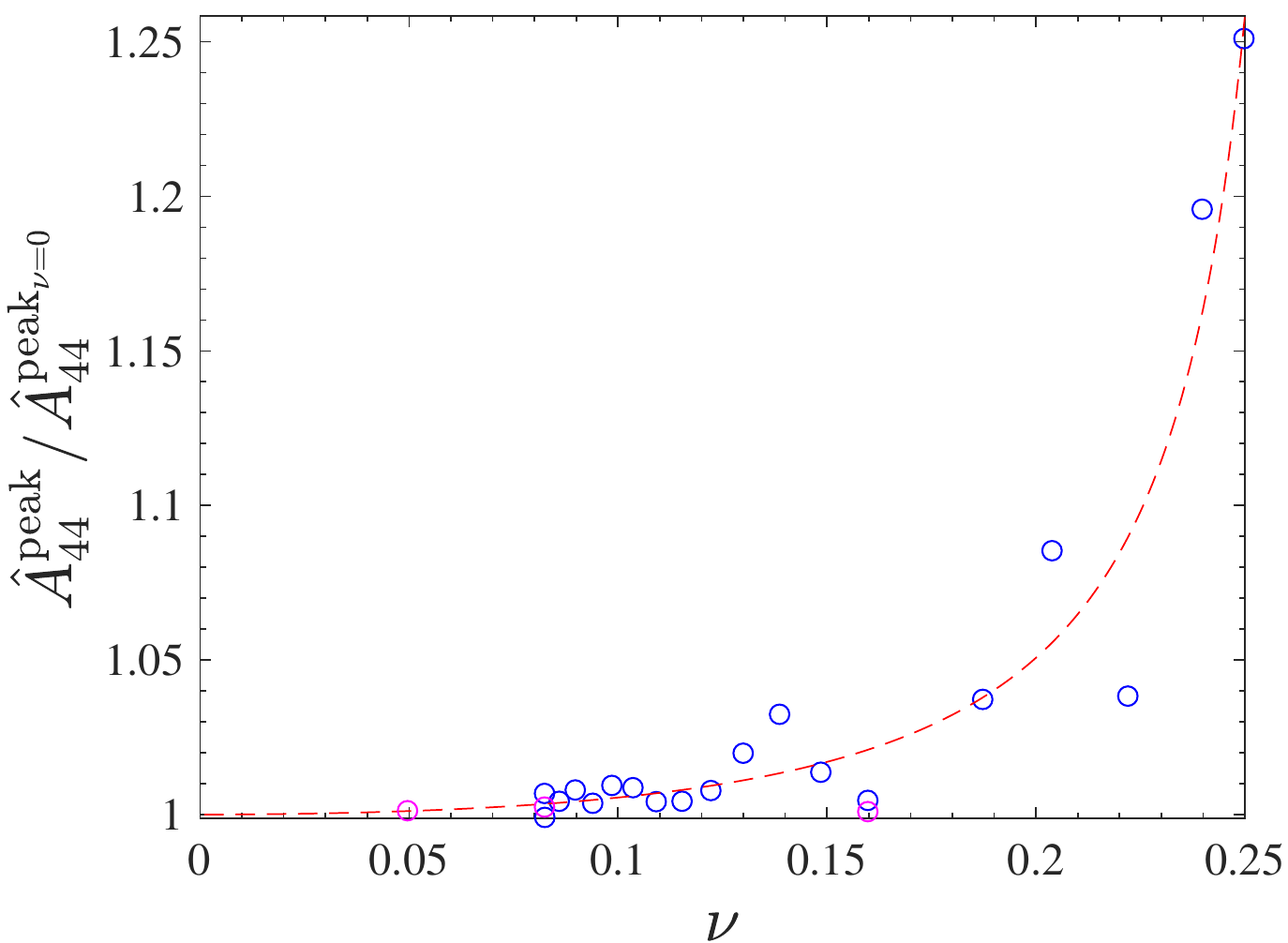}
  \hspace{1mm}
  \includegraphics[width=0.22\textwidth]{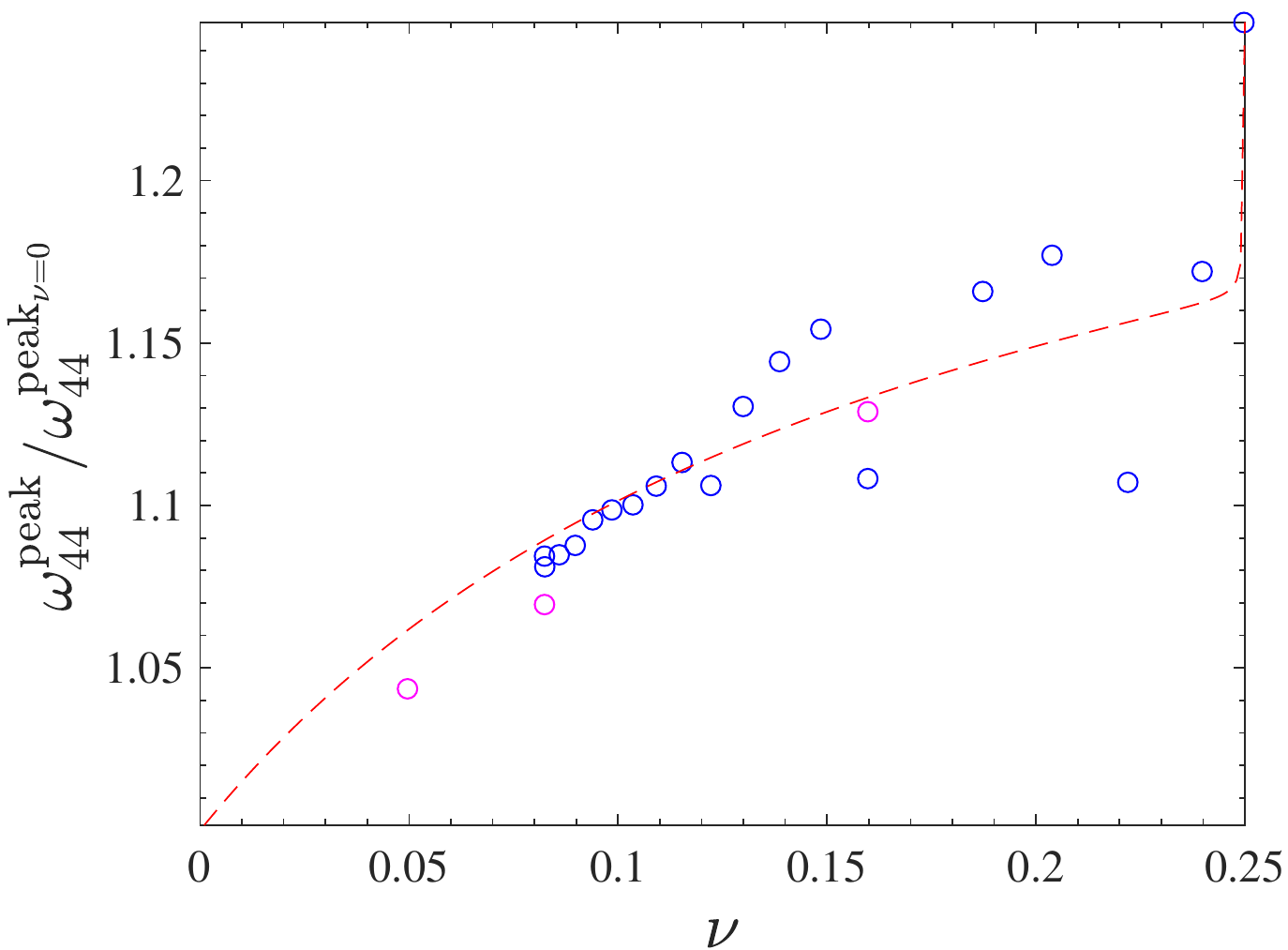}\\
  \includegraphics[width=0.22\textwidth]{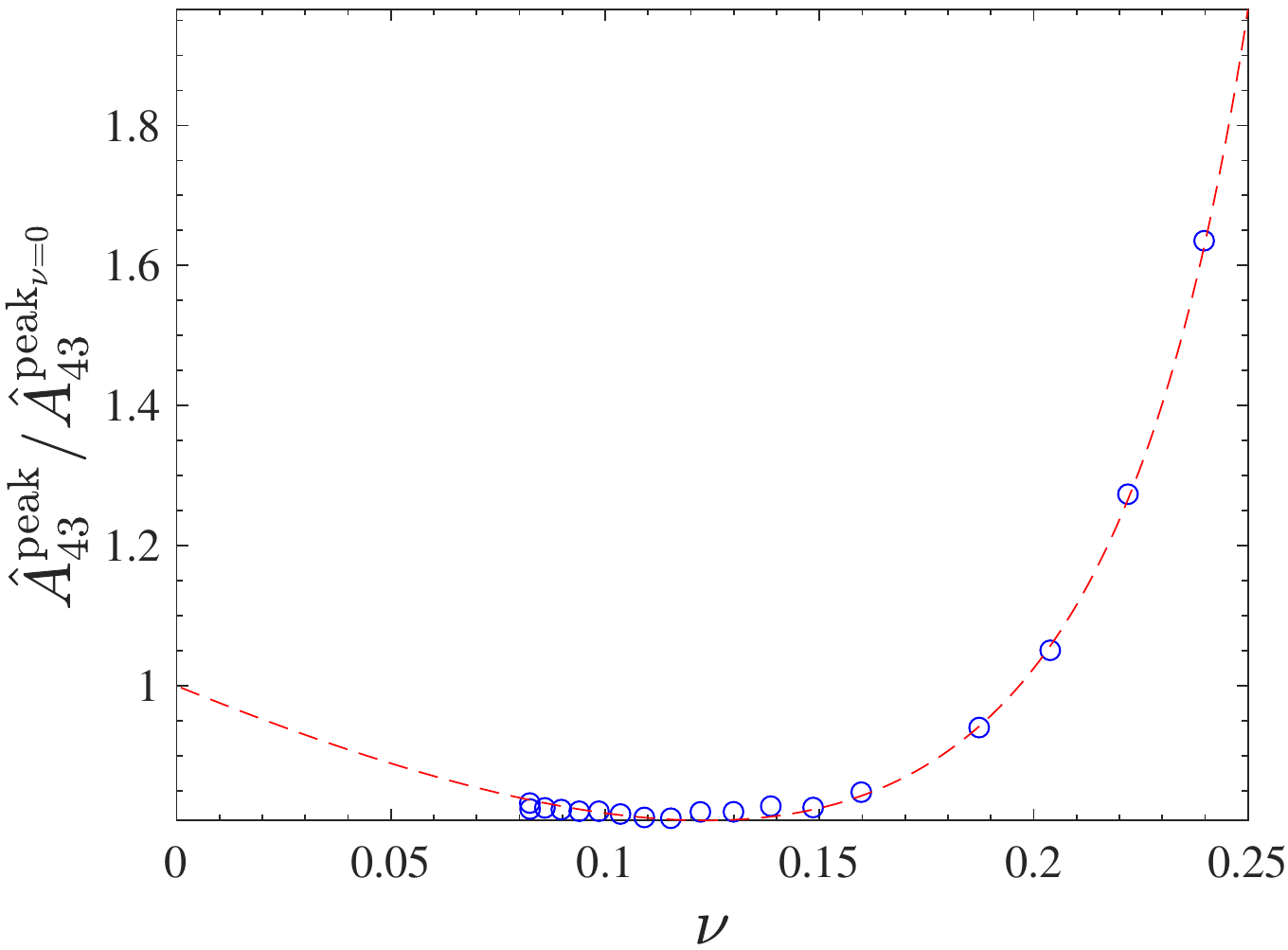}
  \hspace{1mm}
  \includegraphics[width=0.22\textwidth]{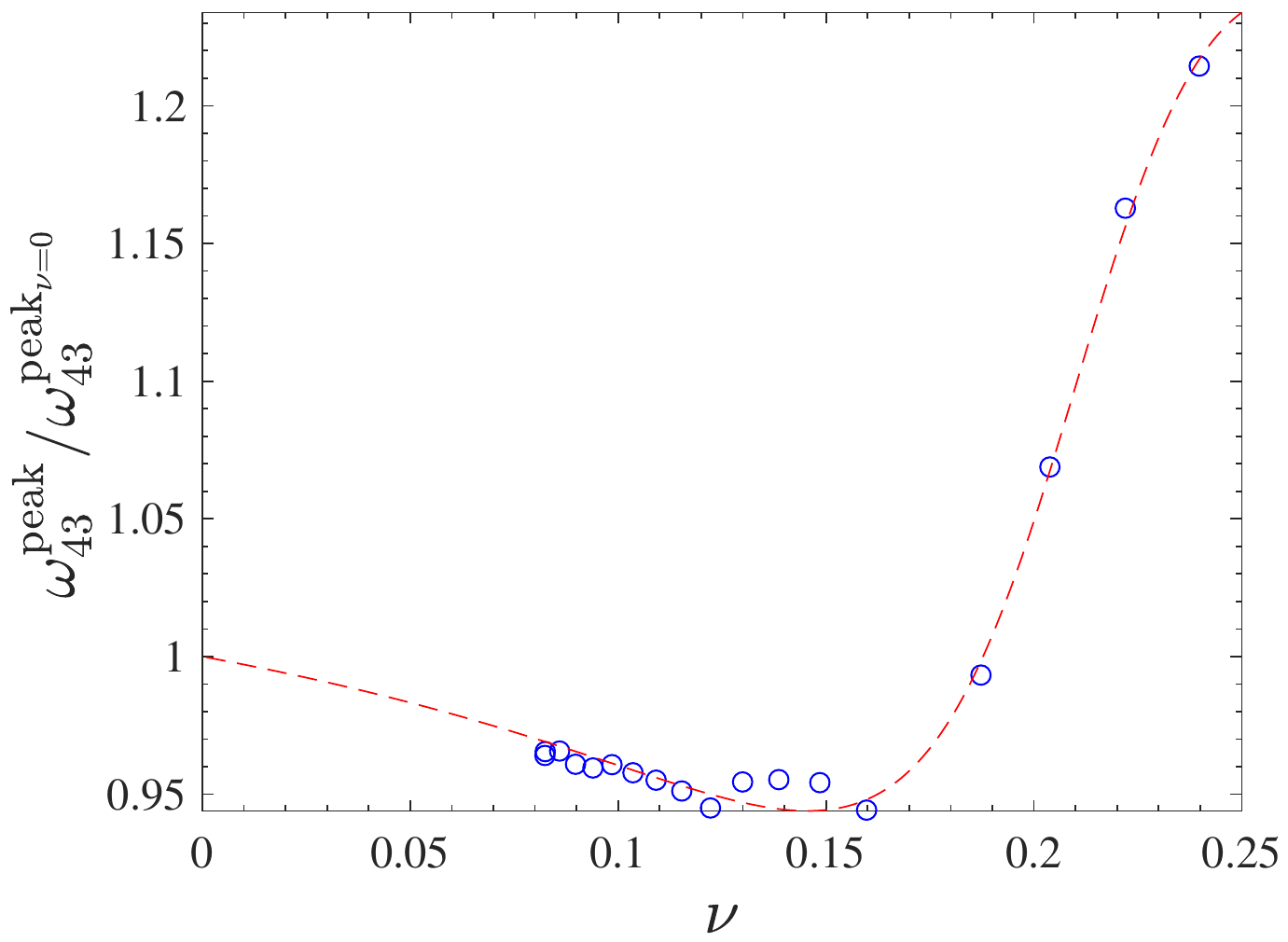}
  \hspace{1mm}
  \includegraphics[width=0.22\textwidth]{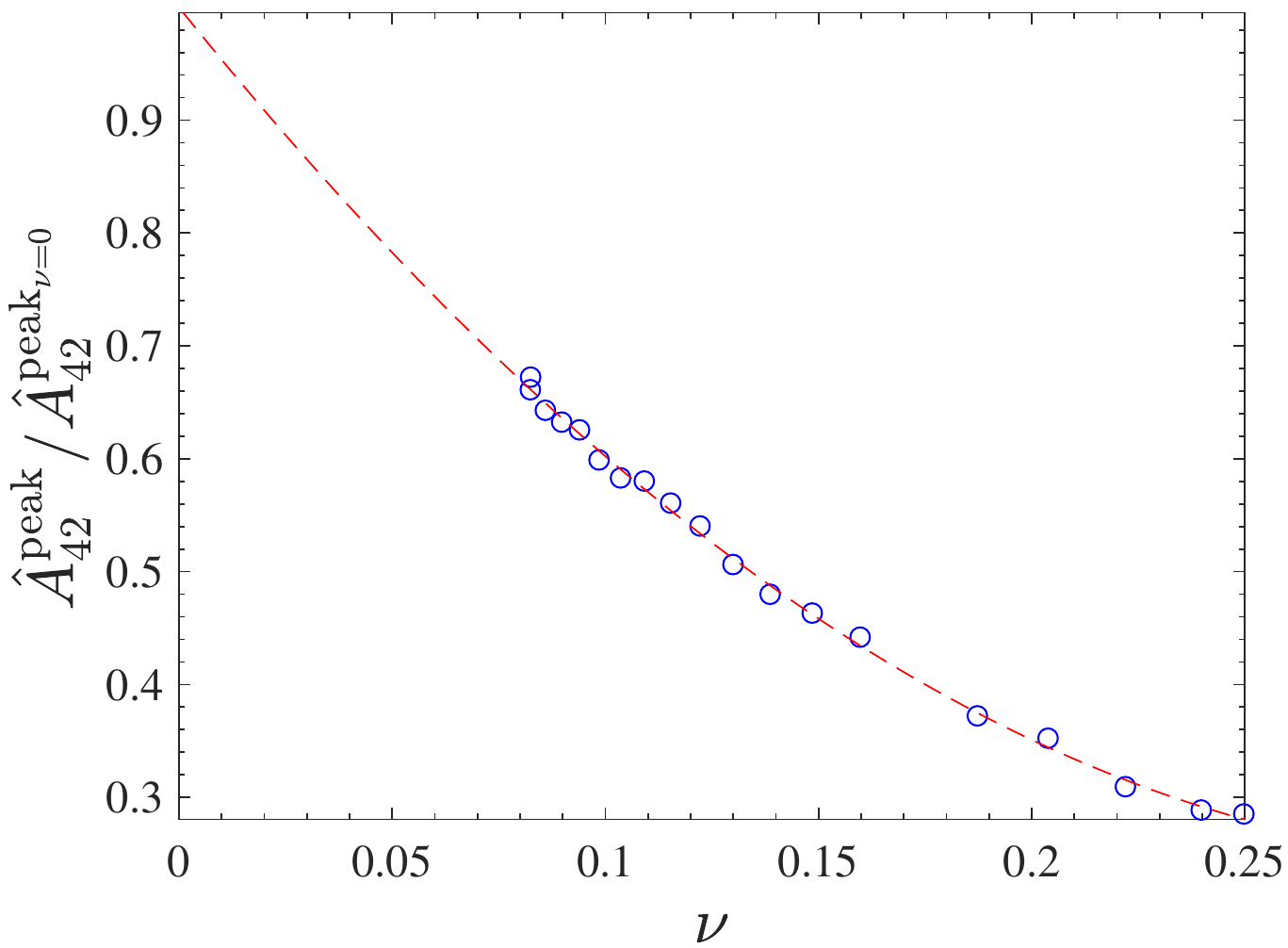}
  \hspace{1mm}
  \includegraphics[width=0.22\textwidth]{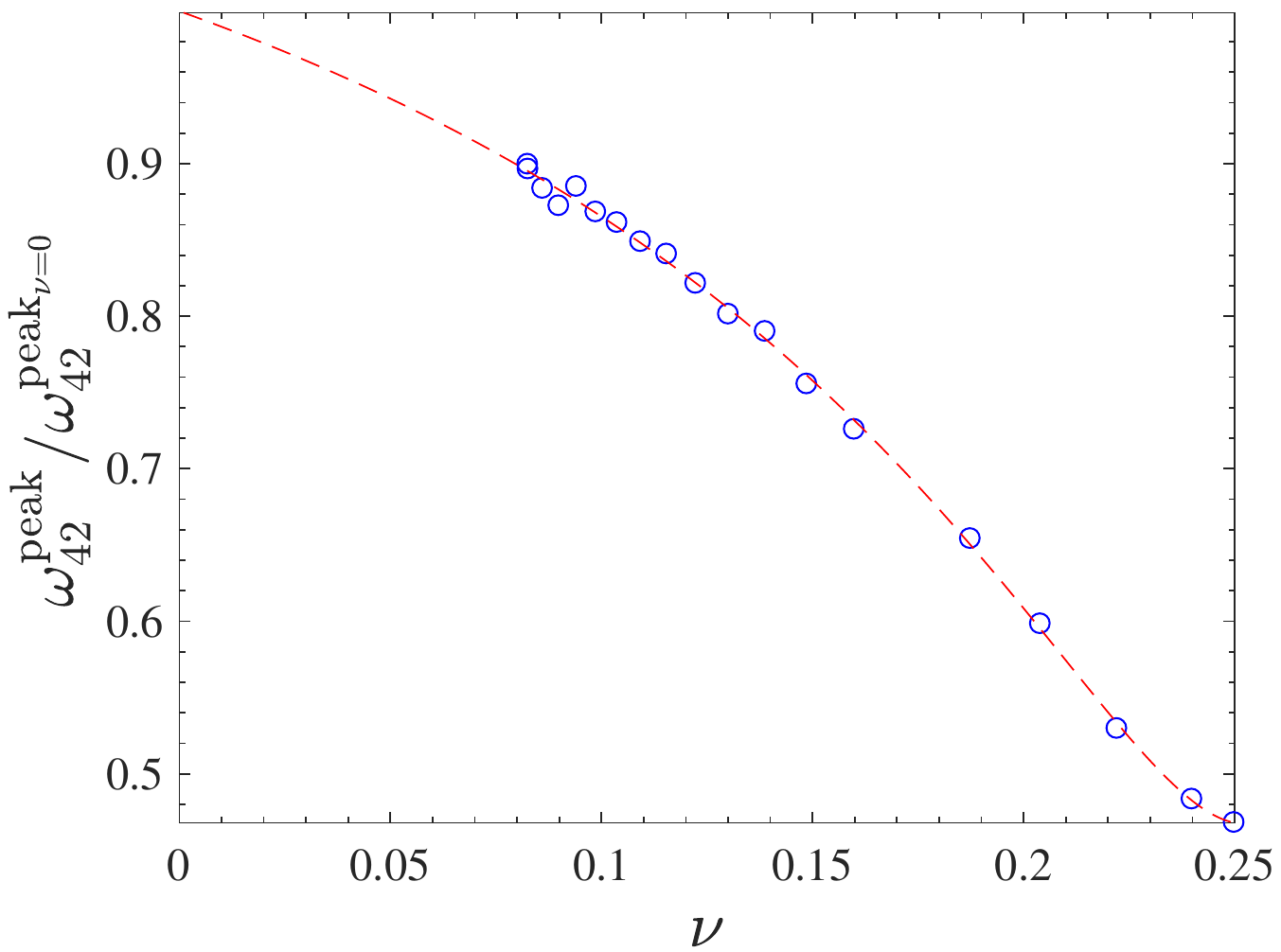}\\
  \includegraphics[width=0.22\textwidth]{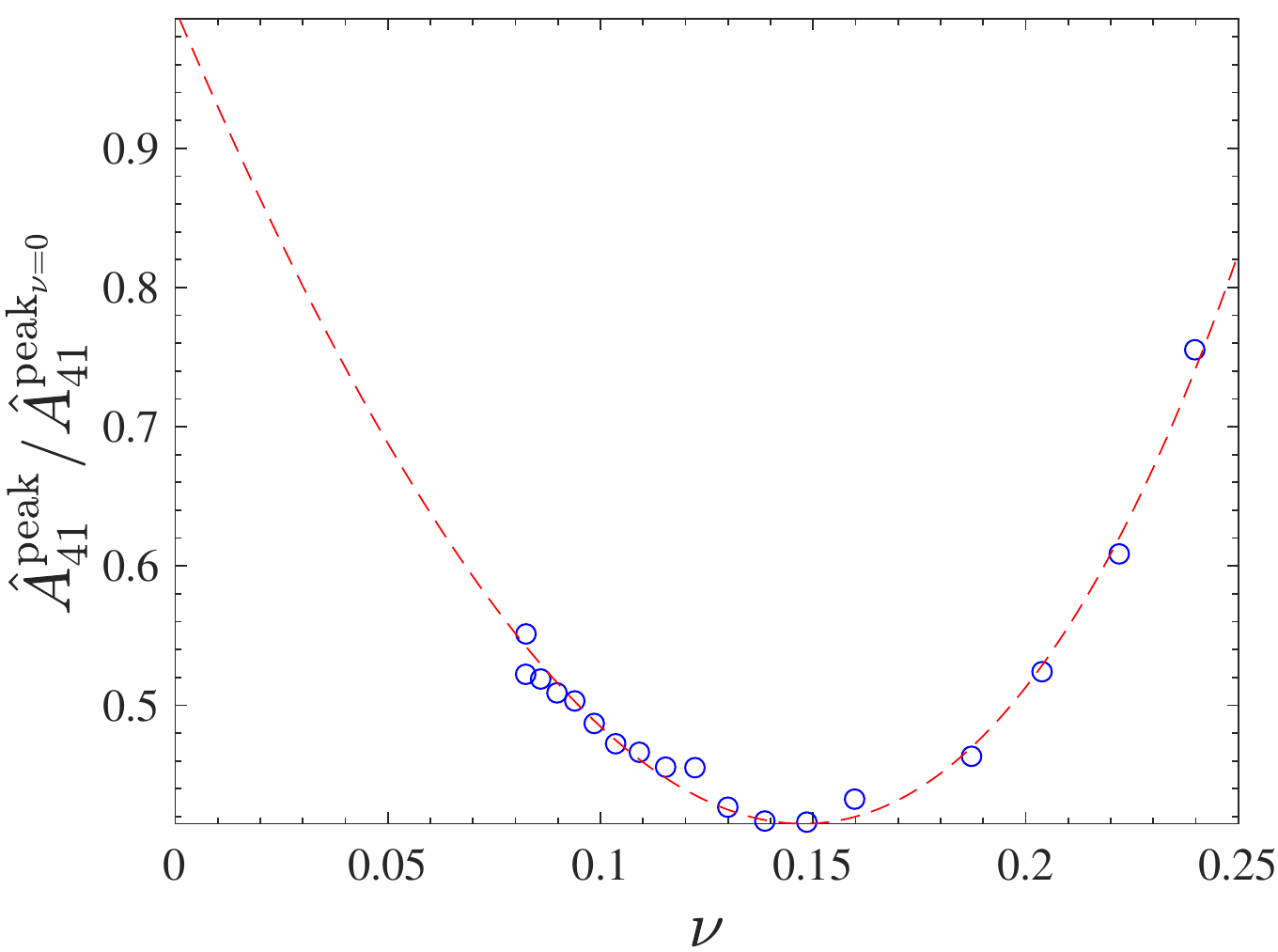}
  \hspace{1mm}
  \includegraphics[width=0.22\textwidth]{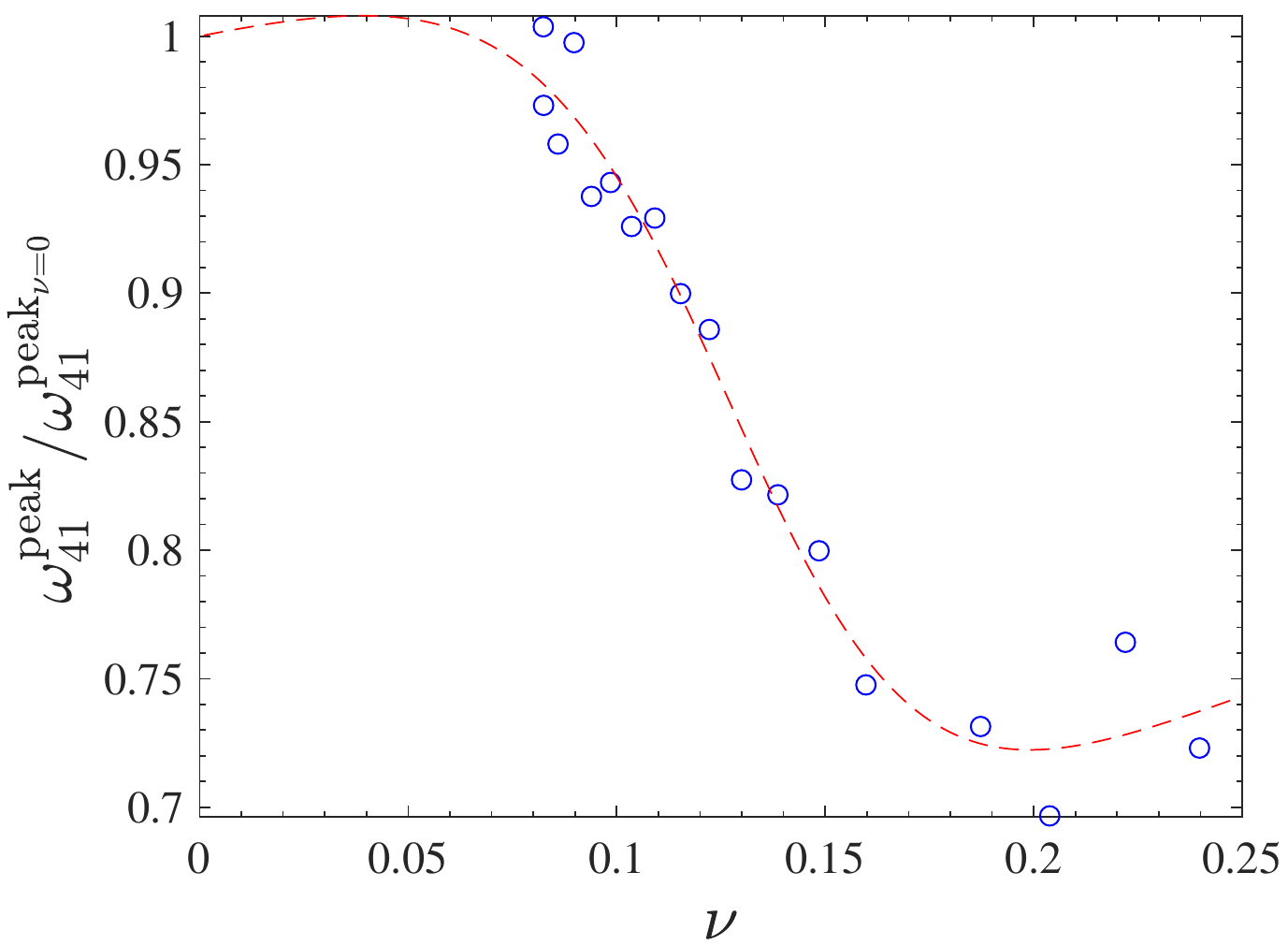}
  \hspace{1mm}
  \includegraphics[width=0.22\textwidth]{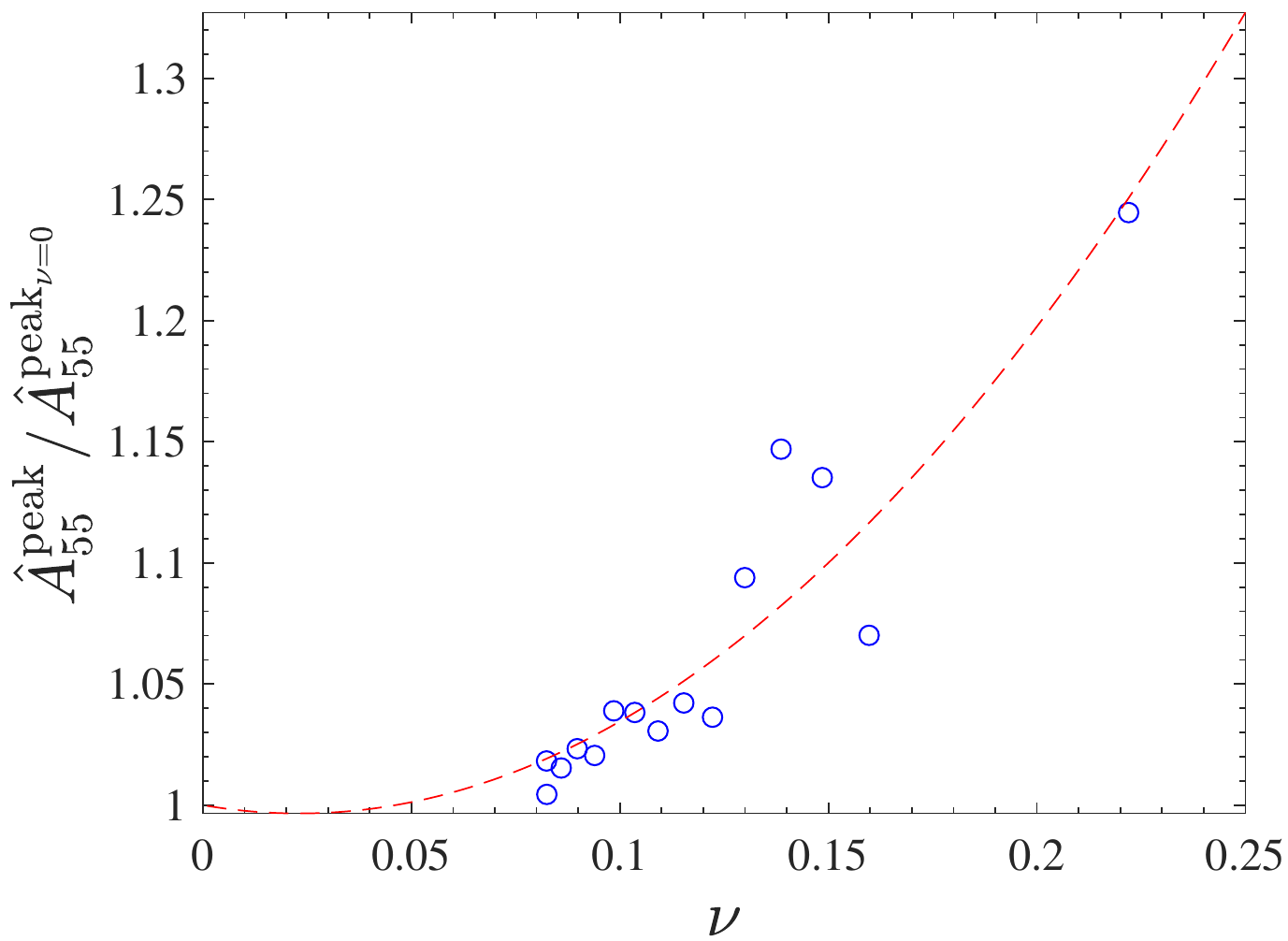}
  \hspace{1mm}
  \includegraphics[width=0.22\textwidth]{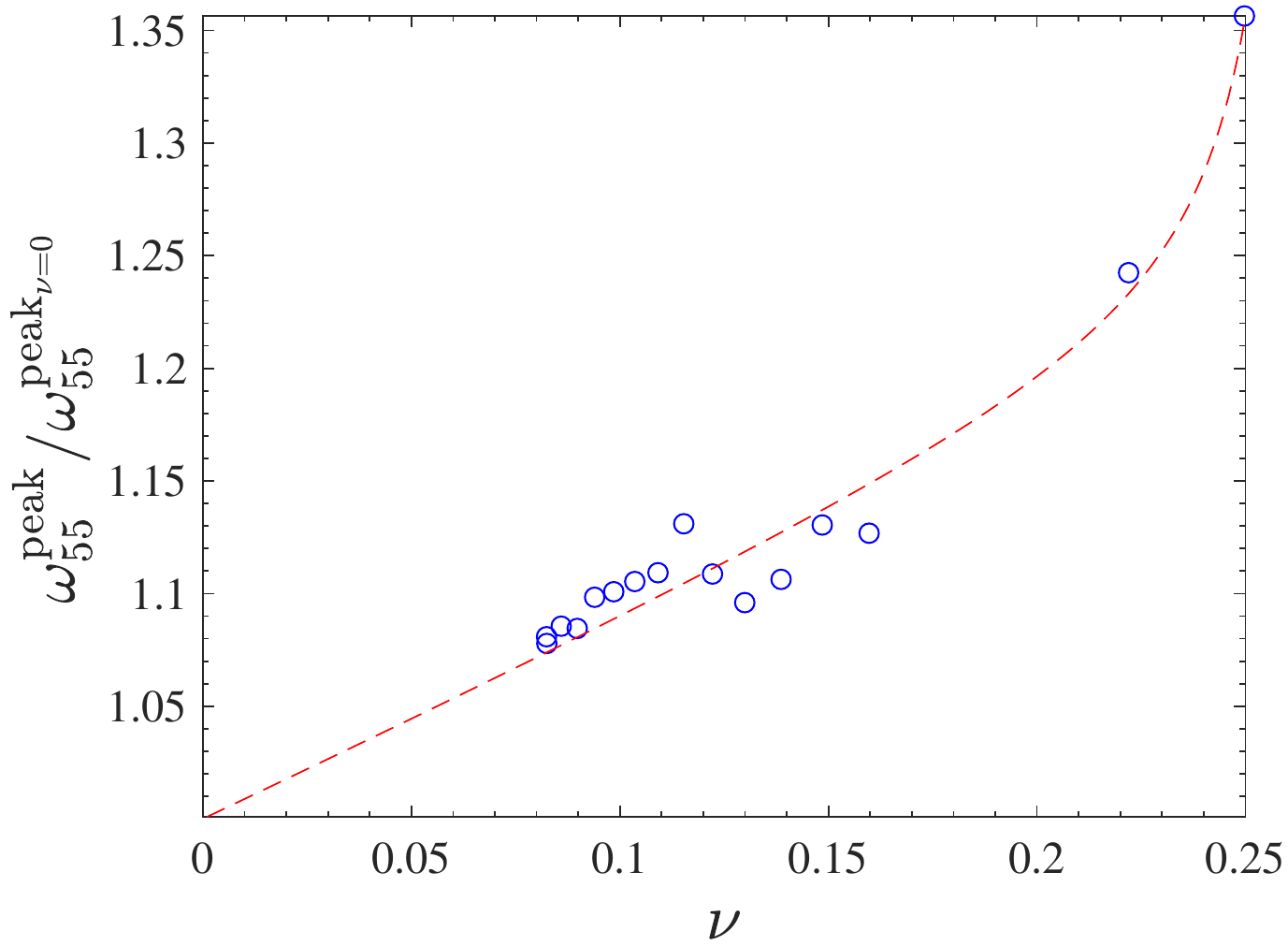}\\
  \caption{\label{fig:orb_peak} Comparison of orbital fits of peak amplitude and frequency 
  	versus SXS and BAM data for the multipoles $\ell \leq 4$, $1\leq m \leq 4$, and $(\ell,m)=(5,5)$.}
\end{figure*}

\begin{table*}
  \caption{\label{tab:fit0}Parameters for the fit of the peak amplitude and frequency of
    all multipoles up to $\ell=m=5$. From left to right, the columns report: the multipolar indices;
    the values of the amplitude and frequency in the test-particle limit, $(\hat{A}^0_\lm,\omega^0_\lm)$; the amplitude fit coefficients $(n_i^{A_\lm},d_i^{A_\lm})$ and
    the frequency fit coefficients $(n_i^{\omega_\lm},d_i^{\omega_\lm})$ for the functions $(\hat{\hat{A}}_{\ell m},\hat{\omega}_\lm)$ defined
    in Eqs.~\eqref{eq:A}-\eqref{eq:omg} and fitted using the rational function template of Eq.~\eqref{eq:flm}. Note that since all $d_2^{A_\lm}$ values are
    found to be equal to zero we do not explicitly report them in the table.}
\begin{ruledtabular}
\begin{tabular}{l l | l l | c c c |cc c c c }
  $\ell$ & $m$ & $\hat{A}^{0}_{\ell m}$  & $\omega^0_{\lm}$ &  $n_1^{A_{\ell m}}$  & $n_2^{A_{\ell m}}$  & $d_1^{A_{\ell m}}$ & ${n}_1^{\omega_\lm}$ 	& ${n}_2^{\omega_\lm}$ 	& ${d}_1^{\omega_\lm}$ 	& $d_2^{\omega_\lm}$\\ \hline
	2   &  2  & $0.295896$      	& $0.273356$   & $-0.041285$& $1.5971$   & \dots    & $0.84074$		& $1.6976$ & \dots 		& \dots\\ 
	    &  1  & $0.106935$    	& $0.290672$   & $9.0912$ 	& $3.9331$   & $11.108$ & $-0.060432$	& $1.9995$ & $0.23248$ 	& \dots\\ \hline
	3   &  3  & $0.051670$     	& $0.454622$   & $0.098379$ & $3.8179$   & \dots   	& $1.1054$    	& $2.2957$ & \dots 		& \dots\\ 
	    &  2  & $0.018168$     	& $0.451817$   & $-6.142$  	& $11.372$   & $-3.6448$& $-9.0214$   	& $21.078$ & $-8.6636$ 	& $19.493$\\ 
	    &  1  & $0.005694$    	& $0.411755$   & $-5.49$    & $10.915$   & \dots   	& \dots   		& $7.5362$ & $-2.7555$ 	& $38.572$\\ \hline
        4   &  4  & $0.014579$  	& $0.635415$   & $-3.6757$  & $0.32156$  & $-3.6784$& $3.2876$   	& $-29.122$& $1.696$ 	& $-22.761$\\ 
	    &  3  & $0.004962$    	& $0.636870$   & $-5.7791$ 	& $12.589$   & $-3.3039$& $-9.0124$   	& $22.011$ & $-8.732$ 	& $20.518$\\ 
	    &  2  & $0.001656$    	& $0.626030$   & $-4.7096$  & $7.3253$   & \dots   	& $-7.0558$   	& $12.738$ & $-6.0595$ 	& $9.3389$\\
	    &  1  & $0.000487$   	& $0.552201$   & $-8.4449$  & $26.825$   & $-1.2565$& $-10.876$   	& $37.904$ & $-11.194$ 	& $42.77$\\ 
        5   &  5  & $0.005227$  	& $0.818117$   & $-0.29628$	& $6.4207$   & \dots    & $-2.8918$   	& $-3.2012$& $-3.773$   & \dots\\
\end{tabular}
\end{ruledtabular}
\end{table*}

\subsection{Overview of the analytical model}
\label{sc:PM_rep}
The description of the ringdown is based on the model introduced in Ref.~\cite{Damour:2014yha}
based on a suitable fit of NR waveform data. The original model was improved in Refs.~\cite{Nagar:2018zoe,Nagar:2016iwa}
and also adopted, with some modifications, in Ref.~\cite{Bohe:2016gbl,Cotesta:2018fcv}.
The idea of Ref.~\cite{Damour:2014yha} is to first (i) factorize in the waveform the contribution of the fundamental
quasi-normal mode and then (ii) to fit the remaining, time-dependent, complex factor. Here we generalize
this procedure, precisely as it was introduced in~\cite{Damour:2014yha}, to all modes up to $\ell=m=5$.
In particular, for each $(\ell,m)$ the fit is done from the peak of each mode. To ease the notation,
we define
\be
h(\tau)\equiv h_\lm(\tau) \quad \text{with}\quad \tau\geq 0 \ ,
\ee
where $\tau \equiv (t-t^{\rm peak}_{\ell m})/M_{\rm BH}$, and $M_{\rm BH}$ is
the mass of the final BH, computed from the fits
of~\cite{Jimenez-Forteza:2016oae}. In the following we assume
that each quantity carries indices $\lm$, that we do not write
explictly except when it is needed to avoid confusion.
Defining the complex frequency of the fundamental
QNM as $\sigma_1\equiv \alpha_1 + \ii \omega_1$,
the QNM-factorized waveform $\bar{h}(\tau)$ is defined by
\begin{align}
h\left(\tau\right)= e^{-\sigma_1\tau-\ii\phi_0}\bar{h}\left(\tau\right).
\end{align}
This is then separated into phase and amplitude as
\begin{align}
\bar{h}(\tau )=A_{\bar{h}}(\tau )e^{\ii\phi_{\bar{h}}(\tau )},
\end{align}
that are separately fitted using the following ans\"atze
\begin{align}
  \label{eq:A_template}
A_{\bar{h}}(\tau ) = &c_1^A \tanh \left(c_2^A\tau +c_3^A \right) +c_4^A,\\
\label{eq:phi_temp}
\phi_{\bar{h}}(\tau ) = & -c_1^{\phi} \mathrm{ln} \left( \frac{1+c_3^{\phi} e^{-c_2^{\phi} \tau } +c_4^\phi e^{-2c_2^\phi \tau }}{1+c_3^{\phi}+c_4^{\phi}}\right).
\end{align}
However, not all of the coefficients will be free parameters, as
we impose five additional constraints so that the fit incorporates
the physically correct behavior both at $\tau=0$ and and at late
times~\cite{Damour:2014yha}. Imposing these conditions yields
\begin{align}
c_2^A = & \frac{1}{2}\alpha_{21},\\
c_4^A = & \hat{A}^{\rm peak} - c_1^A \tanh \left(c_3^A\right),\\
c_1^A = & \hat{A}^{\rm peak} \alpha_1 \frac{\cosh^2\left(c_3^A\right)}{c_2^A},\\
c_1^{\phi} = & \Delta\omega \frac{1+c_3^{\phi}+c_4^{\phi}}{c_2^{\phi}\left(c_3^{\phi}+2c_4^{\phi}\right)},\\
c_2^{\phi} = & \alpha_{21},
\end{align}
with $\Delta\omega = \omega_1 - M_{\rm BH}\omega^{\rm peak}$,
$\hat{A}^{\rm peak}$ is the value of the multipole
amplitude at its peak, as defined
in Eq.~\eqref{eq:Ahat} below (with the $\ell,m$ dependence explicit),
and $\alpha_{21}\equiv \alpha_2-\alpha_1$, i.e. the difference between
the inverse damping times of the first overtone ($\alpha_2$)
and of the fundamental mode $(\alpha_1)$. The three remaining parameters,
$\left(c_3^A,c_3^\phi,c_4^\phi\right)$, that are the only genuine free parameters
of the model, are then fitted directly for any NR dataset.
We then define two kind of fits: (i) we address as {\it primary}
the fit of $(c_3^A,c_3^\phi,c_4^\phi)$ for a given SXS dataset;
(ii) we then call as {\it global interpolating fit} the
one of the coefficients above performed all over the
available NR datasets. Let us highlight a few points.
\begin{enumerate}
\item[(i)] For each mode, one needs accurate fits of the amplitude and frequency at the peak of each multipole.
  Capturing these numbers properly is crucial to accurately reproduce the global post-peak evolution of frequency
  and amplitude.
\item[(ii)] The QNM information can be fitted with extreme precision against the dimensionless spin $\hat{a}_f$
  of the final BH. In practice $\hat{a}_f$ is determined from the fit presented in~\cite{Jimenez-Forteza:2016oae}.
\item[(iii)] The effective post-merger parameters $(c_3^A,c_3^\phi,c_4^\phi)$ are
  very sensitive to noise in the NR waveform and thus extreme precision is not advisable.
  On the other hand they only sub-dominantly impact the waveform.
\item[(iv)] The primary fitting template given in Eq.~\eqref{eq:A_template} does not have enough analytical
  flexibility to accurately fit the waveform amplitude in the extreme-mass-ratio limit and will have to be
  changed in order to fully extend the validity of the model to that regime.
\end{enumerate}

\subsection{Numerical Relativity Data}
\label{sc:NR_hlm}
We use 19 SXS waveforms \cite{Mroue:2013xna,Chu:2015kft,Blackman:2015pia,SXS:catalog}, summarized in Table~\ref{tab:NR_Waveforms}, 
and a single test-particle waveform to perform the fits. The details of the waveform generation
of the latter and further details can be found here~\cite{Harms:2014dqa,Nagar:2016ayt}.
Several quantities characterizing the waveforms are detailed in Table~\ref{tab:NR_Waveforms}.
Most of these are defined in the main text or can be extracted from the ${\tt metadata.txt}$ 
available online~\cite{SXS:catalog}. The only additional quantity that is routinely used to
conservatively assess the accuracy of the waveform is the accumulated phase difference 
between the highest and second-highest resolution up to the peak of the $\ell=m=2$ waveform,
that corresponds to the merger of the two objects.
The waveforms cover the range $1\leq q \leq 10$, corresponding to
$0.08\lesssim \nu\lesssim 0.25$. The waveforms 
are between $15.5$ and $29.1$ cycles long and have eccentricities never exceeding 
$\epsilon\approx 2.1\times 10^{-3}$. For most waveforms,
$|\delta\phi^{\rm NR}_{\rm mrg}|\lesssim 0.5~{\rm rad}$ with the 
exception of SXS:BBH:0063 which reaches $\delta\phi^{\rm NR}_{\rm mrg}\approx 1~{\rm rad}$
which is still an exceptable margin of error.

\subsection{Fits: waveform peak frequency and amplitude.}
\label{sec:peak}
\begin{table*}
\caption{\label{tab:postmerger_phase}
  Fit coefficients of the postpeak functions $(c_3^{A_\lm},c_3^{\phi_\lm},c_4^{\phi_\lm})$ entering Eqs.~\eqref{eq:A_template}-\eqref{eq:phi_temp}.
  Note the rather special functional form needed for $c_3^{A_{32}}$ and $c_3^{A_{44}}$, that is necessary to properly account for
  nearly equal-mass data. In addition, the fits of some multipoles are discontinuous, the interface between the branches being at mass ratios $q=2.5$ or $q=10$. 
  Such mass ratios correspond to the values $\nu=10/49$ and $\nu=10/121$ that appear in the argument of the $\theta$ functions.}
\begin{ruledtabular}
\begin{tabular}{c c c| c | c | c c}
& $\ell$ & $m$ 	& $c_3^{A_\lm}$ &$c_3^{\phi_\lm}$    & $c_4^{\phi_\lm}$\\
\hline\hline
& 2   &  2 & $-0.56187+0.75497\nu$	& $\frac{4.4414-63.107\nu+296.64\nu^2}{1-13.299\nu+69.129\nu^2}$ & $\frac{7.1508-109.47\nu}{1+556.34\nu+287.42\nu^2}$ 	& \\
& 2    &  1 &$\frac{0.23882-2.2982\nu+5.7022\nu^2}{1-7.7463\nu+27.266\nu^2}$  & $\frac{2.6269-37.677\nu+181.61\nu^2}{1-16.082\nu+89.836\nu^2}$ & $\frac{4.355-53.763\nu+188.06\nu^2}{1-18.427\nu+147.16\nu^2}$ 	& \\\hline
& 3   &  3 &$-0.39337+0.93118\nu$ & $3.1017-6.5849\nu$ & $\frac{3.4521-24.153\nu+53.029\nu^2}{1+3.1413\nu}$ 	& \\
& 3    &  2 &$\frac{0.1877-3.0017\nu+19.501\nu^2}{1-1.8199\nu}-e^{-703.67(\nu-2/9)^2}$ & $\frac{0.90944-1.8924\nu+3.6848\nu^2}{1-8.9739\nu+21.024\nu^2}$ & $\frac{2.3038-50.79\nu+334.41\nu^2}{1-18.326\nu+99.54\nu^2}$ 	& \\
& 3    &  1 &$\frac{3.5042-55.171\nu+217\nu^2}{1-15.749\nu+605.17\nu^3}$ & $\frac{-6.1719+29.617\nu+254.24\nu^2}{1-1.5435\nu}\theta\left(\nu-\frac{10}{121}\right)$  & $3.6485+5.4536\nu$ 	& \\
&     &    & &  $\;\;\;\;\;\;-2.2784\ \theta\left(\frac{10}{121}-\nu\right)$ 	& \\\hline
& 4   &  4 &  $-0.25808+0.84605\nu+1.2376e^{-6054.7(\nu-10/49)^2}$ & $\frac{2.3328-9.4841\nu+19.719\nu^2}{1-2.904\nu}$ & $0.94564+3.2761\nu$ 	& \\
& 4    &  3 & $\frac{-0.02833+2.8738\nu-31.503\nu^2+93.513\nu^3}{1-10.051\nu+156.14\nu^3}$& $\frac{2.284-23.817\nu+70.952\nu^2}{1-10.909\nu+30.723\nu^2}$ & $\frac{2.4966-6.2043\nu}{1-252.47\nu^4}$ 	& \\
& 4    &  2 & $\frac{0.27143-2.2629\nu+4.6249\nu^2}{1-7.6762\nu+15.117\nu^2}$ &  $\frac{2.2065-17.629\nu+65.372\nu^2}{1-4.7744\nu+3.1876\nu^2}$ & Eq.~\eqref{eq:c4phi42} 	& \\
& 4    &  1  & $11.47+10.936\nu$ & $ (-6.0286+46.632\nu)\theta\left(\nu-\frac{10}{121}\right)$  & $1.6629+11.497\nu$ 	& \\
&     &     & &$\;\;\;\;\;\;-2.1747\ \theta\left(\frac{10}{121}-\nu\right)$ & 	& \\\hline
& 5   &  5  & $\frac{-0.19751+3.607\nu-14.898\nu^2}{1-20.046\nu+108.42\nu^2}$ &$0.83326+10.945\nu$ & $\frac{0.45082-9.5961\nu+52.88\nu^2}{1-19.808\nu+99.078\nu^2}$ 	& 
\end{tabular}
\end{ruledtabular}
\end{table*}
Let us turn now to discussing the fits of amplitude and frequency
at each multipole peak. For consistency with previous work,
we shall use from now on the Regge-Wheeler-Zerilli normalized~\cite{Nagar:2005ea}
strain waveform
\be
\label{eq:RWZ}
\Psi_{\ell m}\equiv h_{\ell m}/\sqrt{(\ell+2)(\ell +1)\ell (\ell -1)},
\ee
so that the multipolar waveform is separated in amplitude and phase as
\be
\Psi_{\ell m}=A_{\ell m}e^{-{\rm i}\phi_{\ell m}},
\ee
and the frequency is $\omega_{\ell m}=\dot{\phi}_{\ell m}$.
We then define $t^{\rm peak}_\lm$ as the time where each $A_{\lm}$ peaks,
i.e. $\dot{A}_\lm(t_{\rm peak})=0$ and then we measure the values
of amplitude and frequency at
$t_{\rm peak}$, $(A_{\ell m}^{\rm peak},\omega_\lm^{\rm peak})$.
We hence define
\begin{align}
  \omega_{\ell m}^{\rm peak}  &\equiv\omega_{\ell m}(t)\vert_{t=t_{\ell m}^{\rm peak}},\\
  A^{\rm peak}_{\ell m} &\equiv A_{\ell m}(t)\vert_{t=t_{\ell m}^{\rm peak}}.
\end{align}
To build analytical fits of $(A_{\ell m}^{\rm peak},\omega_\lm^{\rm peak})$ we proceed as follows.
First, we factor out from the peak values the leading-order $\nu$ behavior. This is given by the 
function
\be
c_{\ell + \epsilon}(\nu) =X_2^{\ell + \epsilon -1} + (-)^{\ell+\epsilon}X_1^{\ell + \epsilon -1}
\ee 
and we define
\be
\label{eq:Ahat}
\hat{A}_{\lm}\equiv  A_\lm/c_{\ell +\epsilon}(\nu) \ .
\ee
As a second step, we also factor out the  the test-particle
values $(\hat{A}_\lm^0,\omega_\lm^0)$, that are known
with high accuracy (see Table~3 of Ref.~\cite{Harms:2014dqa}). In pratice, the quantities to
be fitted are $(\hat{\hat{A}}^{\rm peak}_\lm,\hat{\omega}_\lm^{\rm peak})$ defined as
\begin{align}
\label{eq:A}
A^{\rm peak}_{\ell m}&=c_{\ell + \epsilon}(\nu)\hat{A}^{\rm 0}_{\ell m} \hat{\hat{A}}^{\rm peak}_{\ell m},\\
\label{eq:omg}
\omega^{\rm peak}_{\ell m}&= \omega^{\rm 0}_{\ell m}\hat{\omega}^{\rm peak}_{\ell m}.
\end{align}
Figure~\ref{fig:orb_peak} illustrates the behavior of the NR quantities $(\hat{\hat{A}}_\lm,\hat{\omega}_\lm)$
versus $\nu$. Whenever possible, we show together SXS and BAM datapoints to illustrate the consistency
between results obtained with very different computational infrastructures. One finds that the 
datapoints can be easily fitted with a rational function with the general form
\be
\label{eq:flm}
f_{\ell m} = \frac{1+n_1^{f_\lm}\nu+n_2^{f_\lm}\nu^2}{1+d_1^{f_\lm}\nu + d_2^{f_\lm}\nu^2},
\ee
where $f_\lm$ is either $\hat{\hat{A}}_\lm^{\rm peak}$ or $\hat{\omega}_\lm^{\rm peak}$. 
The fit coefficients are listed in Table~\ref{tab:fit0}. 
All fits have been done with \texttt{fitnlm} of \texttt{MATLAB}. 
The coefficients were set to zero by default if \texttt{fitnlm} returned a significant
p-value\footnote{The p-value of \texttt{fitnlm} indicates the probability of 
a specific coefficient to be zero as can be inferred from the data. 
In the following we simply refer to this quantity as the p-value.}, i.e. $\gtrapprox 0.3$.

\begin{figure*}[t]
  \center
  \includegraphics[width=0.185\textwidth]{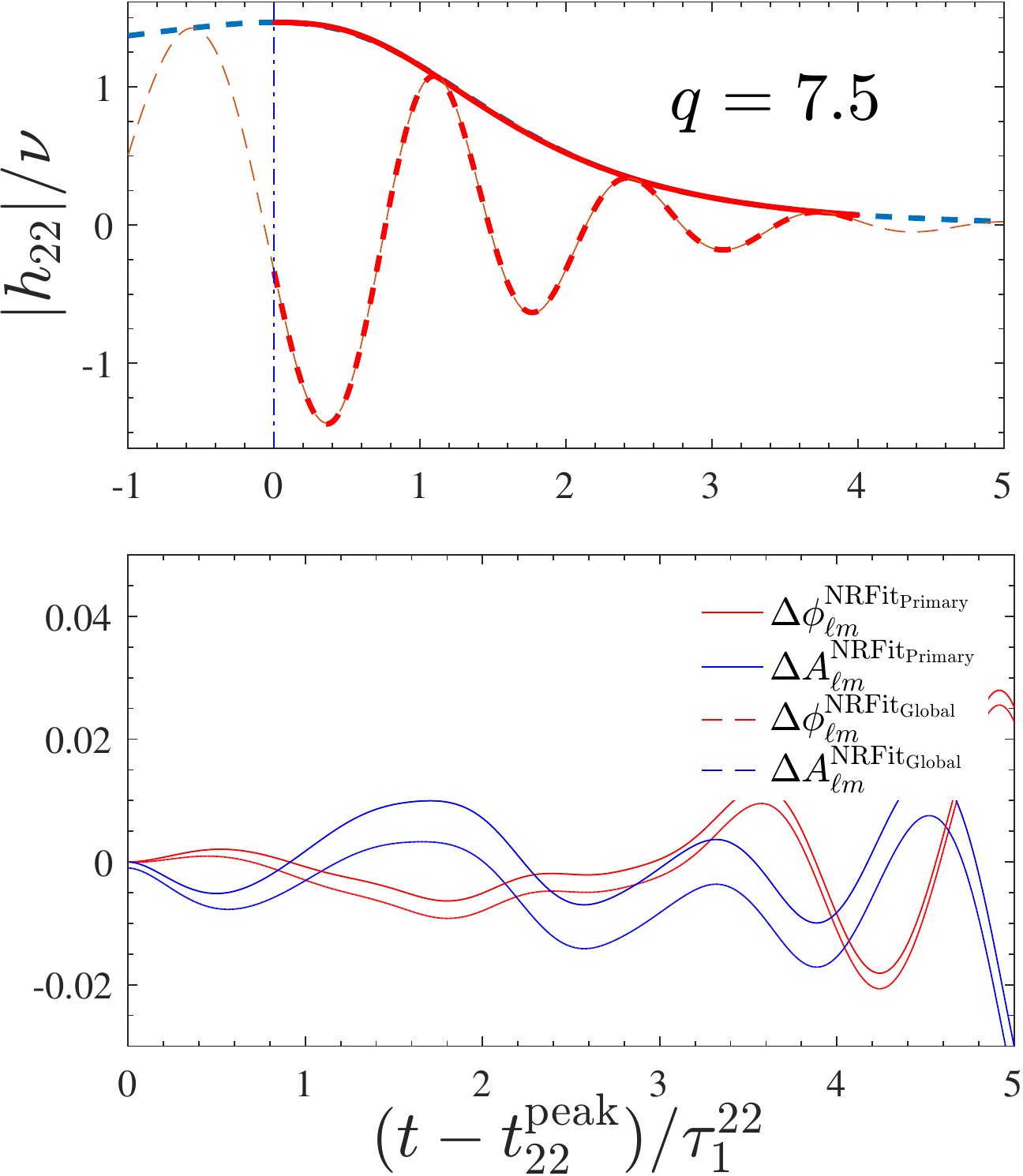}
  \includegraphics[width=0.19\textwidth]{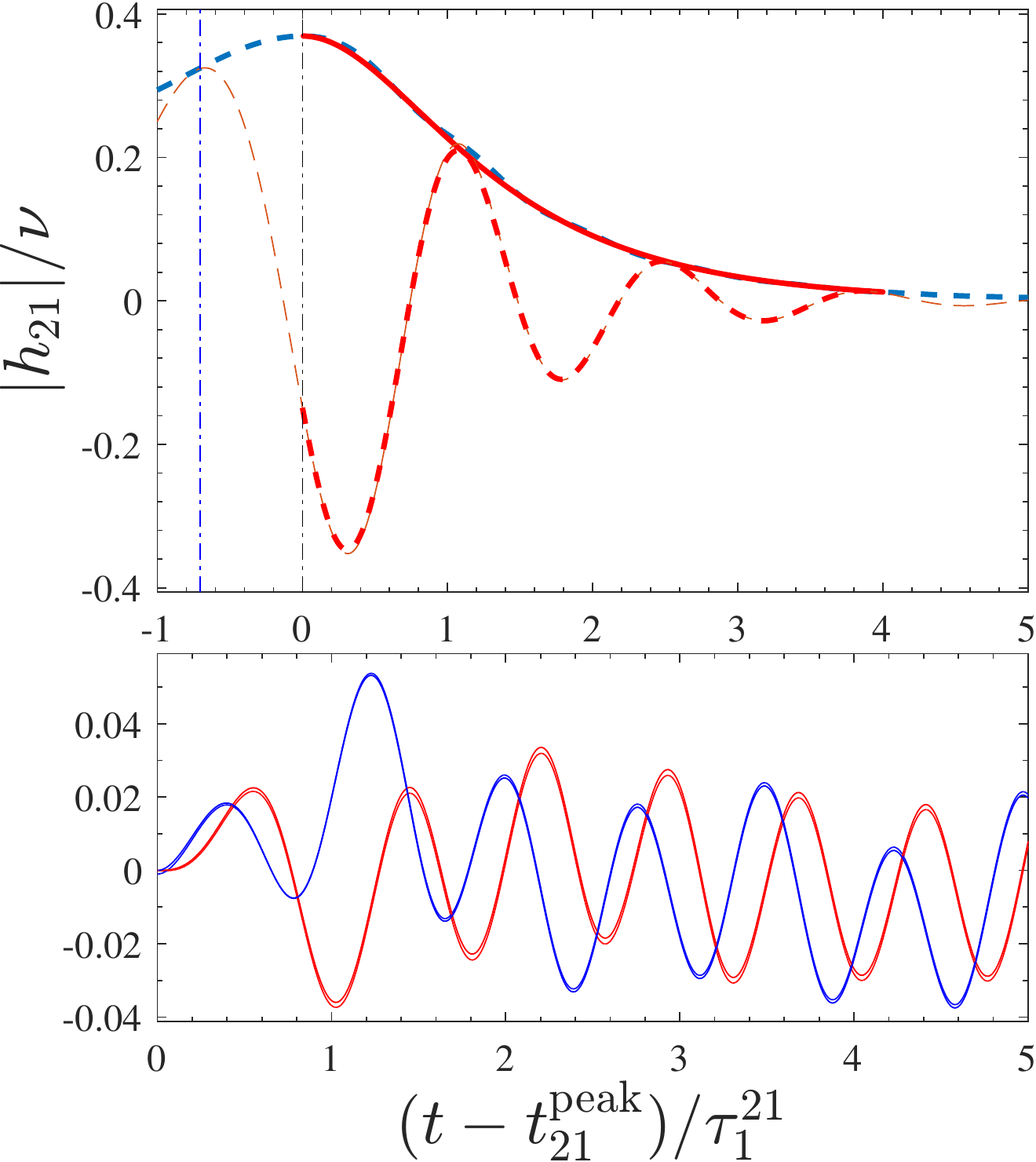}
  \includegraphics[width=0.19\textwidth]{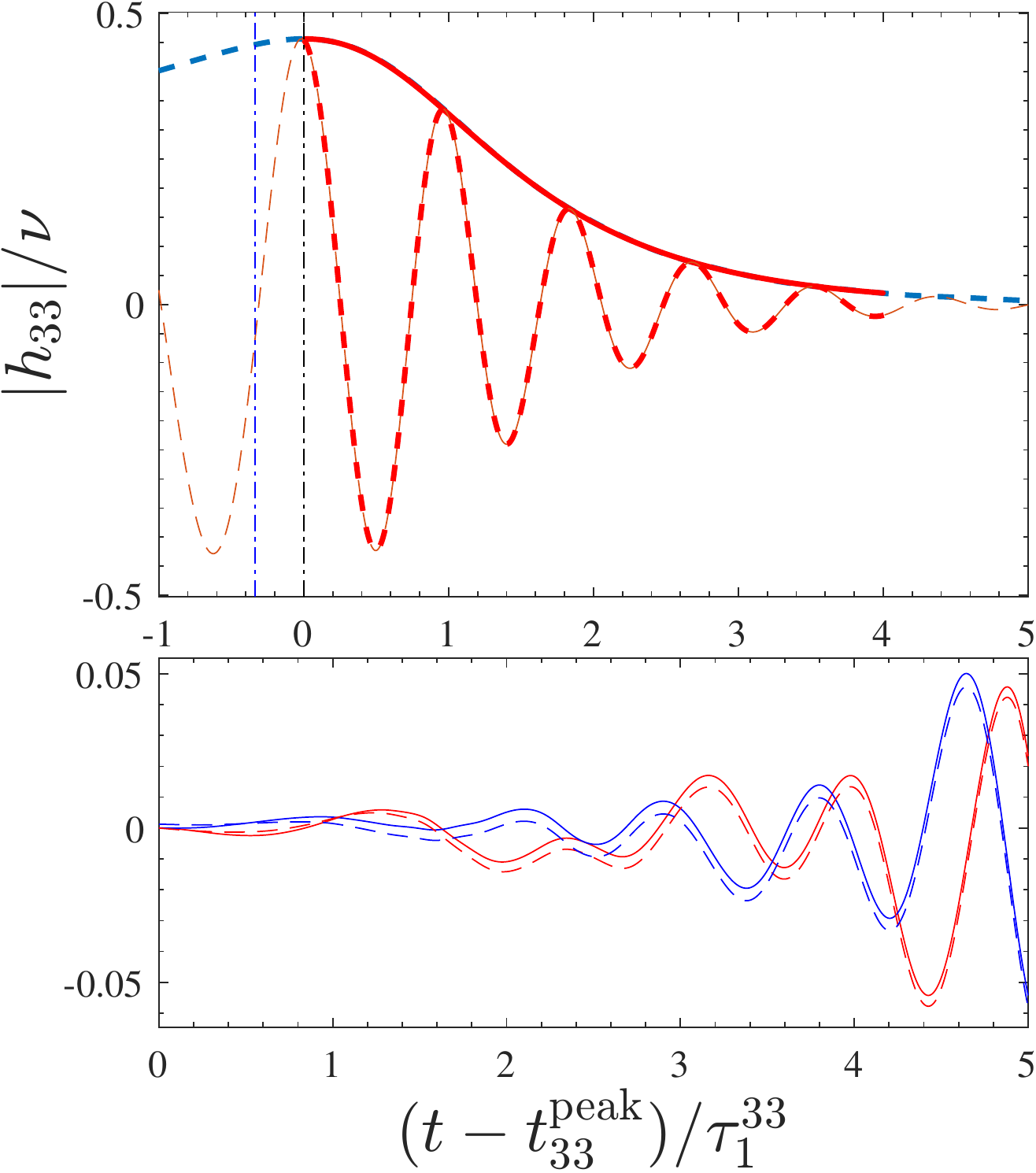}
  \includegraphics[width=0.185\textwidth]{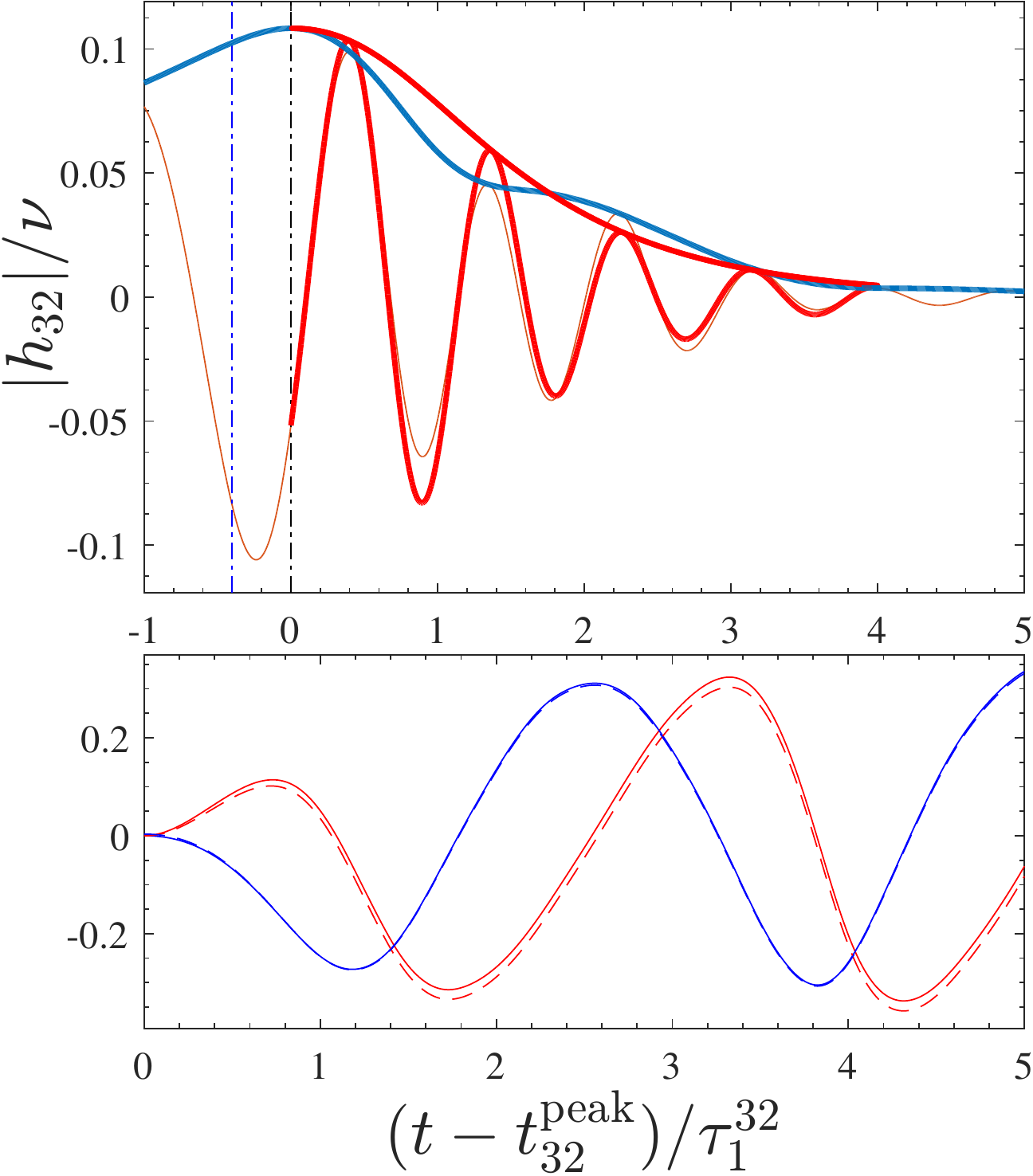}
  \includegraphics[width=0.19\textwidth]{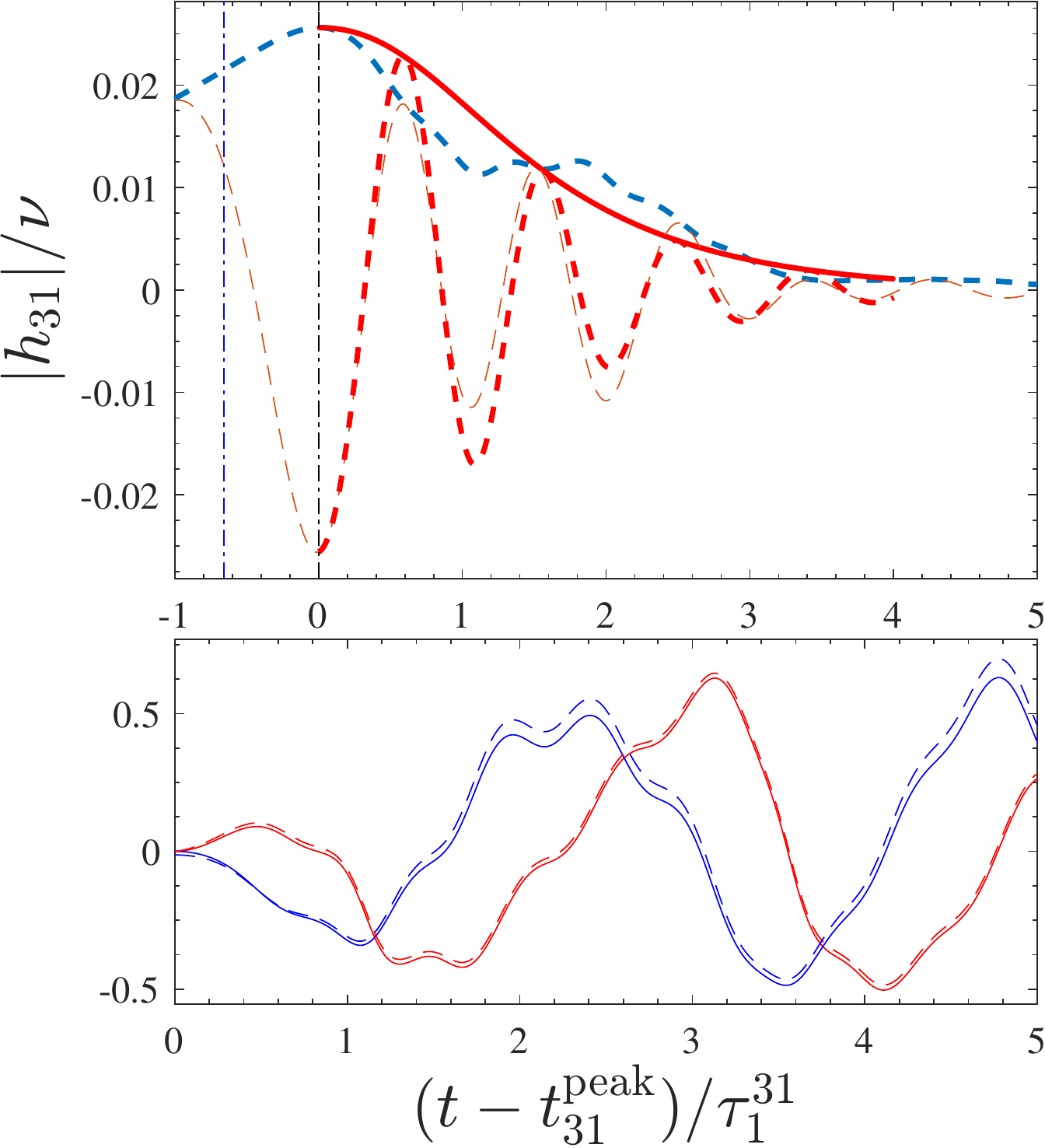}\\
  \includegraphics[width=0.19\textwidth]{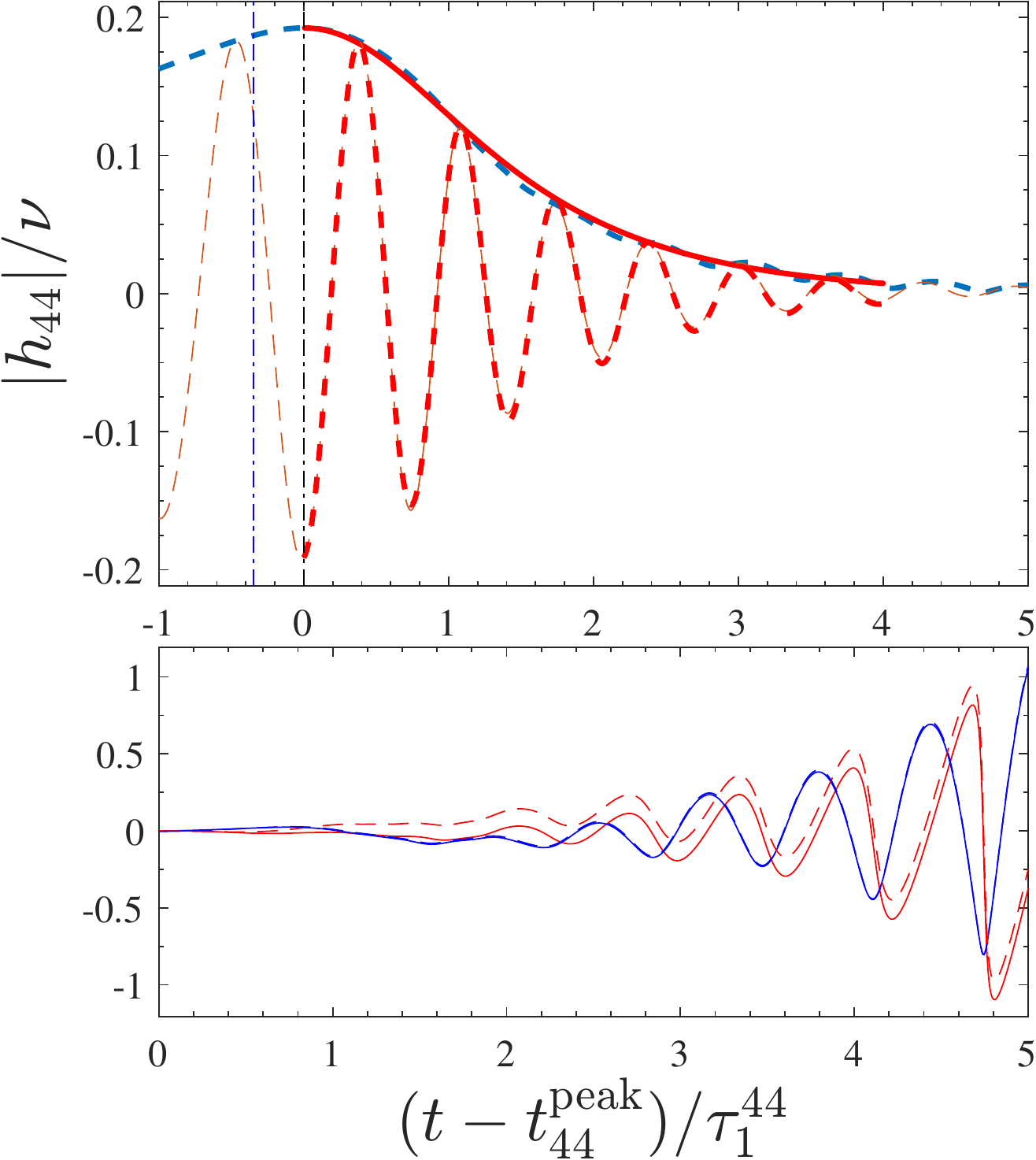}
  \includegraphics[width=0.19\textwidth]{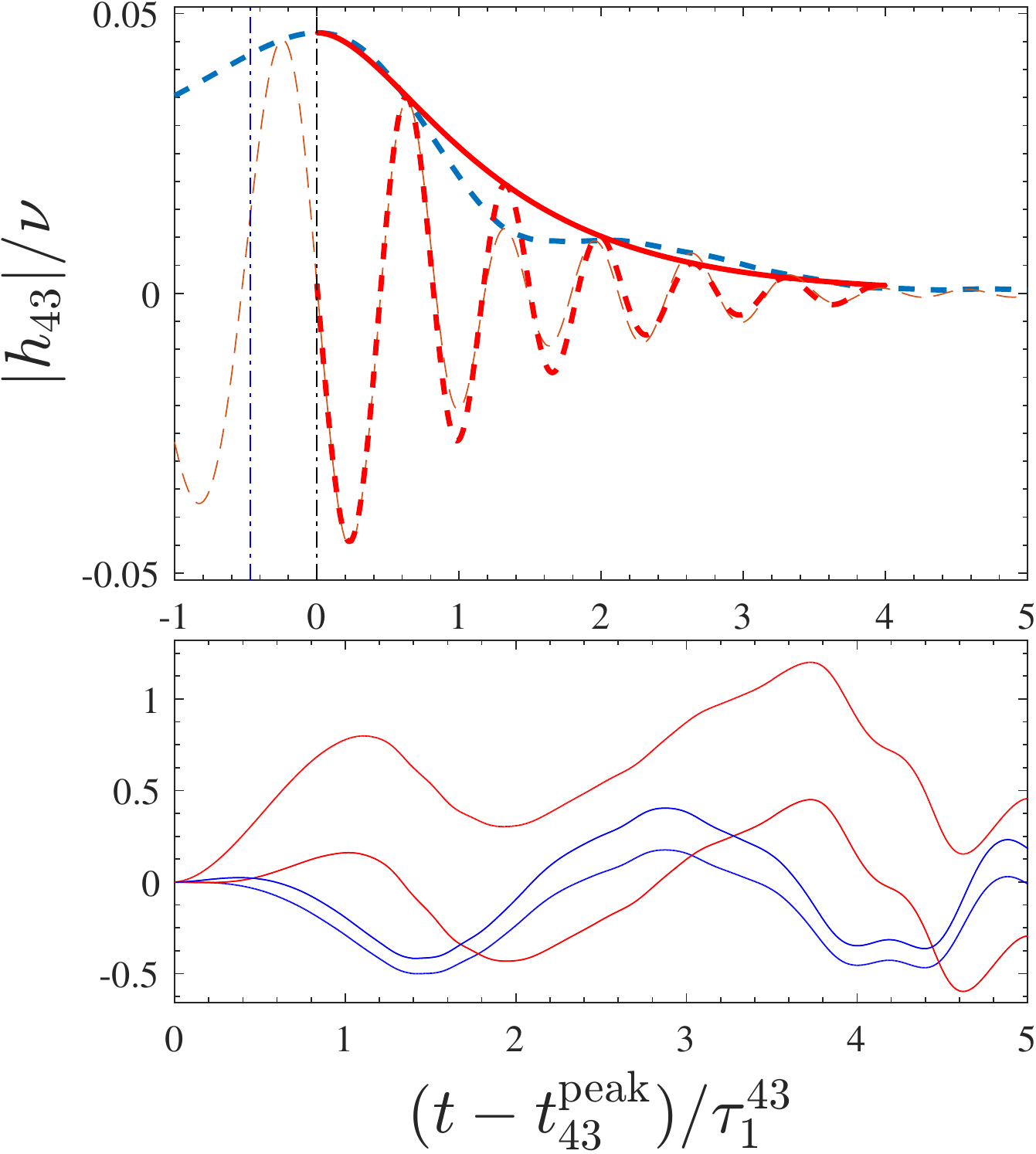}
  \includegraphics[width=0.192\textwidth]{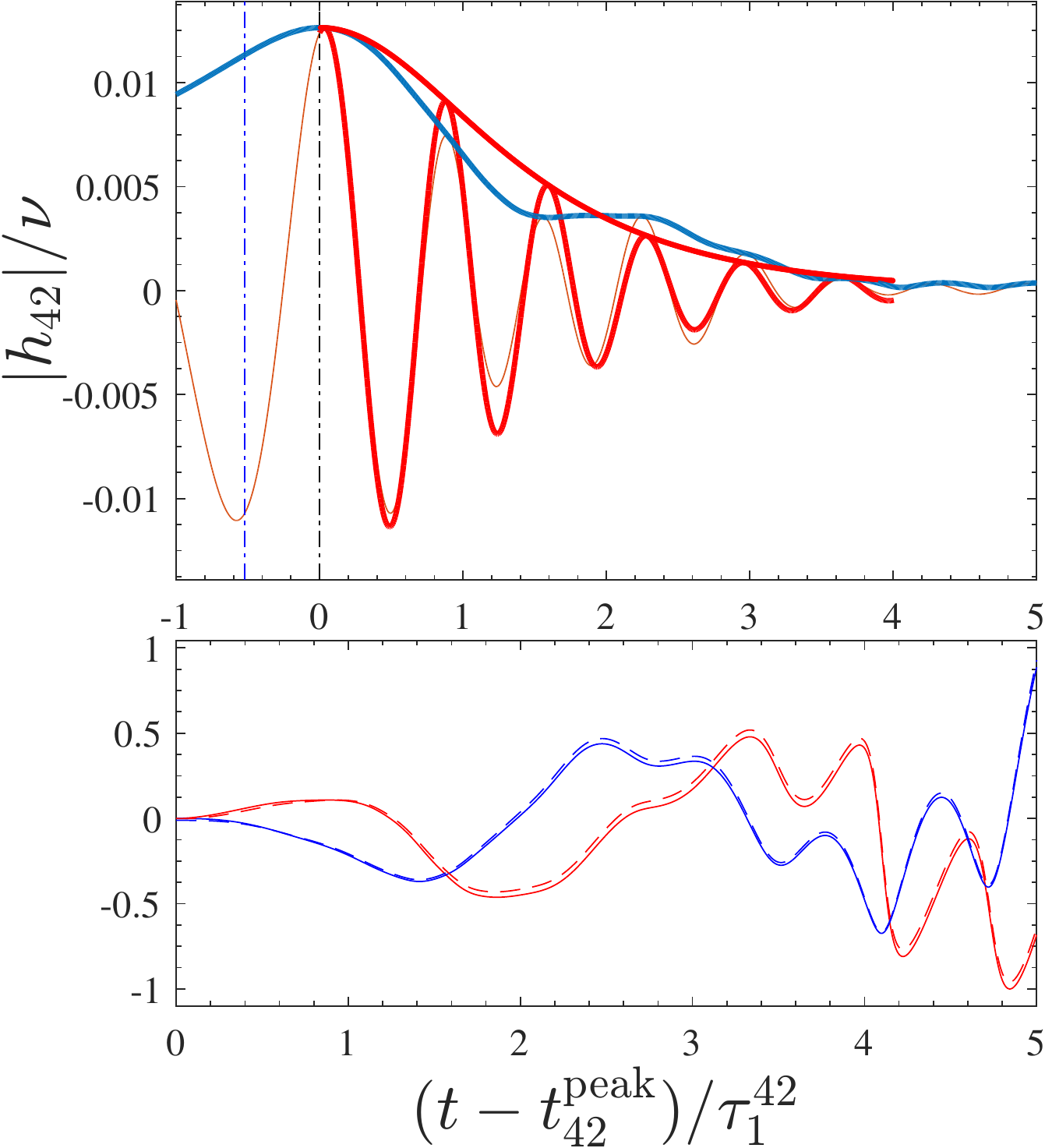}
  \includegraphics[width=0.185\textwidth]{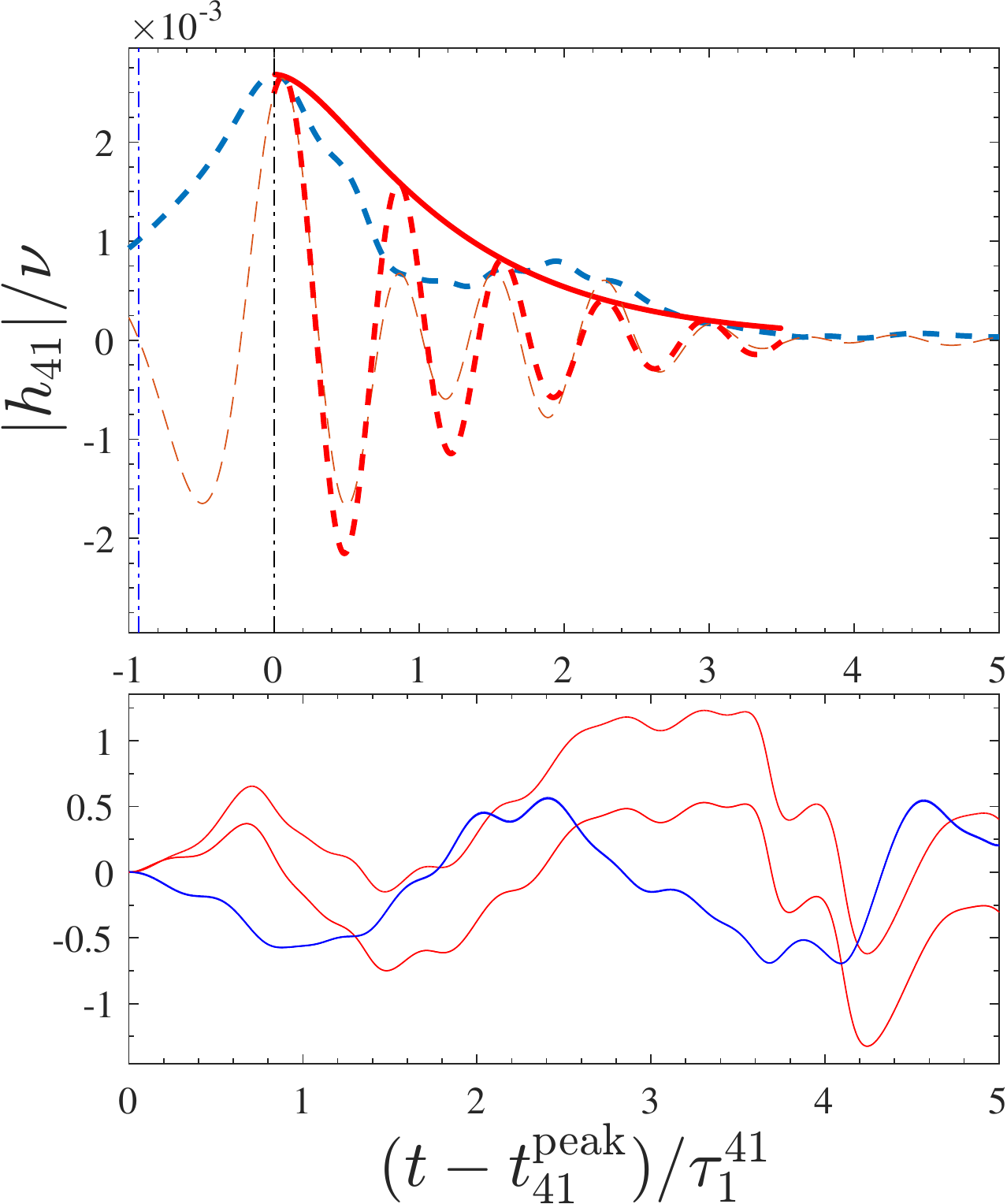}
  \includegraphics[width=0.19\textwidth]{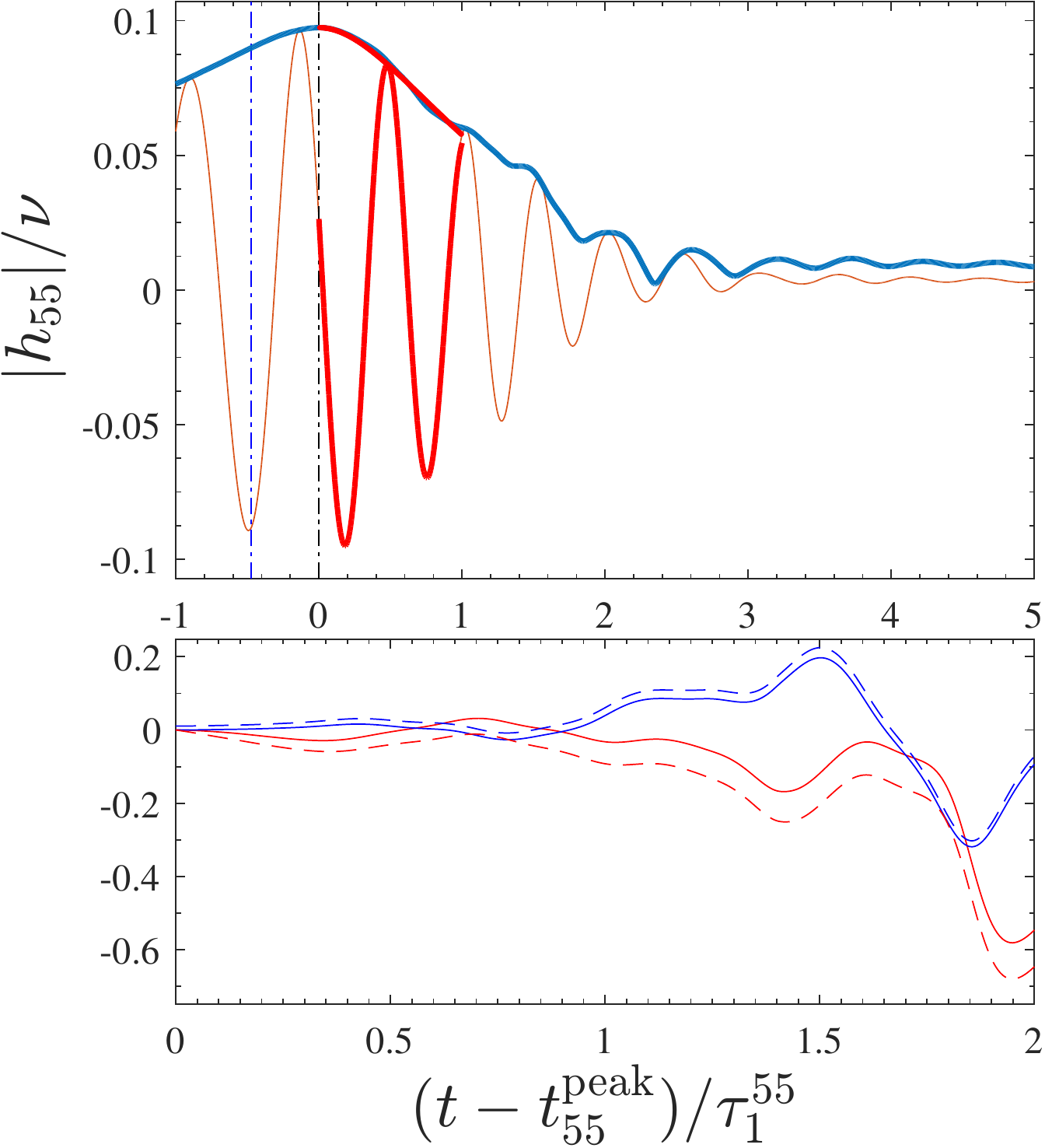}
  \caption{\label{fig:primary_fit} Performance of the primary postpeak fit and and of the global 
  interpolating fit on dataset SXS:BBH:0299 with mass ratio $q=7.5$. We here consider the 
  modes  $(\ell,m)=(2,2),(2,1),(3,3),(3,2),(3,1),(4,4),(4,3),(4,2),(4,1)$ and $(5,5)$.
  The panel of each mode is divided into two subpanels. In the top one, the tick-red lines  
  represent the waveform (amplitude, solid, and real part, dashed) obtained evaluating the 
  primary fit of the parameters of Eqs.~\eqref{eq:A_template}-\eqref{eq:phi_temp}. By contrast, 
  the thin orange line is the real part of the NR waveform, while the corresponding modulus is depicted 
  as a blue, dashed, thick line. The vertical line in black marks $t_{\rm peak}^{\lm}$, while 
  the blue one is $t_{22}^{\rm peak}$. For each mode, the time $t-t_{\lm}^{\rm peak}$ 
  is expressed in units of $\tau_1^{\lm}\equiv 1/\alpha_1^{\lm}$, the damping time of the corresponding 
  $(\ell,m)$ fundamental QNM. In the bottom subpanel we show the phase difference 
  between the NR and the fit for the phase (red online) and the {\it fractional} amplitude
  difference (blue online). The differences with the primary fit are shown as solid lines, 
  while those with the the global, interpolating, fit as dashed lines. One sees a more than
  acceptable consistency between the performances of the two fits.}
\end{figure*}

\subsection{Fits: postpeak waveform and ringdown}
\label{sec:postpeak_nospin}
As mentioned above, for the postpeak multipolar waveform we adopt
a 2-step fitting procedure: (i) for each NR dataset considered, 
we perform a primary fit, to determine the parameters
$(c_3^{A_\lm},c_3^{\phi_\lm},c_4^{\phi_\lm})$
that describe the postpeak behavior for each NR waveform at our disposal; 
then (ii) these parameters are fitted versus $\nu$ to
get the global interpolating fits. 
The $\ell=m=2$ postpeak fits
are informed using all the 19 datasets in Table~\ref{tab:NR_Waveforms}.
By contrast, only subsamples are used for the higher modes, depending
on the accuracy of the corresponding waveform.
More precisely we use the following datasets (numbering follows Table~\ref{tab:NR_Waveforms}):$\{2-16,18,19\}$ for $(2,1)$;  $\{2-19\}$
for $(3,3)$; $\{1-15,17,18\}$  for $(3,2)$; $\{2-11,13,14,17,18\}$ for $(3,1)$; $\{1-11,13,15-19\}$ for $(4,4)$; 
$\{2-9,13-19\}$ for $(4,3)$; $\{1-8,10-14,17-18\}$ for $(4,2)$
and $\{3,6,9-19\}$ for $(5,5)$.
For each $(\ell,m)$, the primary fit is always performed over a time interval
$\Delta \tau_\lm$. We choose   $\Delta \tau_\lm=4\tau^\lm_1=4/\alpha^\lm_1$
for $\ell=m=2$ as well as for all other multipoles (except $\ell=m=5$) for
datesets $\{10-19\}$ (corresponding to $q\geq 6$). 
For datasets $\{1-9\}$(corresponding to $q\leq 5.5$) and for the $\ell =m=5$ mode all over, 
we use $\Delta \tau_\lm=\tau^\lm_1$. 
This choice was partly driven by data-quality issues and partly by
the presence of mode mixing (see bottom, right panel of Fig.~\ref{fig:primary_fit} 
as an example of the data-quality issues in the $(5,5)$ mode).
The performances of both the primary and global fits are illustrated
in Fig.~\ref{fig:primary_fit}, that refers to the illustrative dataset
SXS:BBH:0299, with $q=7.5$. For this specific comparison we
are plotting $h_\lm/\nu$ instead of $\Psi_\lm$. For convenience,
for each $(\ell,m)$, time is expressed using the variable
$(t-t_\lm^{\rm peak})/\tau_1^\lm$. As mentioned above, the rightmost,
bottom, panel of the figure illustrates how the numerical noise shows up
already at $t-t_{55}^{\rm peak}\approx \tau^{55}_1$. The temporal interval
where the primary fit is performed is highlighted with the thick-red lines
(solid for the modulus, dashed for the real part) in the top part of each
panel. The fact that our post-peak templates lacks, by design, of the
possibility of accommodating any type of mode mixing results apparent
from inspecting the figure (especially the amplitude and phase differences,
that are displayed in the bottom part of each panel). For example,
it is well known that the ringdown part of the $(2,1)$ modes is mostly 
represented by a superposition of the fundamental modes 
with $\sigma^{\pm}_1=\alpha^{21}_1 \pm \ii\omega^{21}_1$; similarly, 
the $(3,2)$ multipole incorporates both the $\ell=2$, $m=2$ and $\ell=3$, $m=2$ QNMs 
frequencies of the final black hole [similar considerations hold for $(3,1)$, $(4,3)$ and $(4,2)$]
because the waveform is expanded in spin-weighted spherical harmonics and not
along the basis of the spheroidal harmonics that is naturally associated to the
finally formed black hole. The lack of modelization of this effect is responsible
of the fact that, for these modes, the fit residuals show a constant-amplitude
oscillation, instead of being (essentially) flat as they are supposed to be in the
case of the $\ell=m$ modes, where the effect of mode mixing is usually
largely suppressed (though it may increase with the mass ratio, see Ref.~\cite{Bernuzzi:2010ty}).
Note however that the corresponding plots are still showing oscillations that grow with
time. This effect mostly comes from numerical noise that gets amplified by the
radius-extrapolation procedure. We also need to highlight that in the bottom
part of each panel we show both the residuals with the primary fits (solid lines)
and with the global interpolating fits (dashed lines). The plot proves, on average,
a more than acceptable consistency and reliability of the global interpolating fit.
In conclusion, the postmerger template we are using here gives a simple and effective,
although certainly physically incomplete, representation of the actual physics.
We shall assess below the influence of this approximation on standard measures of merit.
\begin{table*}
	\caption{\label{tab:QNM_hlm} Parameters of the fitting function given by Eq.~\eqref{eq:QNM_hlm} used
	to fit the QNM parameters entering the phenomenological description of the postmerger waveform. We list here the fundamental QNM frequency
	$\omega^\lm_1$ and (inverse) damping time $\alpha_1^\lm$ as well as the difference $\alpha_{21}^\lm=\alpha^\lm_2-\alpha^\lm_1$.
	}
\begin{ruledtabular}
\begin{tabular}{l | l l | c | c c c | c c c }
	    ${Y^\prime}$			& $\ell$ 	& $m$ 	 	& ${Y^\prime}_0$ 		& $b^{Y^\prime}_1$ 		& $b^{Y^\prime}_2$ 	& $b^{Y^\prime}_3$	& $c^{Y^\prime}_1$ 		& $c^{Y^\prime}_2$ 		& $c^{Y^\prime}_3$ 	 \\ \hline\hline
	    $\omega_1^\lm$	& 2   	& 2     	& $0.373672$ 	& $-1.5367$ 	& $0.5503$ 	& \dots 	& $-1.8700$ 	& $0.9848$ 		& $-0.10943$ \\
	    			&   	& 1 		& $0.373672$ 	& $-0.79546$	& $-0.1908$ & $0.11460$ & $-0.96337$ 	& $-0.1495$ 	& $0.19522$ \\ \hline
	    			& 3 	& 3 		& $0.599443$ 	& $-1.84922$ 	& $0.9294$ 	& $-0.07613$& $-2.18719$ 	& $1.4903$ 		& $-0.3014$\\
	    			& 3   & 2  		& $0.599443$ 	& $-0.251$ 		& $-0.891$ 	& $0.2706$ 	& $-0.475$ 		& $-0.911$ 		& $+0.4609$\\
	    			& 3  & 1 		& $0.599443$ 	& $-0.70941$ 	& $-0.16975$& $0.08559$ & $-0.82174$ 	& $-0.16792$ 	& $0.14524$\\ \hline
	    			& 4	& 4 		& $0.809178$ 	& $-1.83156$ 	& $0.9016$ 	& $-0.06579$& $-2.17745$ 	& $1.4753$ 		& $-0.2961$\\
	    			& 4 & 3 		& $0.809178$ 	& $-1.8397$ 	& $0.9616$ 	& $-0.11339$& $-2.0979$ 	& $1.3701$ 		& $-0.2675$\\
	    			& 4 & 2 		& $0.0941640$ 	& $-1.44152$ 	& $0.0542$ 	& $0.39020$ & $-1.43312$ 	& $0.1167$ 		& $0.32253$\\
	    			& 		& 1 		& $0.0941640$ 	& $1.1018882$ 	& $-0.88643$& $-0.78266$& $1.1065495$ 	& $-0.80961$ 	& $-0.68905$\\ \hline\hline
	    $\alpha_1^\lm$	& 2   	& 2     	& $0.08896$ 	& $-1.90036$ 	& $0.86200$ & $0.0384893$ 	& $-1.87933$& $0.88062$ & \dots\\
	    			&  		& 1 		& $0.0889623$ 	& $-1.31253$ 	& $-0.21033$& $0.52502$ 	& $-1.30041$& $-0.1566$ & $0.46204$\\\hline
	    			& 3		& 3 		& $0.0927030$ 	& $-1.8310$ 	& $0.7568$ 	& $0.0745$ 		& $-1.8098$ & $0.7926$ 	& $0.0196$\\
	    			&  		& 2 		& $0.0927030$ 	& $-1.58277$ 	& $0.2783$ 	& $0.30503$ 	& $-1.56797$& $0.3290$ 	& $0.24155$\\
	    			&  		& 1 		& $0.0927030$ 	& $-1.2345$ 	& $-0.30447$& $0.5446$ 		& $-1.2263$ & $-0.24223$& $0.47738$\\ \hline
	    			& 4 	& 4 		& $0.0941640$ 	& $-1.8662$ 	& $0.8248$ 	& $0.0417$ 		& $-1.8514$ & $0.8736$ 	& $-0.0198$\\
	    			&  		& 3 		& $0.0941640$ 	& $-1.7177$ 	& $0.5320$ 	& $0.1860$ 		& $-1.7065$ & $0.5876$ 	& $0.120939$\\
	    			&  		& 2 		& $0.190170$ 	& $-1.38840$ 	& \dots 	& $0.39333$ 	& $-1.37584$& $0.0600017$& $0.32632$\\
	    			&  		& 1 		& $0.190170$ 	& $1.0590157$ 	& $-0.8650630$& $-0.75222$ 	& $1.0654880$& $-0.7830051$& $-0.65814$\\ \hline\hline
	    $\alpha_{21}^\lm$& 2   	& 2     	& $0.184953$ 	& $-1.89397$ 	& $0.88126$ 	& $0.0130256$& $-1.83901$ 	& $0.84162$ & \dots\\
	    			&  		& 1 		& $0.184952$ 	& $-1.1329$ 	& $-0.3520$ 	& $0.4924$ 	& $-1.10334$ 	& $-0.3037$ & $0.4262$\\\hline
	    			& 3		& 3 		& $0.188595$ 	& $-1.8011$ 	& $0.7046$ 		& $0.0968$ 	& $-1.7653$ 	& $0.7176$ 	& $0.0504$\\
	    			&  		& 2 		& $0.188595$ 	& $-1.5212$ 	& $0.1563$ 		& $0.3652$ 	& $-1.4968$ 	& $0.1968$ 	& $0.3021$\\
	    			&  		& 1 		& $0.188595$ 	& $-1.035$ 		& $-0.3816$ 	& $0.4486$ 	& $-1.023$ 		& $-0.3170$ & $0.3898$\\ \hline
	    			& 4 	& 4 		& $0.190170$ 	& $-1.8546$ 	& $0.8041$ 		& $0.0507$ 	& $-1.8315$ 	& $0.8391$ 	& $-0.0051$\\
	    			&  		& 3 		& $0.190170$ 	& $-1.6860$ 	& $0.4724$ 		& $0.2139$ 	& $-1.6684$ 	& $0.5198$ 	& $0.1508$\\
	    			&  		& 2 		& $0.809178$ 	& $-0.6644$ 	& $-0.3357$ 	& $0.1425$ 	& $-0.8366$ 	& $-0.2921$ & $0.2254$\\
	    			&  		& 1 		& $0.809178$ 	& $-0.68647$ 	& $-0.1852590$ 	& $0.0934997$ & $-0.77272$ 	& $-0.1986852$& $0.1485093$\\
\end{tabular}
\end{ruledtabular}
\end{table*}
\begin{table}[t]
\caption{\label{tab:Dt_hlm_fits} 
The fit parameters to analytically represent the time lag between the peak of the $(\ell,m)$ 
waveform multipole and the peak of the $(2,2)$ mode, Eq.~\eqref{eq:Dt_lm}. The coefficients
refer to the functional form of Eqs.~\eqref{eq:Dt_peak_lm}-\eqref{eq:Dt_hat_peak_lm}. }
\begin{ruledtabular}
\begin{tabular}{l l | l c c  c c}
	$\ell$ & $m$ 	& $\Delta t^{\rm 0}_{\ell m}$ 	& $n_1^{\Delta t_{\lm}}$ & $n_2^{\Delta t_\lm}$ & ${d}_1^{\Delta t_\lm}$ & $d_2^{\Delta t_\lm}$\\ \hline
	2   &  1  	& $11.7900$        		& $-3.764$  	& $6.9051$ 	& \dots 		& \dots \\ \hline 
	3   &  3  	& $3.49238$        		& $-0.11298$ 	& $5.0056$ 	& \dots 		& \dots \\ 
	3   &  2  	& $ 9.22687$        	& $-11.398$  	& $33.244$ 	& $-8.1976$ 	& $19.537$\\ 
	3   &  1  	& $12.9338$        		& \dots   		& $-25.615$ & $0.88803$ 	& $16.292$ \\\hline 
	4   &  4  	& $5.28280$        		& $-8.4686$  	& $18.006$ 	& $-6.7964$ 	& $11.368$ \\ 
	4   &  3  	& $9.59669$        		& $-11.345$  	& $38.813$ 	& $-7.5049$ 	& $22.399$\\ 
	4   &  2  	& $11.9225$        		& $-3.8284$  	& $-12.399$ & \dots 	 	& \dots \\
	4   &  1  	& $13.1116$        		& $-9.6225$  	& $38.451$  & $-7.7998$  	& $32.405$ \\ \hline
	5   &  5  	& $6.561811$ 			& $-12.198$ 	& $40.327$  & $-11.501$  	& $39.431$ 
\end{tabular}
\end{ruledtabular}
\end{table}

\begin{figure}[t]
  \center
  \includegraphics[width=0.45\textwidth]{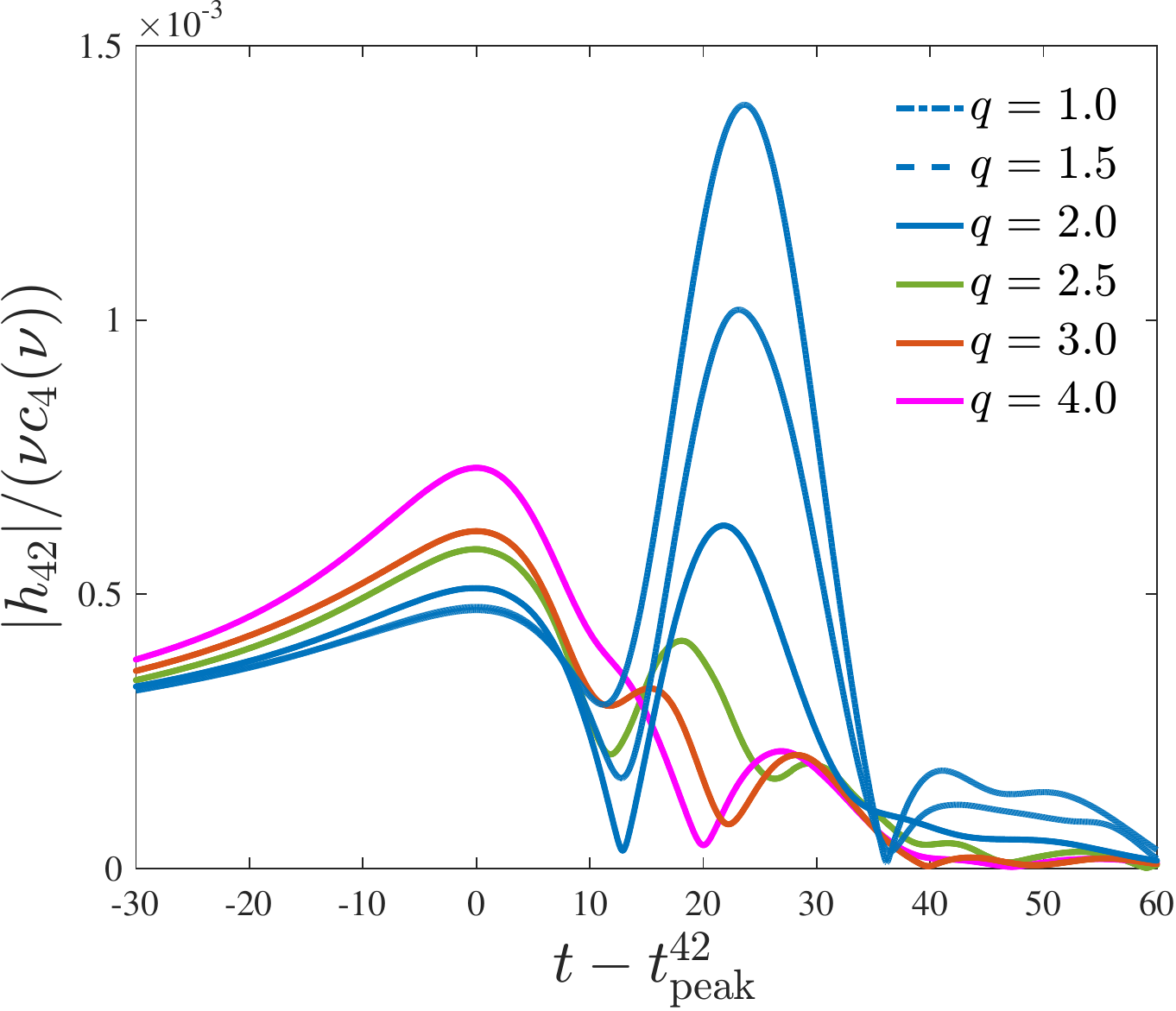}
  \caption{\label{fig:42peak} NR data for $q=7.5$: behavior of the post-peak
    amplitude for $(4,2)$, with a large secondary peak due to mode-mixing.}
\end{figure}

The fits of $(c_3^{A_\lm},c_3^{\phi_\lm},c_4^{\phi_\lm})$ were obtained
using the function \texttt{fitnlm} of \texttt{MATLAB}. The functional
form of the fitting function was adapted multipole by multipole,
so to have enough flexibility to reduce as much as possible the
differences in phase and amplitude, without, however, overfitting the data.
We mostly use rational functions that are explicitly listed
in Table~\ref{tab:postmerger_phase}.
Inspecting the table, one notes that the derivatives with respect to $\nu$
of $c_3^{\phi_{31}}$ and of $c_3^{\phi_{41}}$ are discontinuous
at $\nu=10/121$ (corresponding to $q=10$).
Similarly, the derivative with respect to $\nu$ for $c_4^{\phi_{42}}$
turns out to be discontinuous at both $\nu=10/49$ (i.e. $q=2.5$)
and at $\nu=3/16$ (i.e. $q=3$) as it turned out convenient to
have it represented by the following piece-wise function
\begin{align}
\!\!\!\frac{132.56-1155.5\nu+2516.8\nu^2}{1-3.8231\nu}&\qquad \text{if}\quad q \leq 2.5\nonumber,\\
  \label{eq:c4phi42}
\!\!\!  -554.18\nu + 120.23                             &\qquad \text{if}\quad 2.5< q < 3 , \\
\!\!\!  \frac{-0.58736+16.401\nu}{1-4.5202\nu}          &\qquad \text{if} \quad q\geq 3\nonumber .
\end{align}
The need of such functional representation is related to our approximation
of neglecting mode-mixing effects, whose impact depends on the mass ratio.
Indeed, one finds (see the illustrative Fig.~\ref{fig:42peak})
that the qualitative features of the $(4,2)$ multipole in the range
$1\leq q \leq 2.5$ are peculiar: the postmerger waveform amplitude
has a double-peaked structure due to mode-mixing, and in this range
of mass ratios the amplitude of the second peak is {\it larger}
than that of the first one. The lowering of the second peak is
rather abrupt with the mass ratio and occurs somewhere in the
interval $2 < q < 2.5$, where we do not have additional NR simulations.
Although such feature is nothing more than an artifact related to
having expressed the waveform in the basis of spherical harmonics
(instead of the natural spheroidal one), it is just approximately
represented by our, rather simplified, fit.

We also need fits of the QNMs quantities 
$(\omega^\lm_1,\alpha^\lm_1)$ and $\alpha_{21}^\lm\equiv \alpha_2^\lm-\alpha_1^\lm$
for all multipoles considered. Each of these parameters is represented as a
function of the dimensionless spin of the final black hole, $\hat{a}_f$, 
that reads
\begin{align}
\label{eq:QNM_hlm}
{Y^\prime}_{\ell m}\left(\hat{a}_f\right)  = {Y^\prime}_0\,\frac{1+b_1^{Y^\prime}\hat{a}_f+b_2^{Y^\prime}\hat{a}_f^2+b_3^{Y^\prime}\hat{a}_f^3}{1+c_1^{Y^\prime}\hat{a}_f+c_2^{Y^\prime}\hat{a}_f^2+c_3^{Y^\prime}\hat{a}_f^3}.
\end{align}
The values of $(\alpha^\lm_1,\omega_1^\lm,\alpha_{21}^\lm)$ to be fitted were obtained
as follows: first, we computed the value of the final spin $\hat{a}_f$
using the NR-informed fit presented in~\cite{Jimenez-Forteza:2016oae}; 
then, we interpolated the tables of Ref.~\cite{Berti:2005ys}. The coefficients of 
the fits above are collected in Table~\ref{tab:QNM_hlm}. All fits were done 
with \texttt{fitnlm} of \texttt{MATLAB} and coefficients have been set to zero
explicitly if the p-value was significant.
\begin{figure}[t]
  \center
  \includegraphics[width=0.45\textwidth]{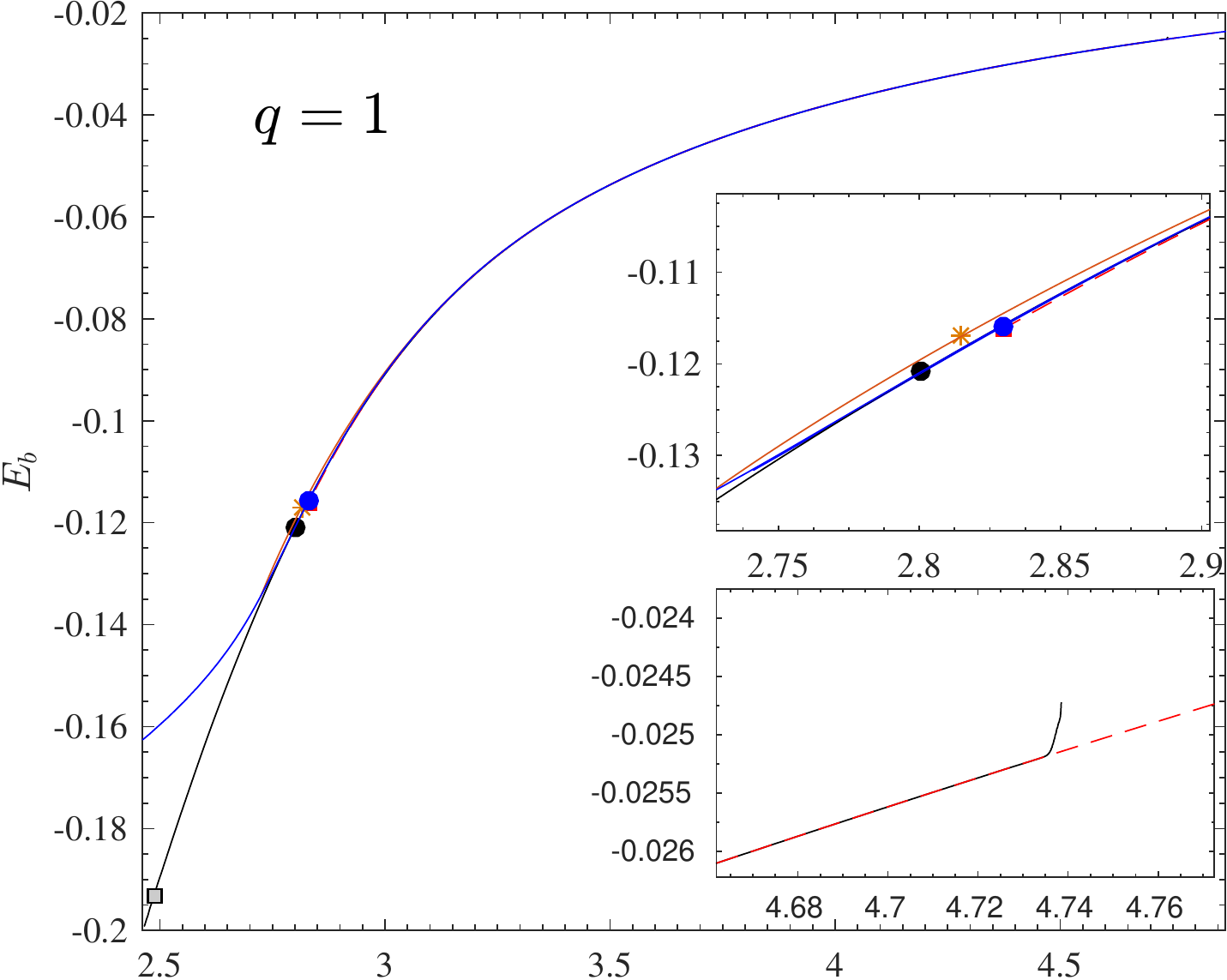}\\
  \vspace{5mm}
  \includegraphics[width=0.45\textwidth]{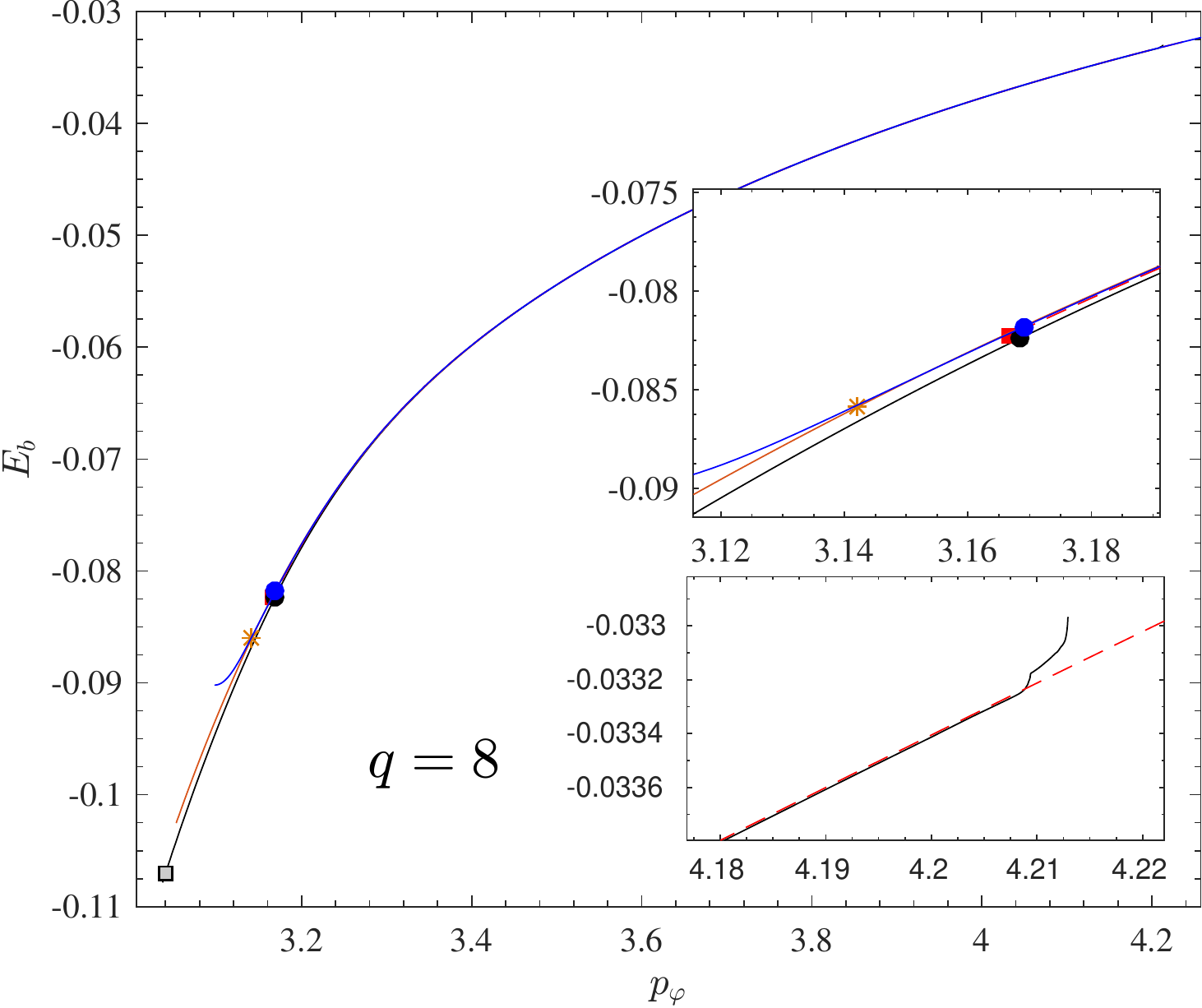}
  \caption{\label{fig:EvsJ} Energetics: comparison between three different EOB models
    and the SXS NR data computed in Ref.~\cite{Nagar:2015xqa} for two different binary
    configurations with mass ratio $q=1$ and $q=8$.
    Blue line: \TEOBiResumSM{};  Red (dashed) line: the \TEOBResumS{} model of 
    Ref.~\cite{Nagar:2015xqa} with $a_6^c(\nu)$ given by Eq.~\eqref{eq:a6c_old};  
    Orange line: {\tt SEOBNRv4}; black line, NR curve. The filled markers indicate 
    the merger points. The figure highlights the consistency between
    the two avatars of the \TEOBResumS{} model. Note that the discrepancy
    between {\tt SEOBNRv4} and the NR curve is much larger than the numerical
    uncertainty on the curve~\cite{Damour:2011fu,Nagar:2015xqa}.}
\end{figure}
\begin{figure}[t]
  \center
\includegraphics[width=0.45\textwidth]{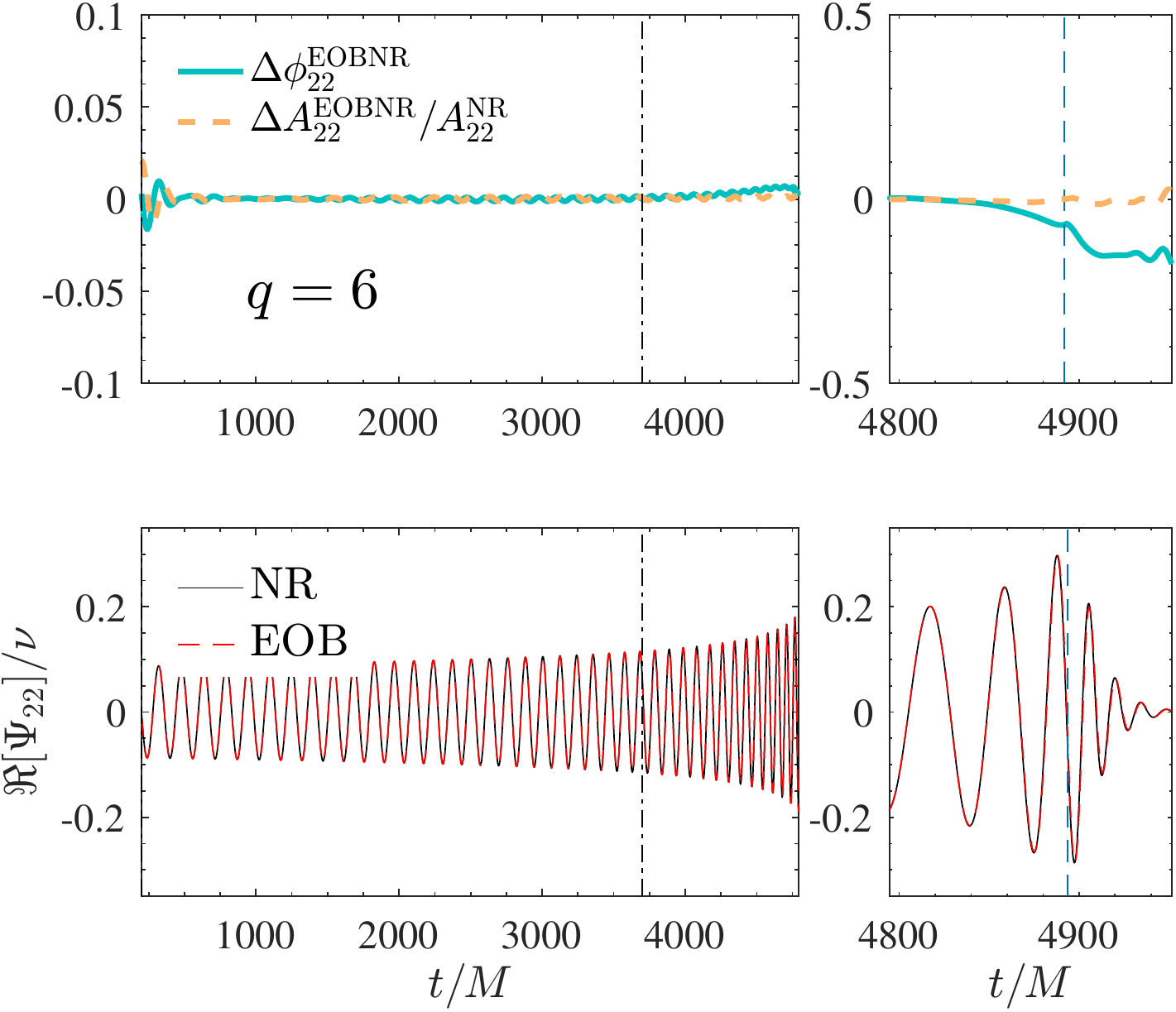}
\caption{Illustrative EOB/NR time-domain comparison for $q=6$
  (corresponding to SXS:BBH:0166) using \TEOBiResumSM{}. Top row: phase
  difference and relative amplitude differences. Bottom row: comparison
  between the real part of the waveform.}
  \label{fig:q6_l2}
\end{figure}
\begin{figure}[t]
  \center
\includegraphics[width=0.5\textwidth]{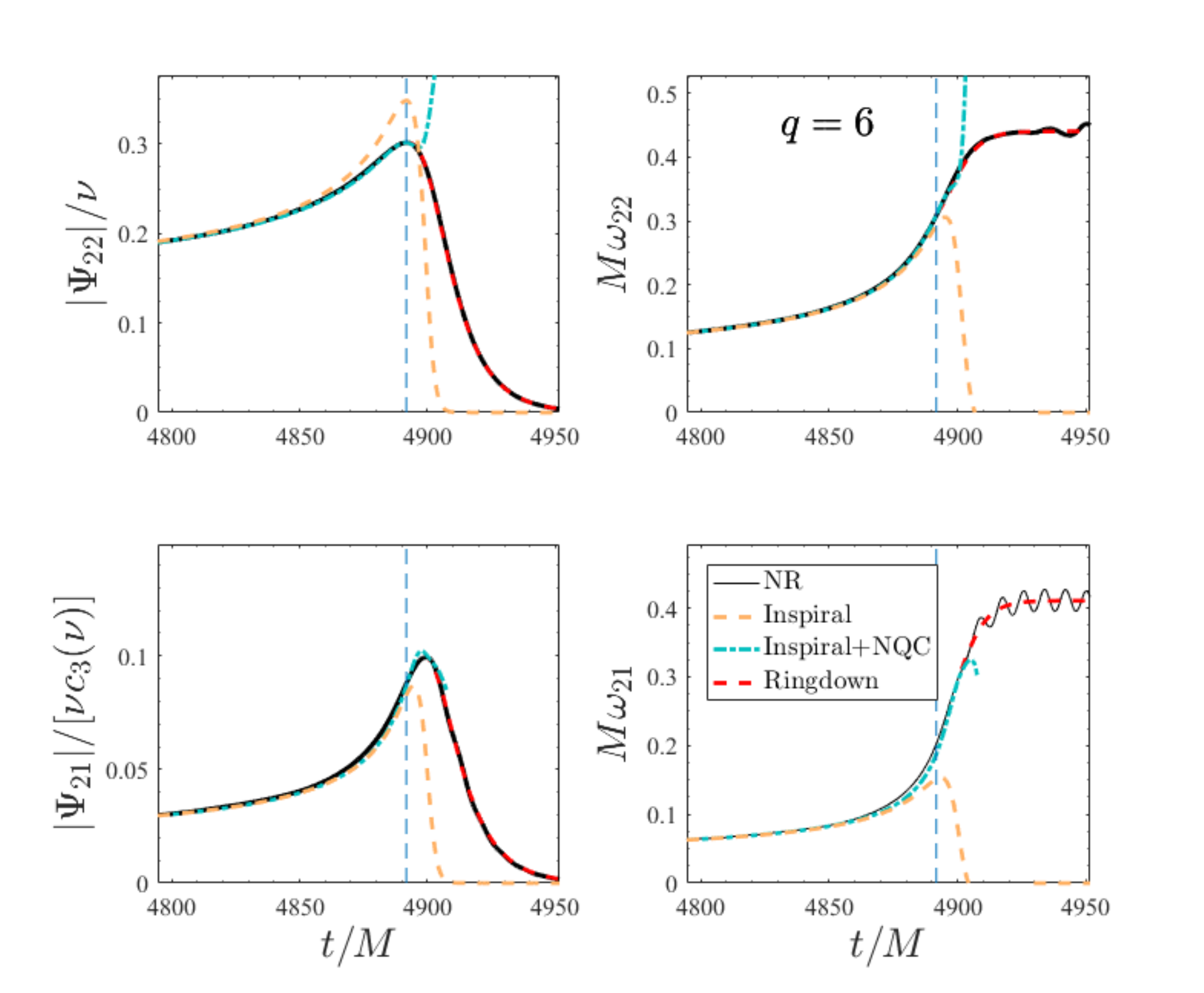}
\caption{\label{fig:q6_Aomg} Mass ratio $q=6$, SXS:BBH:0166.
  Complement to the phasing comparison of Fig.~\ref{fig:q6_l2}: frequency
  and amplitude for $(2,2)$ (top) and $(2,1)$ (bottom) multipoles obtained with \TEOBiResumSM.
  Orange (dashed) lines: purely analytical EOB waveform ($\hat{h}_{\lm}^{\rm NQC}=1$).
  Blue lines: NQC-improved waveform with NQC parameters determined by matching
  to a single NR waveform point. Red lines: postpeak-ringdown part. 
  Vertical line: location of the peak of the $(2,2)$ mode. Mode-mixing is not incorporated 
  in the analytical ringdown description, so the EOB frequency for the $(2,1)$ mode 
  saturates to the plateau and,  differently from the numerical one, does not
  exhibit any oscillation. Note the rather accurate representation of the frequency
  and amplitude already achievable, essentially up to merger, using the
  purely analytical (non-NQC corrected), EOB waveform.}
\end{figure}

The last step required to complete the postmerger model is to extract, from the NR simulations,
the time lag, as a function of $\nu$, between the peak of each multipolar mode and the $(2,2)$ one, i.e. 
\begin{align}
\label{eq:Dt_lm}
\Delta t_{\ell m}^{\rm NR}=t_{\ell m}^{\rm peak}-t_{22}^{\rm peak}.
\end{align}
To represent this as function of $\nu$, as usual we factor
out the test-particle values $\Delta t_\lm^0$ (see Table~3 of Ref.~\cite{Harms:2014dqa}), as
\begin{align}
\label{eq:Dt_peak_lm}
\Delta t_{\ell m}^{\rm NR}= \Delta t^0_{\ell m} \hat{\Delta} t_{\ell m},
\end{align}
and fit the correction $\hat{\Delta} t_\lm$ versus $\nu$. The
fits are done with the following, general, functional form
\begin{align}
\label{eq:Dt_hat_peak_lm}
\hat{\Delta} t_{\ell m} = \frac{1+n_1^{\Delta t_\lm}\nu+n_2^{\Delta t_\lm}\nu^2}{1+d^{\Delta t_\lm}_1\nu+d^{\Delta t_\lm}_2\nu^2}.
\end{align}
The coefficients of the fits, together with the values of $\Delta t_\lm^0$, are listed
in Table~\ref{tab:Dt_hlm_fits}.
The fits have been done with \texttt{fitnlm} of \texttt{MATLAB}. Coefficients have 
been set to zero explicitly if the p-value was significant. 

\begin{figure}[t]
\begin{center}
  \includegraphics[width=0.45\textwidth]{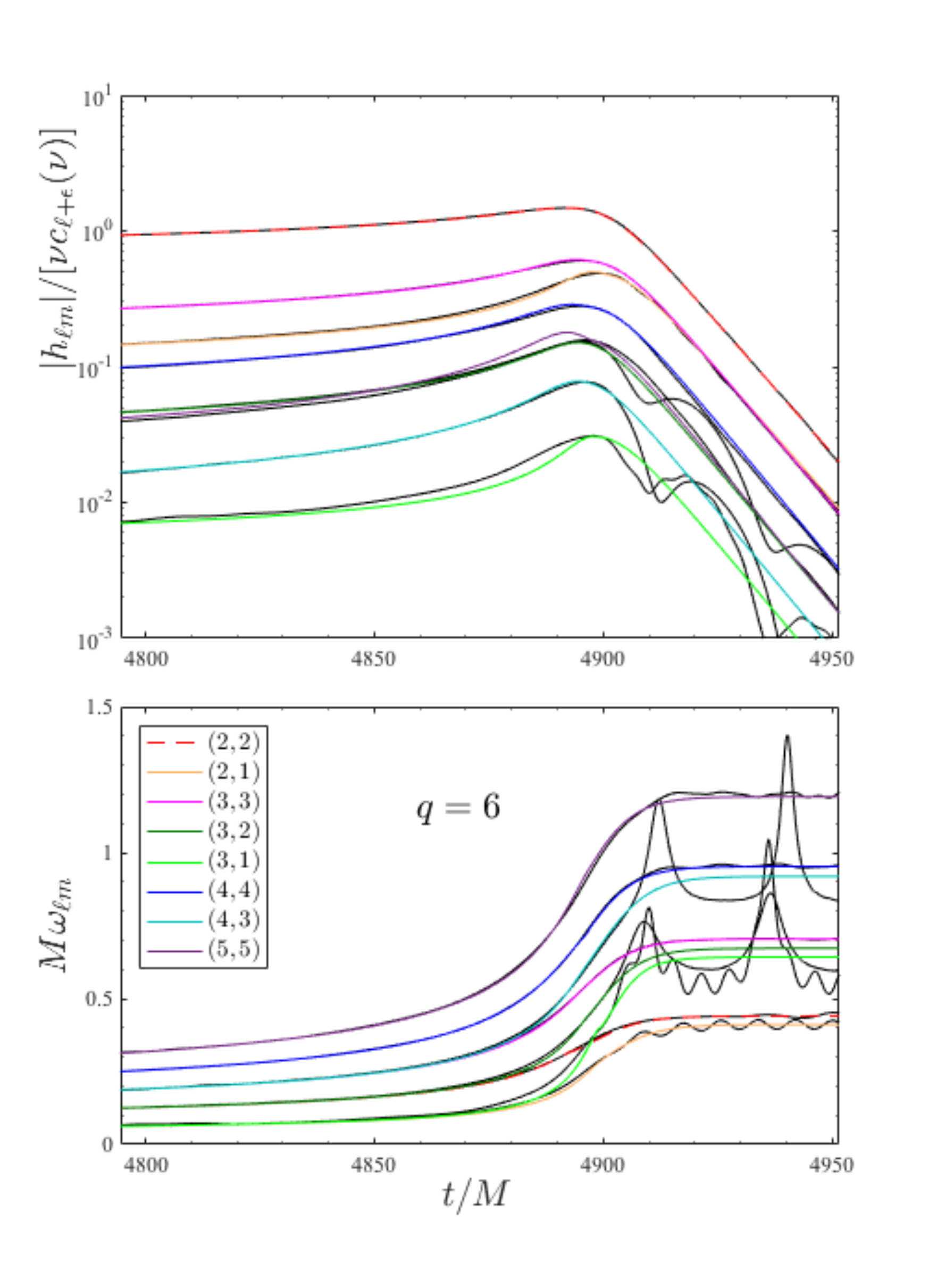}
  \caption{Mass ratio $q=6$, SXS:BBH:0166 dataset (black lines). 
    EOB/NR comparison between amplitudes $| h_{\ell m} (t) |$ (top panel) 
    and frequencies $\omega_{\ell m} (t)$ (bottom panel).
    For readability, here the modes have been normalized by
    $\nu c_{\ell + \epsilon} (\nu)$.}
\label{fig:q6_all}
\end{center}
\end{figure}

\begin{figure}[t]
  \center
  \includegraphics[width=0.23\textwidth]{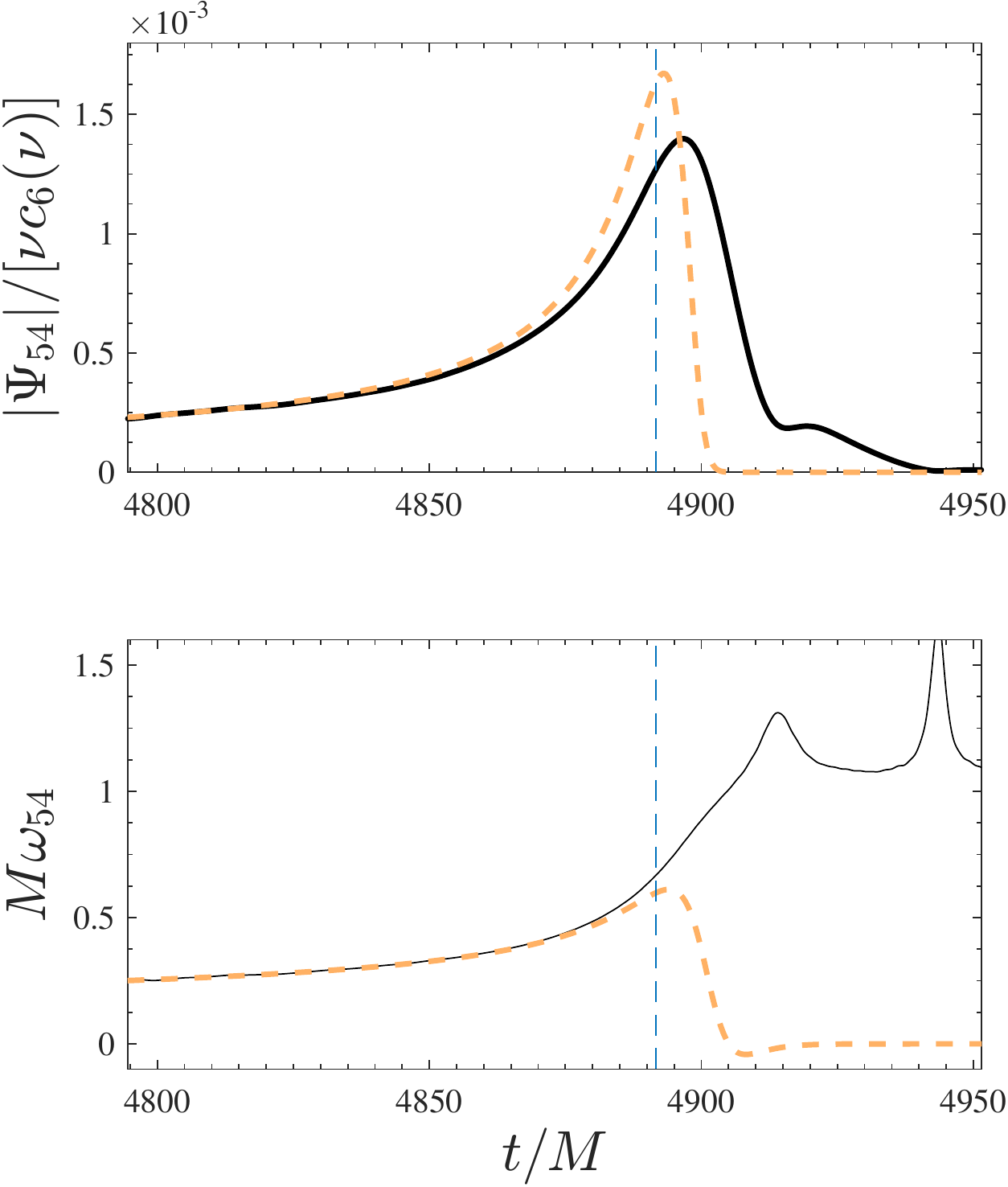}
  \includegraphics[width=0.23\textwidth]{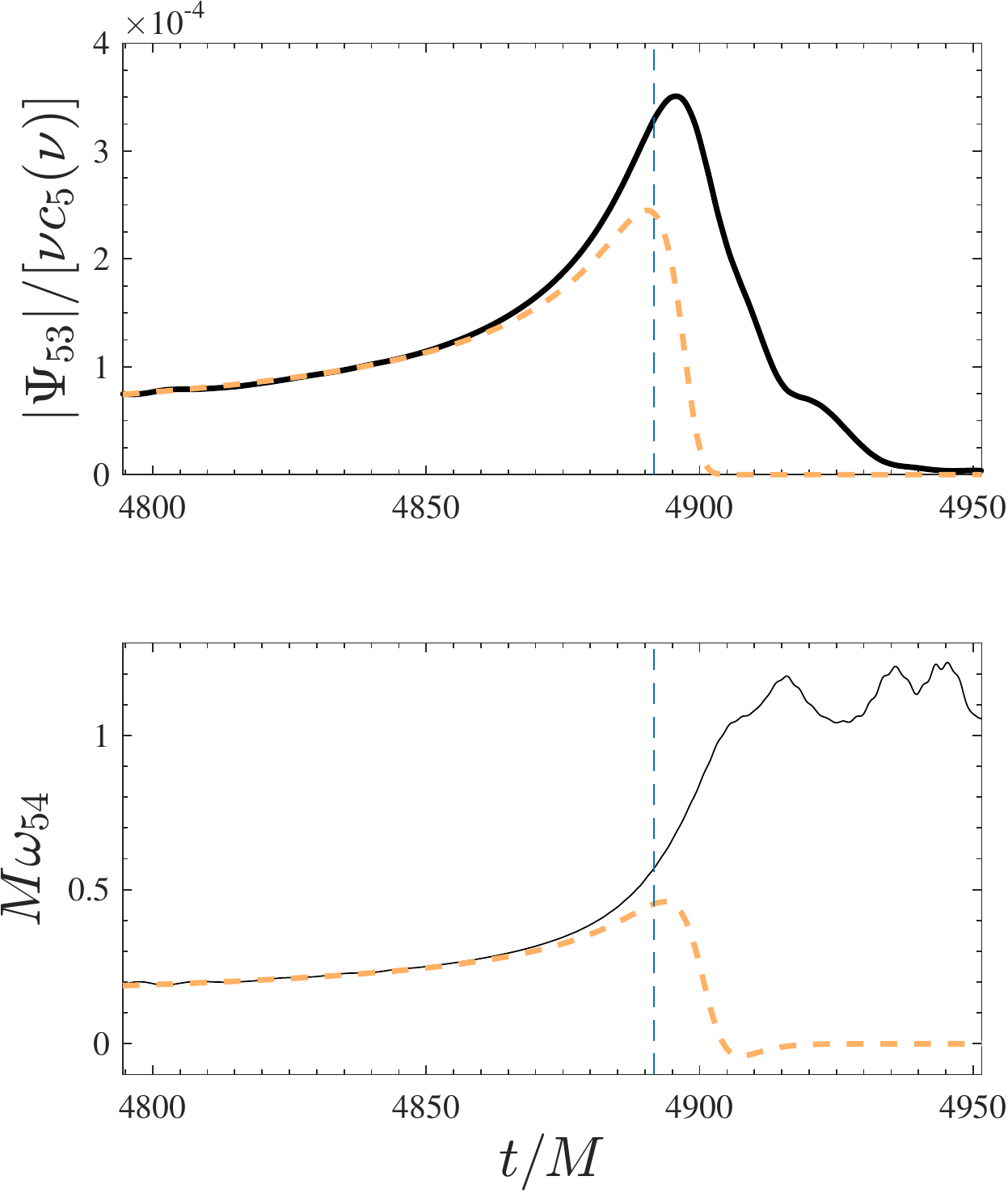}\\
  \includegraphics[width=0.23\textwidth]{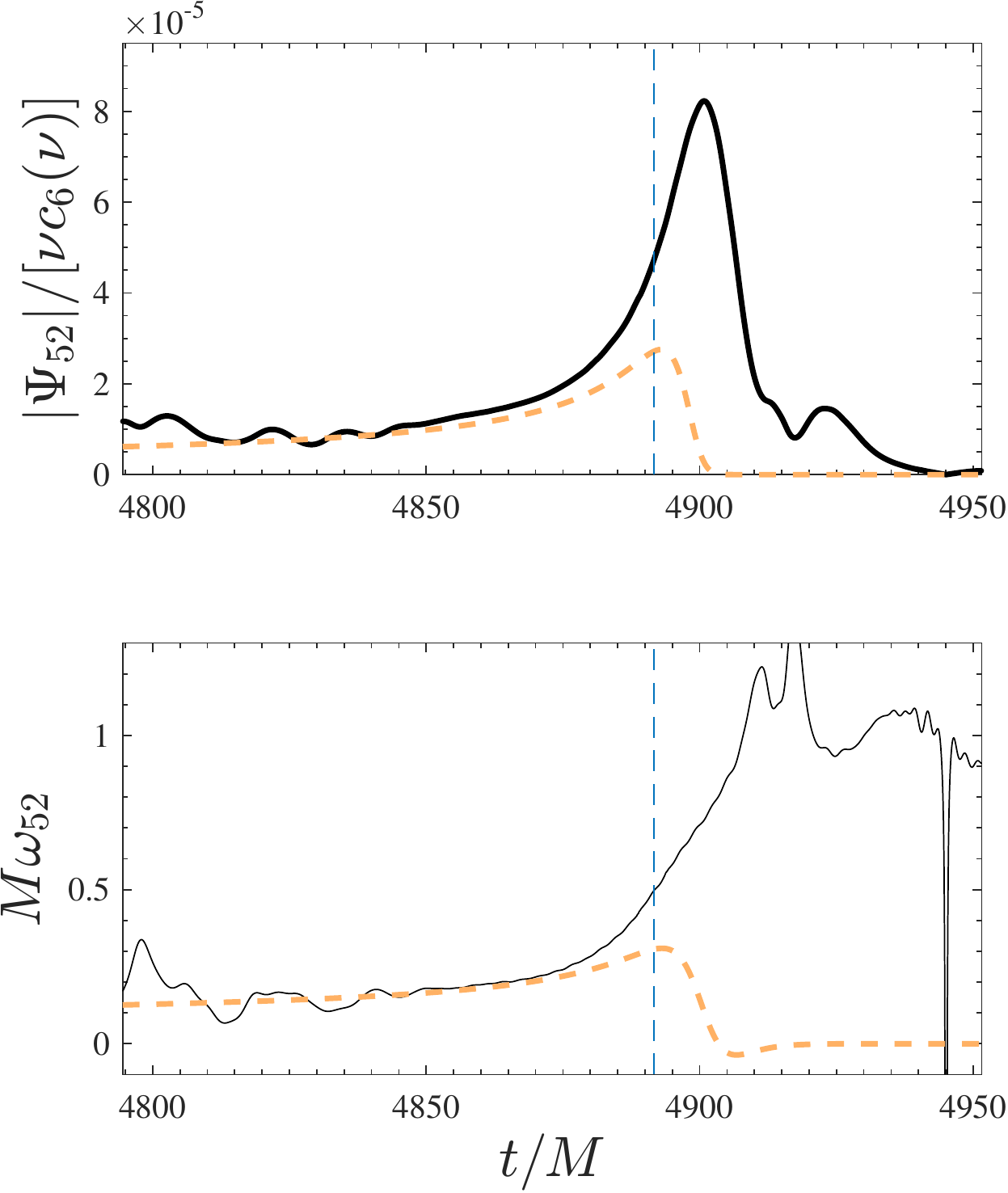}
  \includegraphics[width=0.24\textwidth]{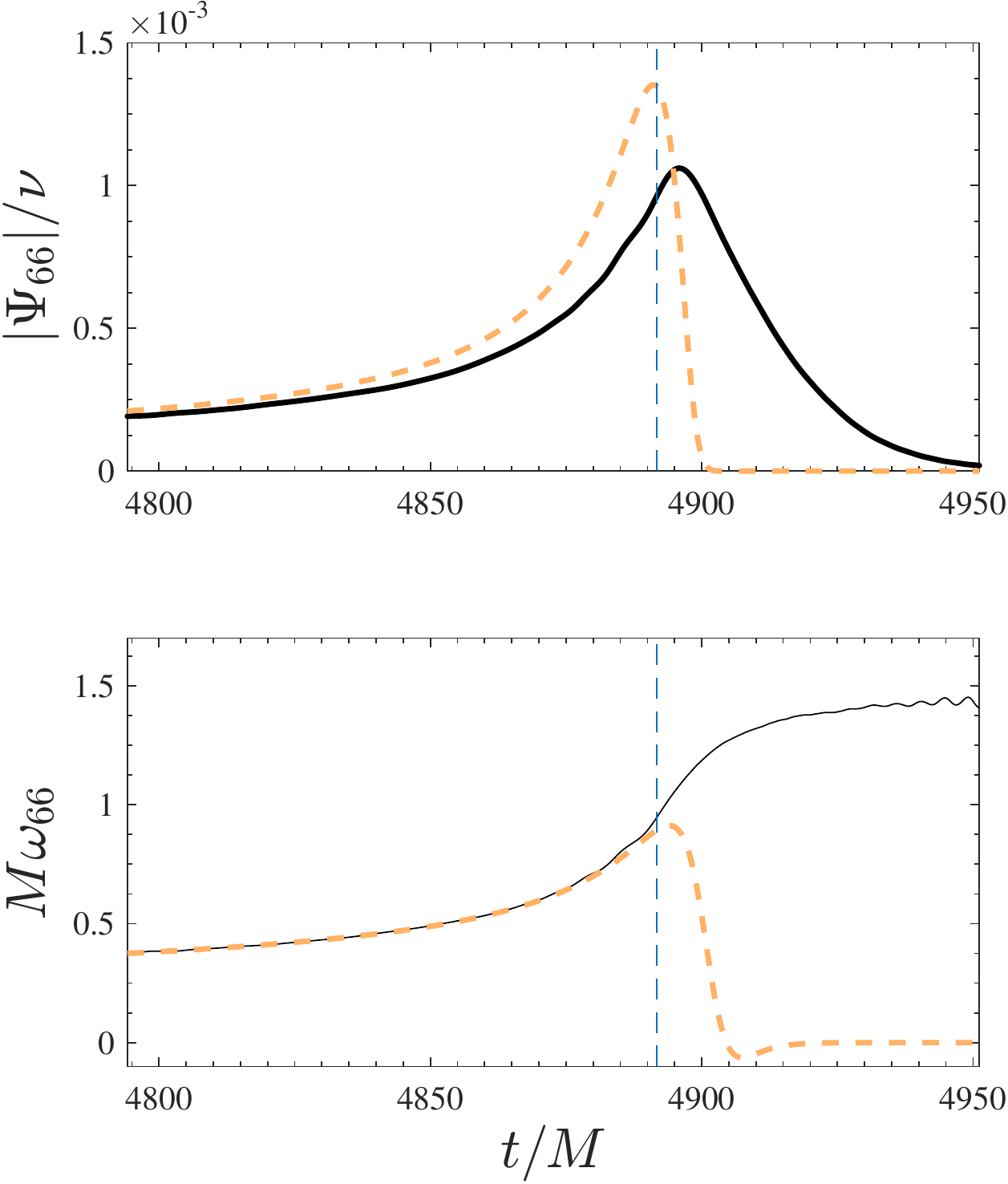} 
  \caption{\label{fig:l5_l6}Mass ratio $q=6$, SXS:BBH:0166. Performance of the bare EOB
    waveform amplitude and frequency (without NQC factor and post-peak description),
    orange, dashed lines, for the subdominant $\ell=5$ multipoles and for $\ell=m=6$.
    The solid, black, lines are the NR multipoles. The vertical line in each panel
    marks the location of the $\ell=m=2$ waveform peak, i.e. the merger.
    It is remarkable the EOB/NR good agreement between frequencies up to
    the merger points, especially for the multipoles with the largest values of $m$.}
\end{figure}

\section{Energetics}
\label{sec:energy}
Now that we have discussed all the building blocks of our nonspinning
waveform model(s), let us turn to discussing its performance towards 
all the available NR data. To simplify the discussion, we 
focus {\it only} on \TEOBiResumSM{}, as it will serve, in a forthcoming study, 
as baseline for constructing a multipolar waveform model for spin-aligned binaries.
In this section, we briefly discuss the energetics of the model.
We do this by means of the  gauge-invariant relation between
the binding energy and angular momentum computed both from
NR data and in \TEOBiResumSM{}. This analysis was extensively
done already for previous versions of our EOB model~\cite{Damour:2011fu,Nagar:2015xqa}
and more recently also for the {\tt SEOBNRv4} model~\cite{Ossokine:2017dge}.
As an illustrative example, we focus  Fig.~\ref{fig:EvsJ} on the
case of two mass ratios, $q=1$ and $q=8$, where 
$E_b\equiv (E-M)/\mu$ is the binding energy per unit mass.
For the EOB case, one has $E\equiv \mu \hat{H}_{\rm EOB}$ computed
along the EOB dynamics. The NR curves (black online) are precisely
those obtained in Ref.~\cite{Nagar:2015xqa}, 
to which we refer the reader for additional details. 
The curves obtained with \TEOBiResumSM{} are shown in blue, 
while those of the EOB model of ~\cite{Nagar:2015xqa} 
(i.e., \TEOBResumS{} as in Ref.~\cite{Nagar:2018zoe}) in red.
The figure illustrates the excellent mutual consistency between
the two models despite the modifications in the radiation reaction
and in the determination of $a_6^c(\nu)$. The location of the merger is
indicated by the markers, of the same color of the corresponding line.
Note that the \TEOBiResumSM{} curves are extended also {\it after}
the merger, but should not be trusted there, as they are obtained from
the pure relative dynamics augmented with the NQC factor that is not
trustable after the NQC point. Moreover, the ringdown losses are not
included in these curves. The correct extension of the EOB $E_b(p_\varphi)$ curve {\it beyond}
merger requires these details to be taken into account and is postponed
to future work. Finally, in the same figure we also show, as orange lines,
the $E_b(p_\varphi)$ curves obtained from {\tt SEOBNRv4}~\cite{Bohe:2016gbl}.
Note that, despite this model being publicly available through the LIGO {\tt LALSuite}
\cite{lalsuite} library, the corresponding code is not giving, by default, the
evolution of the dynamics. Similarly to~\cite{Ossokine:2017dge},
we modified the code in function {\tt XLALSimIMRSpinAlignedEOBModes},
contained in {\tt LALSimIMRSpinAlignedEOB.c}, in order to have
access to this information. In particular 
we were able to obtain an additional output file, containing the full dynamics 
$(t, r, \varphi, p_{r_*}, p_{\varphi}, M\Omega, E)$, when calling {\tt lalsim-inspiral} from command 
line. Then, exploiting the waveform generated by {\tt lalsim-inspiral}
itself, we computed the waveform amplitude, identified the merger time
$t_{\rm mrg}$ corresponding to the amplitude peak, and finally
found $E_b^{\rm mrg} \equiv E_b(t_{\rm mrg})$ and $p_\varphi^{\rm mrg} = p_\varphi(t_{\rm mrg})$.
The changes made to {\tt LALSuite}'s source code, together with the data needed to reproduce 
Fig.~\ref{fig:EvsJ}, are publicly available at~\cite{seob_hacked}.
Inspecting the top and bottom panel of Fig.~\ref{fig:EvsJ}, and in particular the insets,
we conclude that: (i) for $q=1$, {\tt SEOBNRv4} seems to overestimate
the binding energy  during the late stages of the dynamics up to merger
of about $1\%$. Note that, although this 
looks like an acceptably  small number, it is actually {\it larger} than the 
numerical uncertainty on the curve~\cite{Damour:2011fu,Nagar:2015xqa}.
By contrast, (ii) when $q=8$ the various curves look more consistent  among 
themselves, although the {\tt SEOBNRv4} prediction of the merger values
is significantly different from either the NR or the
\TEOBResumS{}/\TEOBiResumSM{} values. We postpone to future investigations
a detailed understanding of the origin of these features of {\tt SEOBNRv4}.

\section{Phasing analysis}
\label{sec:acc}
\subsection{Time-domain phasing analysis}
Let us move now to assessing the quality of the multipolar waveform.
We do so by looking at the usual EOB/NR phase differences as well 
as at comparisons between frequency and amplitudes for the various multipoles. 
As for the case of energetics discussed above, we focus {\it only} on
waveforms generated by \TEOBiResumSM{}.
Aim of this section is to demonstrate the following points: (i) the rather 
remarkable agreement between frequency and amplitude, for all multipoles,
that can be accomplished already with the {\it bare} EOB waveform, even
without the NQC correction factors; (ii) the (rather small) effect brought
by NQC corrections, that is more important on the amplitude than on the
frequency; (iii) the fact that the transition to the ringdown (or postpeak)
phase can be done consistently multipole by multipole, in the sense that
the same procedure can be applied on each mode once the relevant NR information
is taken into account and properly represented; (iv) accurate description
of the postpeak-ringdown phase that is robust and reliable, though still
without mode mixing. We highlight this by selecting a specific
EOB/NR comparison done for $q=6$, that corresponds to SXS:BBH:0166.
\begin{figure*}[t]
\begin{center}
\includegraphics[width=\textwidth]{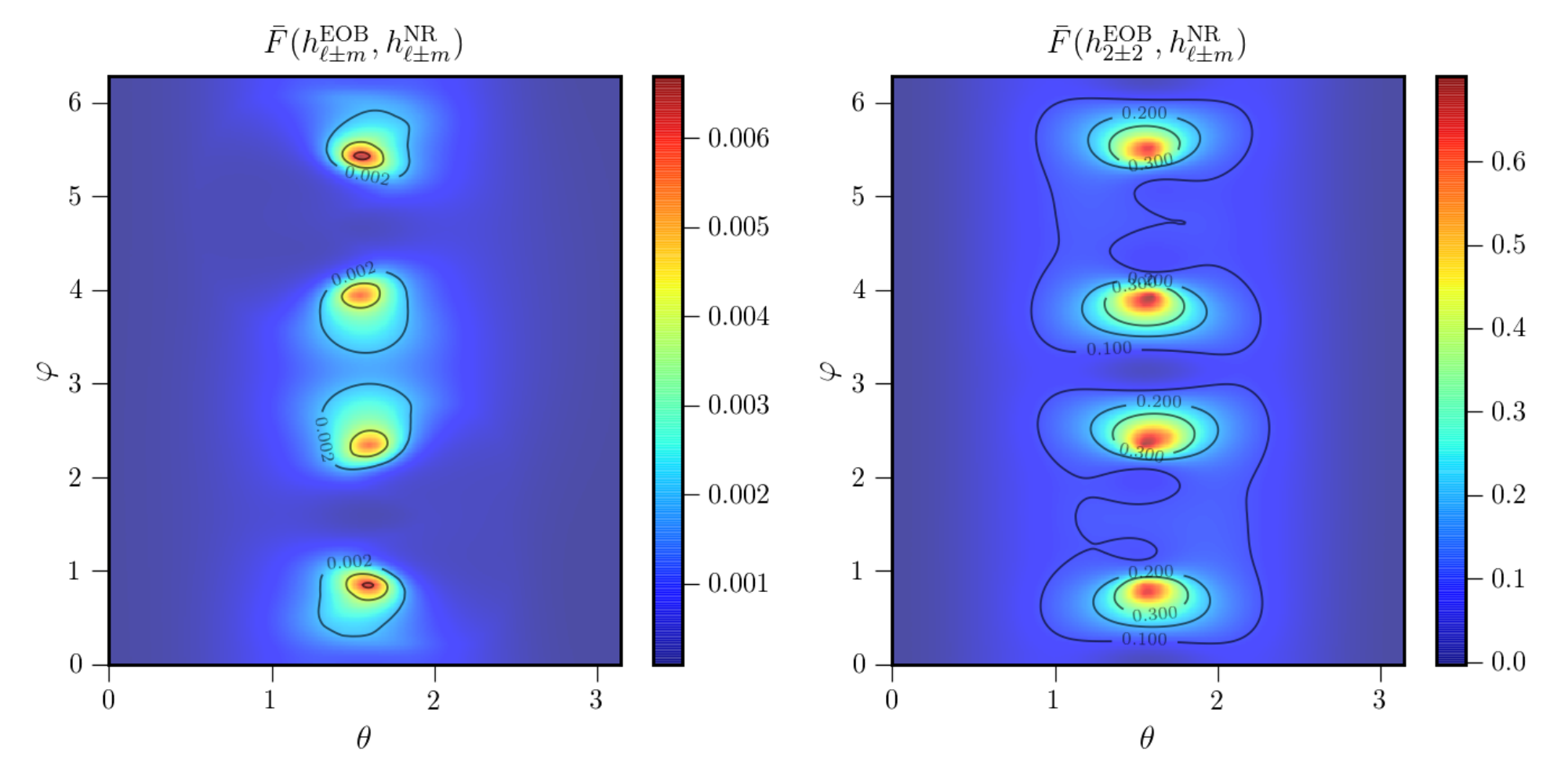}
\caption{Unfaithfulness between the \TEOBiResumSM{} model and SXS:0303, a binary of mass ratio $q=10$,
  assuming a total mass of $M = 100M_{\odot}$ using the Advanced LIGO and Virgo design sensitivity
  PSD (zerodethp). The left plot shows the unfaithfulness between EOB and NR using a subset of
  the most prominent modes: $\lbrace 22, 21, 33, 44, 55 \rbrace$. The right plot shows the
  degredation in the unfaithfulness from using just the dominant 22-mode in the model.
  As expected, the unfaithfulness degrades as we approach edge-on systems where the
  relative contribution of higher modes becomes more pronounced. 
}
\label{fig:EOBvsNR_sky}
\end{center}
\end{figure*}
Figure~\ref{fig:q6_l2} illustrates the (rather good) time-domain
phasing agreement between the $\ell=m=2$ EOB and NR modes. 
From Table~\ref{tab:NR_Waveforms} we see that this dataset was simulated
at one single resolution from the SXS collaboration, so that a specific estimate
of its error bar at merger is not possible. We may however imagine that it is 
of the order of the $q=6.5$ simulation, that  has approximately the same number
of orbits and that starts at approximately the same frequency ($M\omega_0\simeq 0.020$),
i.e. $\sim -0.05$~rad at merger. This value is compatible with the EOB-NR phase
difference at merger that is found in Fig.~\ref{fig:q6_l2}. The vertical dash-dotted 
lines mark the frequency region during the inspiral where the alignment is done
by minimizing the phase difference between two given frequencies~\cite{Damour:2007yf}. 
This figure is complemented by Fig.~\ref{fig:q6_Aomg}, that illustrates the behavior
of EOB and NR amplitudes and frequencies for both the $(2,2)$ mode 
(top row) and the $(2,1)$ mode (bottom row). 
The main role of this comparison is to state clearly the performance of the
purely analytical, EOB-resummed, waveform and pinpoint the role of the NQC correction factor.
Each panel of the figure reports four curves: (i) the NR waveform multipole
(black); (ii) the analytical EOB waveform (orange); (iii) the NQC corrected
EOB waveform (light-blue); and (iv) finally the full EOBNR waveform
completed by the ringdown phase (red), although only the latter appears
separated from the other curves.  The impact of the NQC correction factor 
to the frequency is rather minimal for the $(2,2)$ mode. By contrast, it 
is more important for the $(2,1)$ mode, since it is  able to ``rise'' the 
dashed orange line so to be on top of the black one. As mentioned above, it is worth
noticing that the blue curves are obtained precisely with the same procedure 
for both the $(2,2)$ and $(2,1)$ mode. To do so, for each mode one needs from
NR only the knowledge of four numbers, the values of
$(A_\lm,\dot{A}_\lm,\omega_\lm,\dot{\omega}_\lm)$ at the NQC
extraction point, Eq.~\eqref{tNQC_extr}. In addition, for $(2,1)$ one
also necessitates of $\Delta t^{\rm NR}_{21}$, that allows one 
to locate the postpeak phase of the (2,1) mode at the correct place 
on the EOB time-axis. By {\it correct place} we mean that the EOB post-peak
phase correctly alignes on the NR one thanks to the correct analytical 
representation of $\Delta t^{\rm NR}_{21}$ extracted from NR data: no additional
tuning is needed here and everything falls in place automatically.
Precisely the same approach can be followed for {\it all} other modes,
as illustrated for a few of them in Fig.~\ref{fig:q6_all}, for the amplitudes
of $h_\lm=\sqrt{(\ell+2)(\ell+1)\ell(\ell-1)}\Psi_\lm$ (top panel) and the 
frequencies (bottom). To ease the comparison of all
amplitudes on the same scale, they are shown normalized by
$\nu c_{\ell+\epsilon}(\nu)$ on a logarithmic scale. The details of the
effect of the NQC correction factor are shown in Fig.~\ref{fig:q6_l3l4}
in Appendix~\ref{sec:AppHM}. In addition, as a proof of the robustness
of the procedure, it is possible to complete with peak and postpeak
also the $(4,1)$ and $(5,5)$ modes, that are also explicitly diplayed
in Fig.~\ref{fig:q6_41_55} of Appendix~\ref{sec:AppHM}.
Although these modes are generally considered to be of small
importance\footnote{Note however that if one wished to accurately
  compute the recoil velocity due to the emission of gravitational
waves, these modes have to be taken into account.}, we believe that
it is quite remarkable that the matching procedure originally designed 
for the $(2,2)$ waveform can be applied to them too without any additional
conceptual input. For simplicity, we have decided to not implement the
details of the peak and post-peak structures in all other subdominant
modes beyond $\ell=m=5$. Still, Fig.~\ref{fig:l5_l6} highlights that 
the EOB/NR frequency agreement is already rather good (and in fact
comparable to what found for lower modes, see Appendix~\ref{sec:AppHM}),
especially for the more circularized mode, essentially up to merger.
If the need comes, we expect it will be relatively straightforward
to complete also these modes with the corresponding peak and 
postpeak behavior informed by NR simulations.

\begin{figure}[t]
  \center
\includegraphics[width=0.45\textwidth]{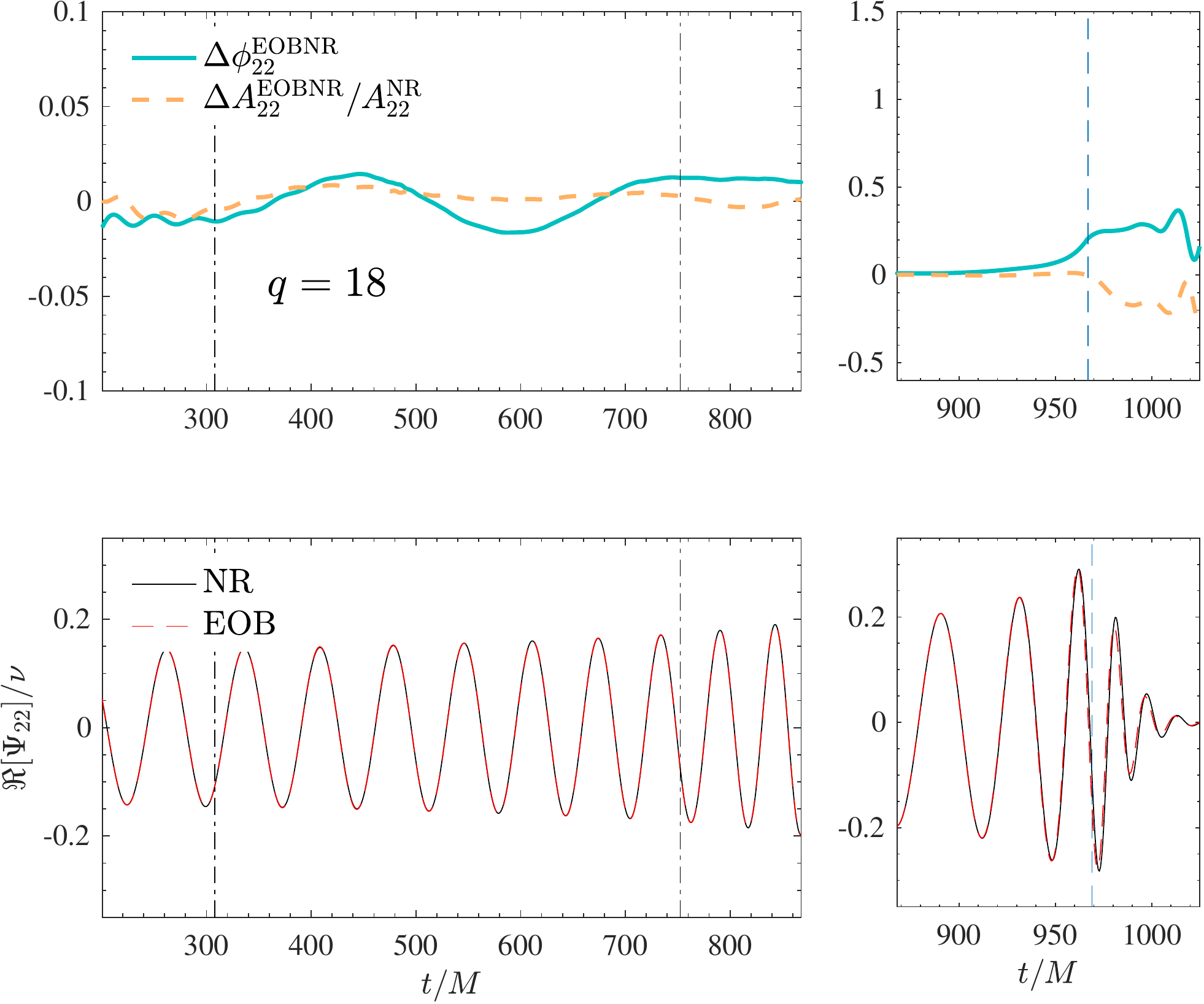} 
\caption{\label{fig:q18_l2}Phasing comparison for $q=18$: comparison between \TEOBiResumSM{} 
and the corresponding NR waveform data obtained with the BAM code. The two waveforms are highly
compatible even if this dataset was not used to inform $a_6^c(\nu)$. Note the effect of the (rather small)
eccentricity that shows up as an oscillation in the phase difference (top panel).}
\end{figure}

Finally, Fig.~\ref{fig:q18_l2} illustrates the robustness of the model up
to mass ratio $q=18$, with a phasing agreement that is of the order of
the accumulated numerical uncertainty typical of these simulations. 
We remind the reader that we didn't use this dataset to inform 
$a_6^c(\nu)$, though we did use it to improve the behavior of 
the ringdown part of the waveform.

\subsection{Unfaithfulness}
\label{sec:barF}
The second comparison we discuss is that of the EOB/NR unfaithfulness.
Given $h_i$ a waveform and $\tilde{h}_i(f)$ its Fourier transform, we define
the following weighted scalar product between two waveforms
\begin{align}
\left\langle h_1 | h_2 \right\rangle = 4 \, \Re \int^{f_{h}}_{f_{l}} \frac{\tilde{h}_1^* (f) \; \tilde{h}_2 (f)}{S_n (f)} \; df ,
\end{align}
where $S_n(f)$ is the power spectral density (PSD) of the detector.
The unfaithfulness $\bar{{F}}$ can then be defined via the
inner product between normalised
($\hat{h} = h / \sqrt{\langle h | h \rangle}$)
waveforms maximised over time and phase shifts
\begin{align}
\bar{{F}} (h_1,h_2) &= 1 - F = 1 - \max\limits_{t_0,\phi_0} \, \langle \hat{h}_1 | \hat{h}_2 \rangle .
\end{align}
Note that the maximization over time and phase shifts has no physical significance, 
corresponding to a change in the merger time and initial phase of the binary. 
In our analysis, we use the zero-detuned high-power (zdethp) power 
spectrum~\cite{Aasi:2013wya,dcc:2974} as a representative 
PSD for aLIGO at design sensitivity. As the NR waveforms are of finite 
length, we use a lower cutoff frequency of $f_l = {\rm{min}} \, (20, f_{\rm{NR}})$~Hz 
and an upper cut-off frequency of  $f_h = 2048$Hz. 
\begin{figure}[t]
  \begin{center}
\includegraphics[width=\columnwidth]{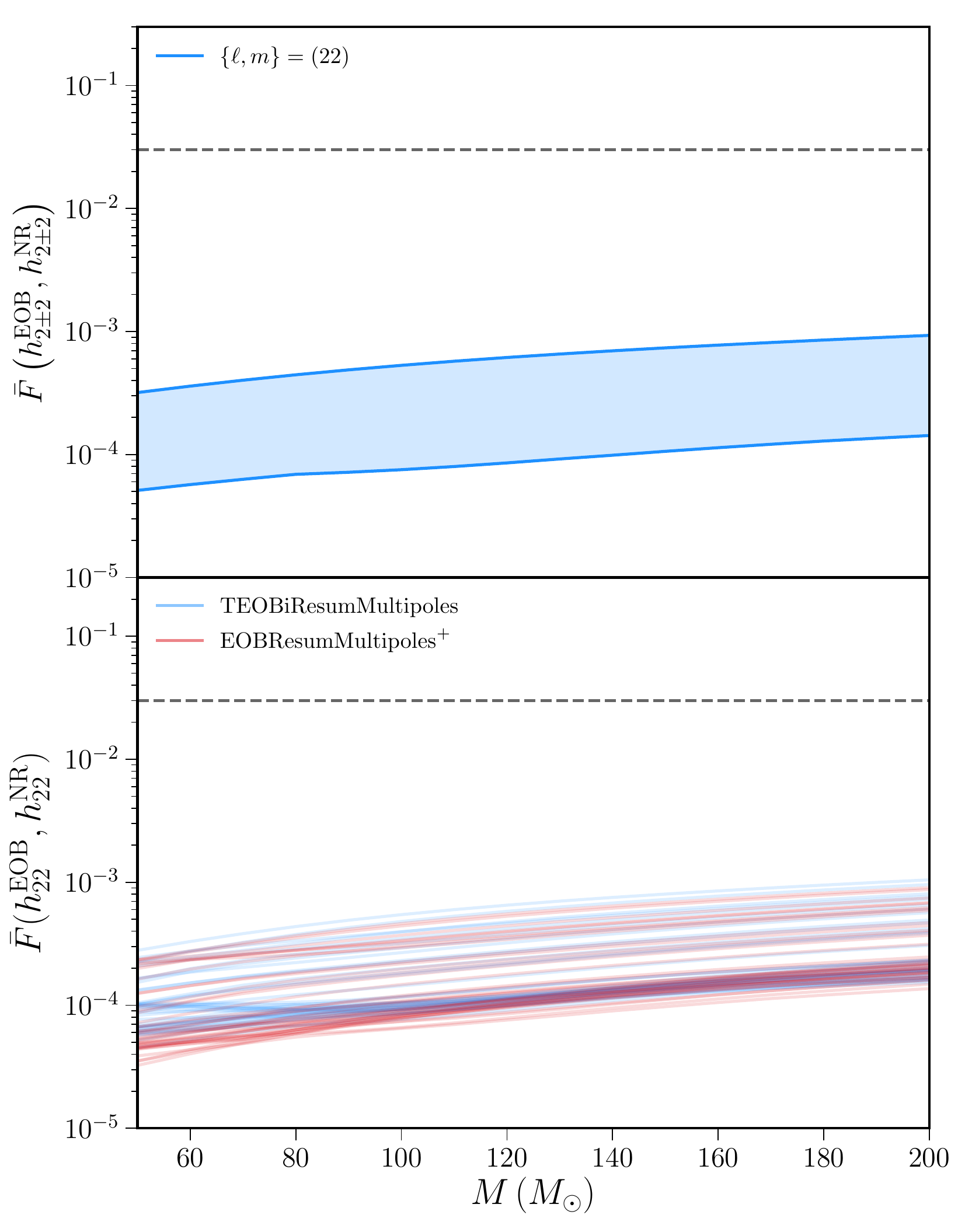}
\caption{ \label{fig:EOBMinMax22} Top panel: Unfaithfulness between SXS simulations and \TEOBiResumSM{} for mass ratios $q \in [2, 10]$. Note that we 
  use the zero-detuned high power Advanced LIGO design sensitivity PSD and assume a face-on configuration. We restrict the analysis to the dominant $(2,2)$
  mode, demonstrating the baseline performance of the model when neglecting higher order modes.
  The model is everywhere below $10^{-3}$ for $M \lesssim 200 M_{\odot}$. Bottom panel: comparison between
  \TEOBiResumSM{} and \EOBResumM{}. The two models, that are relying on different determinations of $a_6^c(\nu)$
  induced by different resummation choices in the radiation reaction, are practically equivalent.}
\end{center}
\end{figure}

\begin{figure}[t]
\begin{center}
\includegraphics[width=\columnwidth]{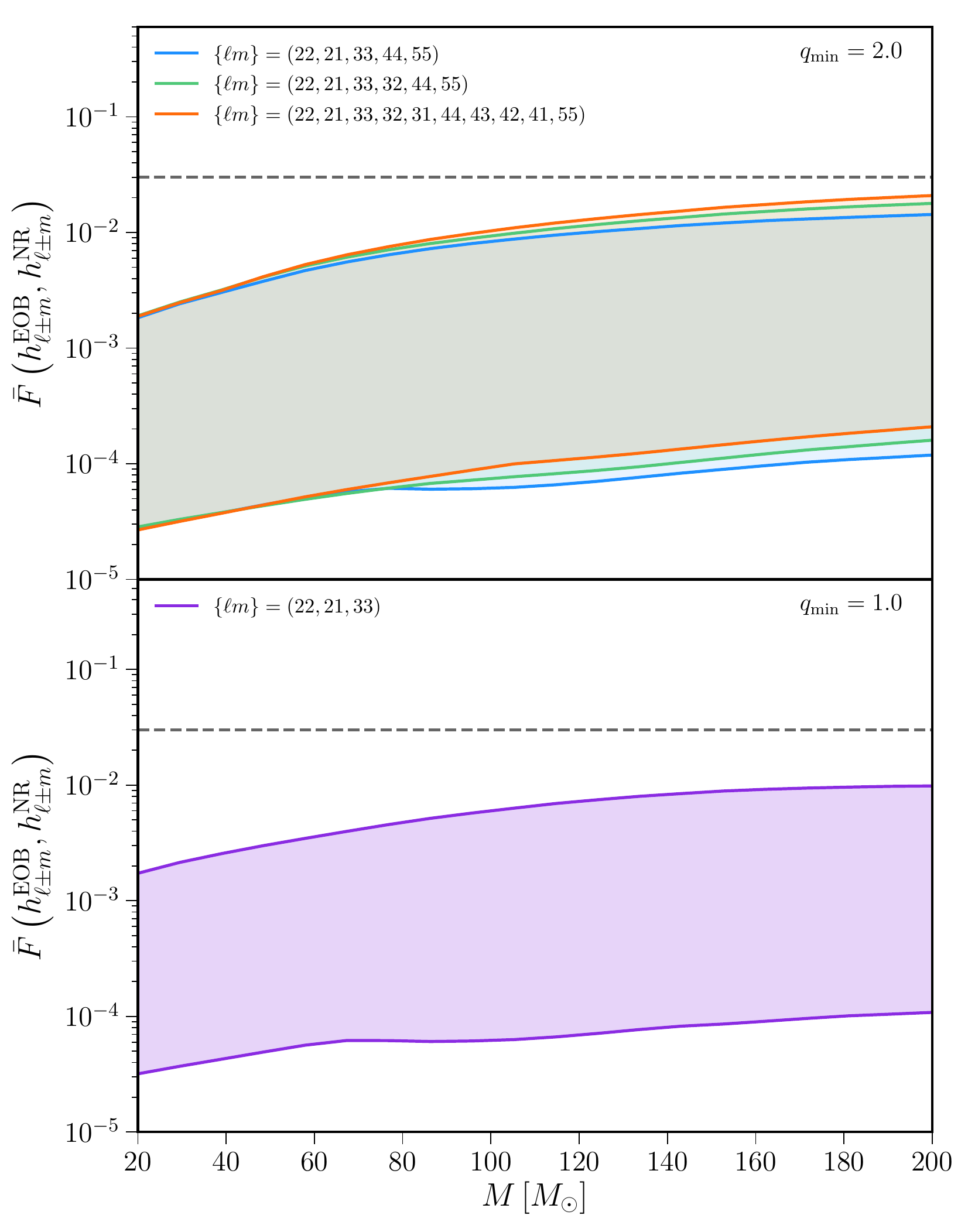}
\caption{Unfaithfulness between \TEOBiResumSM{} and SXS simulations for
  mass ratios using the zero-detuned high power
  Advanced LIGO design sensitivity PSD. We show the minimum and maximum unfaithfulness
  over all angles $(\theta, \varphi)$, demonstrating that the worst case performance
  is always below $3$\% for binaries with a total mass $M \lesssim 200 M_{\odot}$.
  Even though \TEOBiResumSM{} neglects mode-mixing, we do not find a significant
  degradation in performance when considering  $(3,2)$ mode (red curves)
  or all modes up to $\ell = 4$ and the $(5,5)$ mode (green curve). In the top panel, 
  we restrict the analysis to $q \geq 2$ due to the issues highlighted in the text. In the bottom panel, where 
  we neglect the $(4,4)$ mode, we find excellent agreement with NR down to $q=1$. 
}
\label{fig:EOBMinMaxHM}
\end{center}
\end{figure}

\begin{figure}[t]
\begin{center}
  \includegraphics[width=\columnwidth]{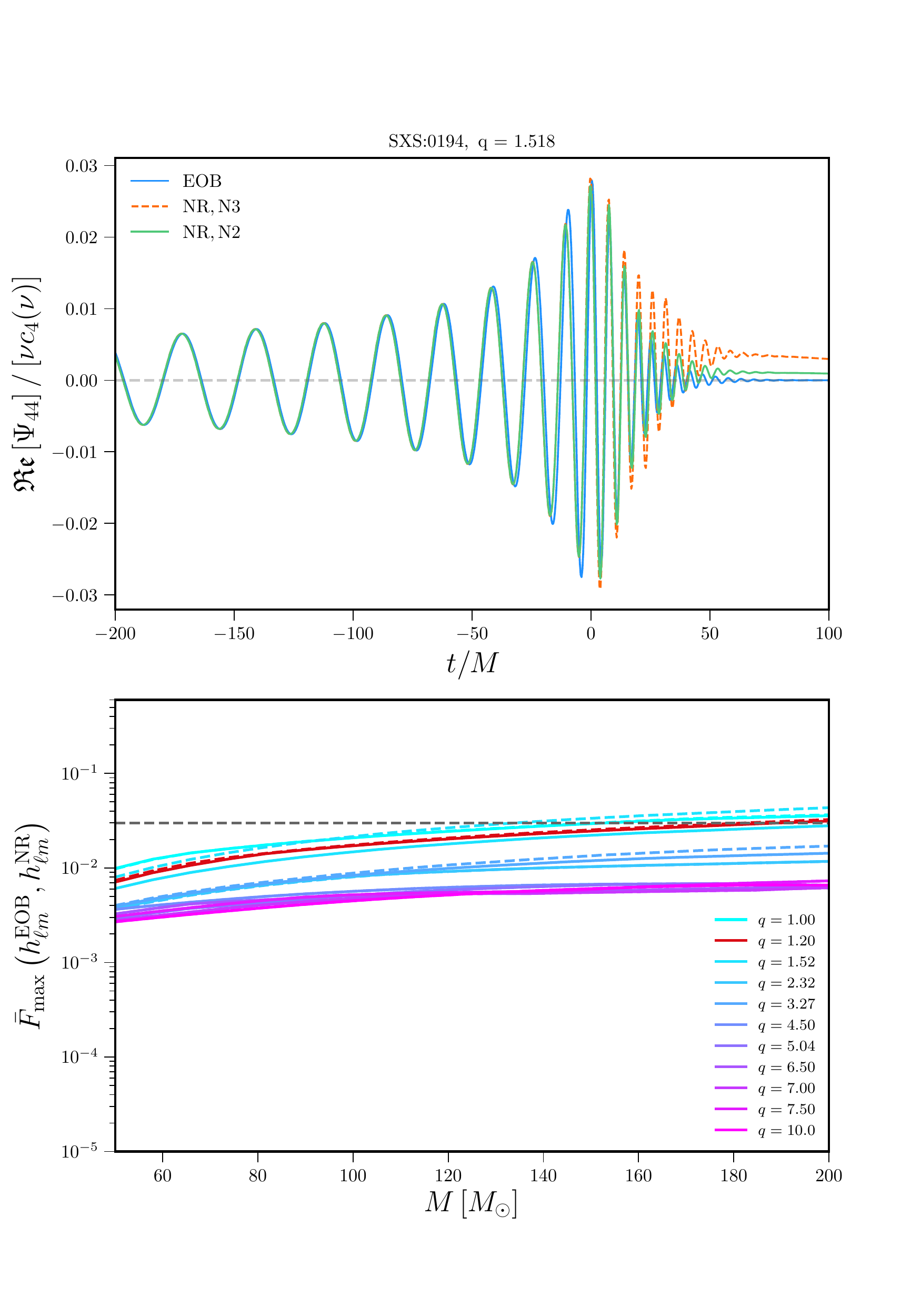}
  \caption{Top panel: systematic effects show up in the $\ell=m=4$ mode
    of SXS:BBH:0194 when extrapolated with the standard choice $N=3$.
    Bottom panel: Unfaithfulness against selected SXS simulations using $N=3$ extrapolated waveforms (dashed) and 
    $N=2$ extrapolated waveforms (solid). As the mass ratio increases, the difference between the two extrapolation
    orders becomes negligible. 
}
\label{fig:worse}
\end{center}
\end{figure}
\begin{figure}[t]
\begin{center}
  \includegraphics[width=\columnwidth]{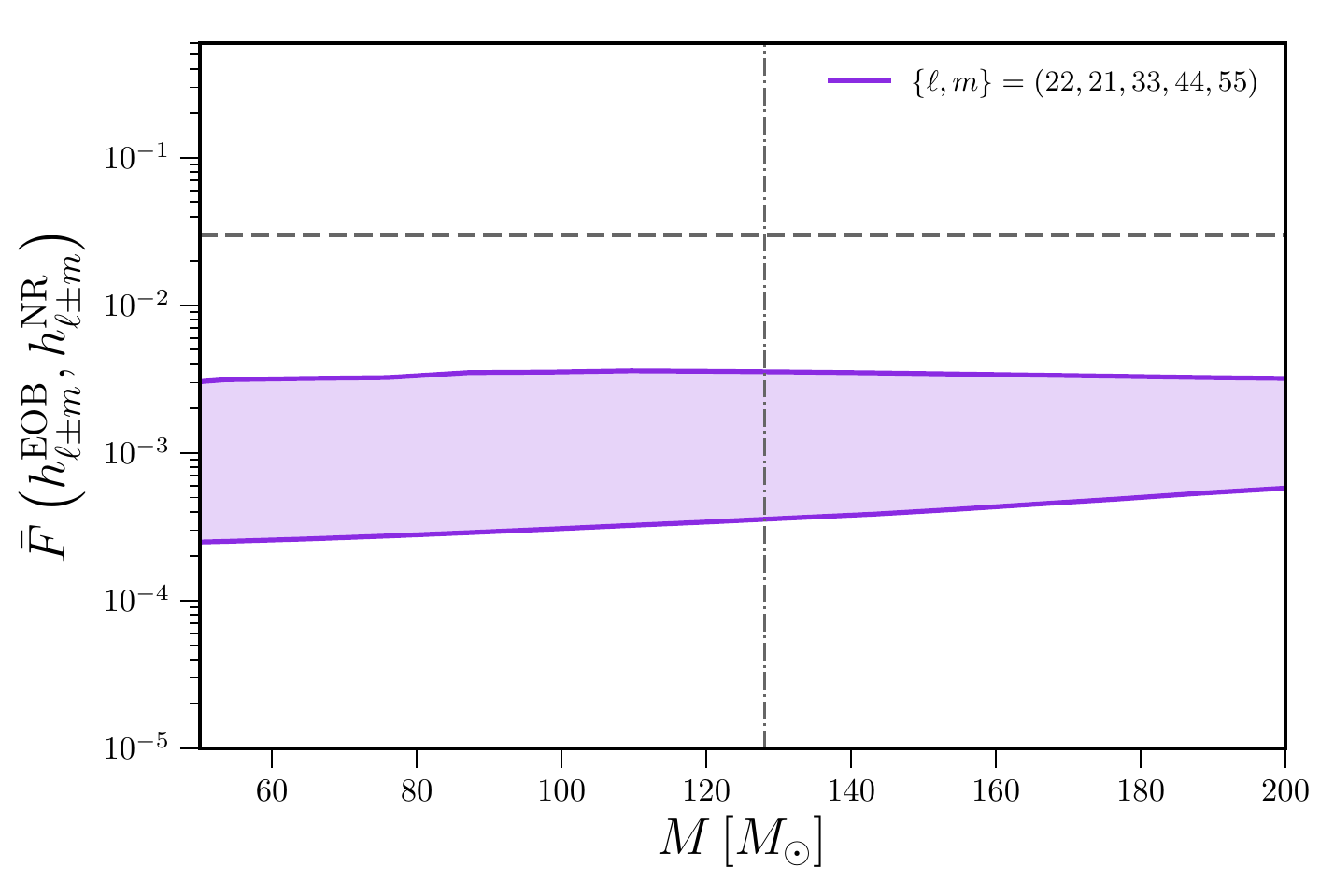}
   \includegraphics[width=\columnwidth]{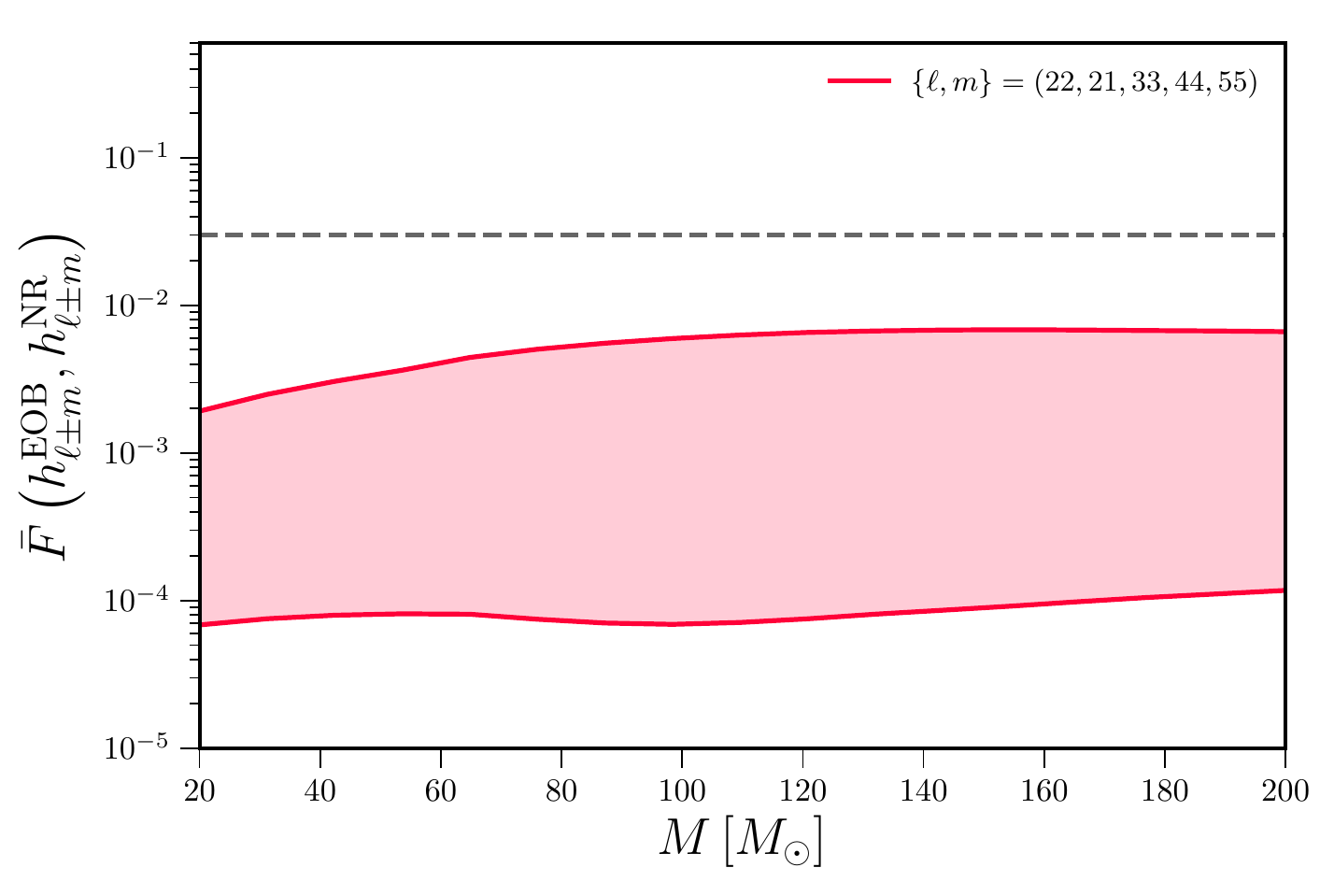}
  \caption{Minimum and maximum unfaithfulness for \TEOBiResumSM{} model against a BAM $q=18$ waveform~\cite{Husa:2015iqa} (top panel)
  and an SXS $q=6$ simulation (bottom panel). In the top panel, the dot-dashed line shows the minimum mass for which the entire 
  NR waveform is in band. The EOB/NR performance for $q=6$ is comparable to (though slightly better than) {\tt SEOBNRv4HM},
  for the same SXS dataset, as deducible by comparison with Fig.~16 of Ref.~\cite{Cotesta:2018fcv}.}
\label{fig:bamq18}
\end{center}
\end{figure}
In Fig.~\ref{fig:EOBvsNR_sky} we illustrate the performance gain obtained by \TEOBiResumSM{} with higher
modes included when comparing to SXS:BBH:0303, a high mass ratio $q=10$ binary. As one may expect,
using only the $(2,2)$-mode in constructing the strain results in an unfaithfulness against the NR
waveforms that is degraded by at least two orders of magnitude. This can be seen in the right panel of Fig.~\ref{fig:EOBvsNR_sky}.
The baseline performance of the dominant $(2,2)$ mode is everywhere below $10^{-3}$ for $M \lesssim 200 M_{\odot}$,
as one sees in the top-panel of Fig.~\ref{fig:EOBMinMax22}. The same figure, in the bottom panel,
illustrates the excellent consistency between \TEOBiResumSM{} and \EOBResumM{}, although the two
models differ both in the description of the radiation reaction and in the determination of $a_6^c(\nu)$.
When considering higher modes, we have verified that, for $\ell\geq 3$, the mode-by-mode EOB/NR
unfaithfulness is slightly, but nonnegligibly, smaller for \TEOBiResumSM{} than for \EOBResumM{},
so that we will not discuss this latter any further.
Including several higher order modes in \TEOBiResumSM{}, we found that the EOB/NR unfaithfulness
is everywhere below $3\%$ for binaries up to $M < 200 M_{\odot}$, as demonstrated in Fig.~\ref{fig:EOBMinMaxHM}.
Note however that we focus here on $q\in [2,10]$. The range $1\leq q \leq 2$ has some issues that
we discuss below. However, when neglecting the $(4,4)$ mode, the model performance is in 
excellent agreement with NR down to $q=1$, as seen in the bottom panel of Fig.~\ref{fig:EOBMinMaxHM}. 
When taking into account the $(3,2)$ mode, the most prominent mode affected
by mode-mixing~\cite{Berti:2014fga}, we find that the performance of the model only slightly
degrades across the whole parameter space and the unfaithfulness robustly remains 
below $3$\% for binaries up to $\lesssim 200 M_{\odot}$.

We investigated the origin of the behavior of $\bar{F}$ for large masses. 
We discovered that it does not come from inaccuracies in the analytical description of 
the ringdown (e.g. the lack of mode mixing effects) but rather from the amplification of 
some NR numerical noise present in the ringdown due to the extrapolation of the waveforms. 
We recall in this respect, that in the {\tt SpEC} code the waveforms are
extracted at finite radius and then extrapolated to infinite distance using polynomials
in the inverse power of the extraction radius. The highest power of this polynomial is
labeled as $N$. In the SXS catalog, data extrapolated with different values of $N$ 
are provided. It is well known that non-optimal values of $N$ may introduce unphysical
features and thus such extrapolation process can be quite delicate~\cite{Boyle:2009vi}. 
Investigating the various datasets present in the SXS catalog for each configuration,
we have found that such amplification of the late-time NR noise shows up {\it only}
for some nearly-equal-mass configuration when the (standard) extrapolation order $N=3$
is used. We recall that the SXS collaboration advises catalog users to employ low 
values of $N$ if one is interested in studying the ringdown and large values of $N$ 
if one is more focused on the inspiral. Here, we use the $N=3$ extrapolation order as a 
reasonable compromise for the whole waveform, although, as mentioned above, {\it we use} 
the $N=2$ extrapolation order to obtain the post-merger fits as the NR 
data is typically cleaner. This phenomenon is illustrated in the top panel of Fig.~\ref{fig:worse}, 
for SXS:BBH:0194. One sees that the $N=3$ extrapolation introduces an unphysical offset 
during the ringdown that is reduced, though not completely eliminated,
when using an $N=2$ extrapolation order. Such systematic effect also shows up
in the unfaithfulness, which is shown in the bottom panel of the Fig.~\ref{fig:worse} 
for several mass ratios. The dashed lines correspond to using $N=3$ data, while
the solid lines are for the $N=2$ data. It is also interesting to note that the differences 
between extrapolation orders becomes progressively negligible as the mass ratio 
increases. In Fig.~\ref{fig:bamq18}, we compare \TEOBiResumSM{} 
against a non-spinning $q=18$ BAM simulation, finding excellent agreement 
against the model. In the bottom panel of the same figure we also show the same 
comparison for the $q=6$ SXS dataset. An analogous comparison is also displayed 
in Fig.~16 of~\cite{Cotesta:2018fcv} for the {\tt SEOBNRv4HM} model,
that incorporates the same number of subdominant modes considered in 
this figure. We find that the performance of \TEOBiResumSM{} on this particular 
SXS dataset is comparable to that of  {\tt SEOBNRv4HM}, though slightly better
especially for low masses. To ease this comparison, in this case we used
$20M_\odot$ as lower-mass boundary. Our analysis illustrates that the 
EOB/NR $\bar{F}$ comparison may be slightly misleading, and \TEOBiResumSM{} 
delivers a faithful  representation of the multipolar waveform also 
for nearly-equal-mass binaries.

\begin{figure*}[t]
\center
\includegraphics[width=0.45\textwidth]{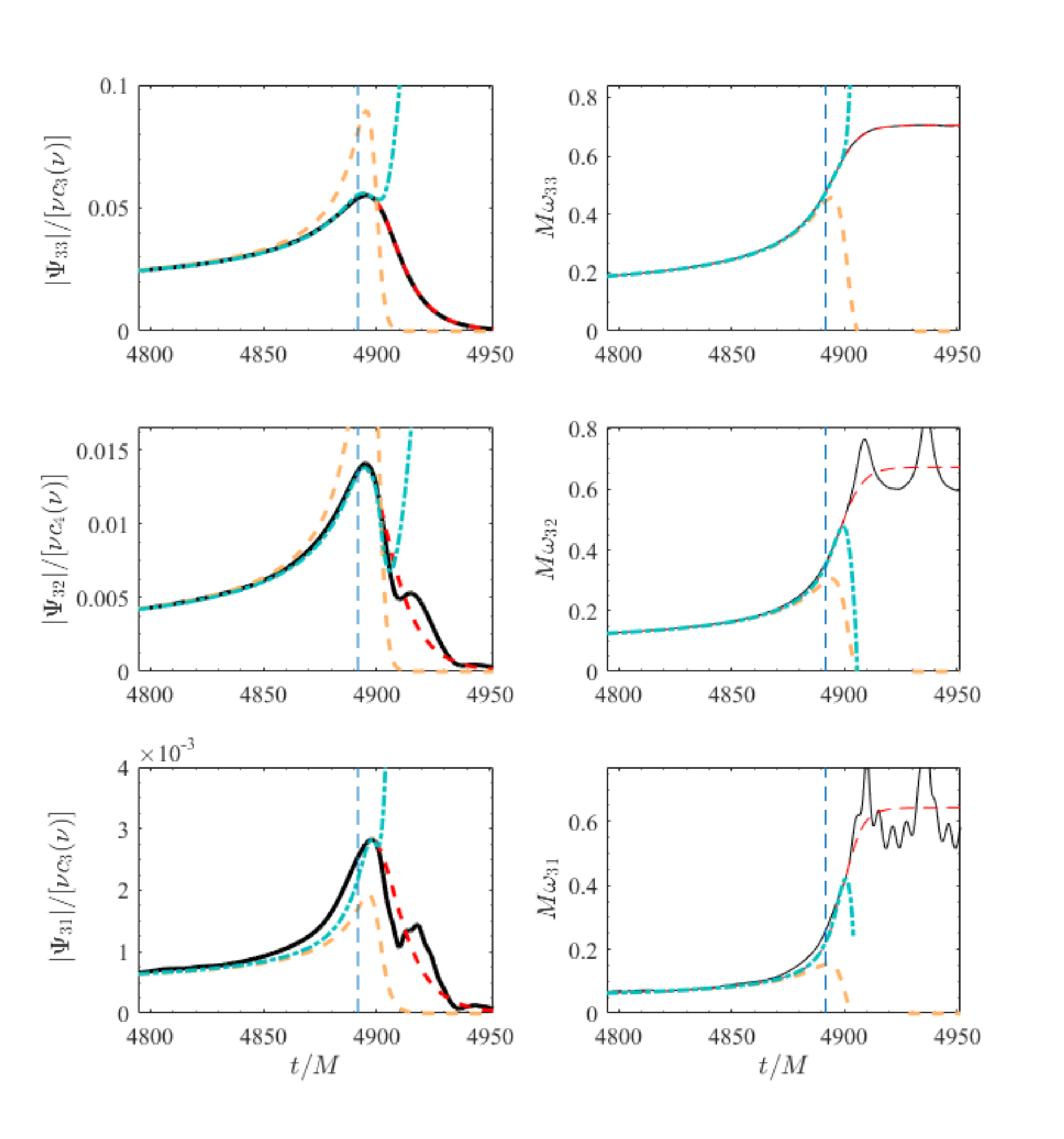}
\includegraphics[width=0.45\textwidth]{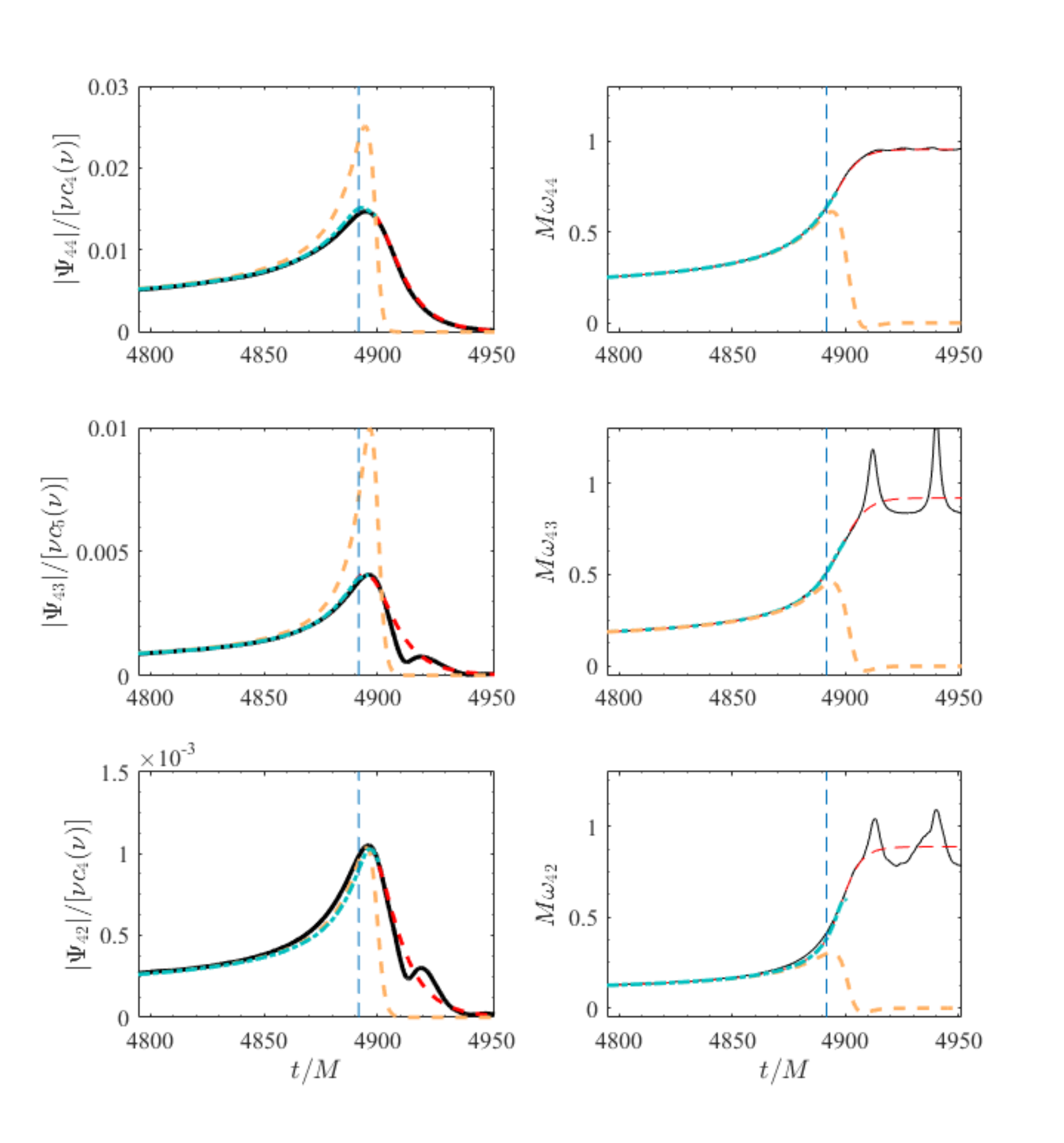}
\caption{\label{fig:q6_l3l4}Mass ratio $q=6$, SXS:BBH:0166. Frequency
  and amplitude comparison, $\ell=3$ and $\ell=4$ multipoles. Black line:
  NR data. Orange (dashed) lines: bare EOB waveform. Blue lines: NQC-modified
  waveform; red line: postmerger-ringdown part. The vertical, dashed, line
  marks the merger location. Note that mode-mixing is not incorporated
  in the analytical ringdown description.}
\end{figure*}

\begin{figure*}[t]
  \center
  \includegraphics[width=0.45\textwidth]{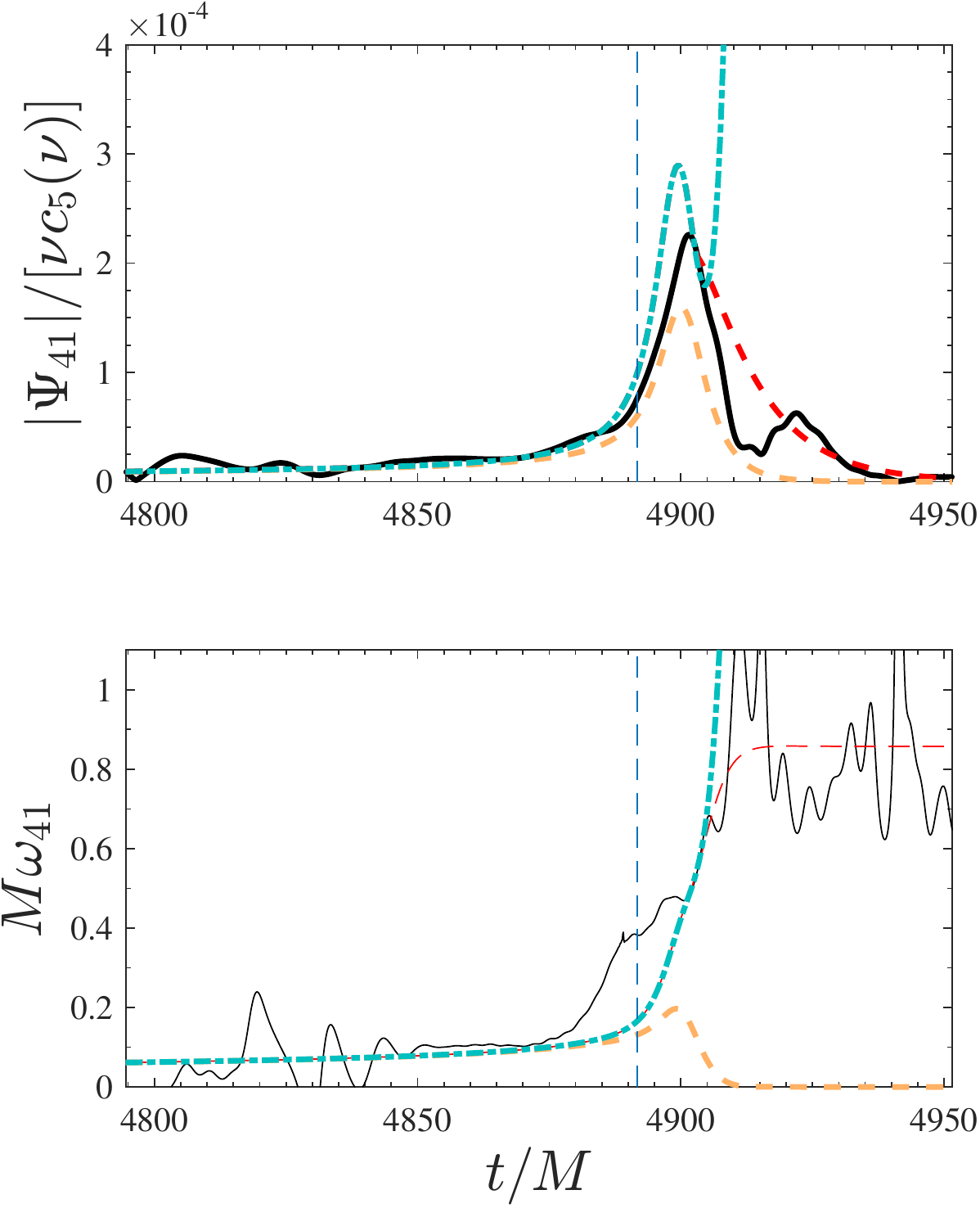}\qquad
  \includegraphics[width=0.45\textwidth]{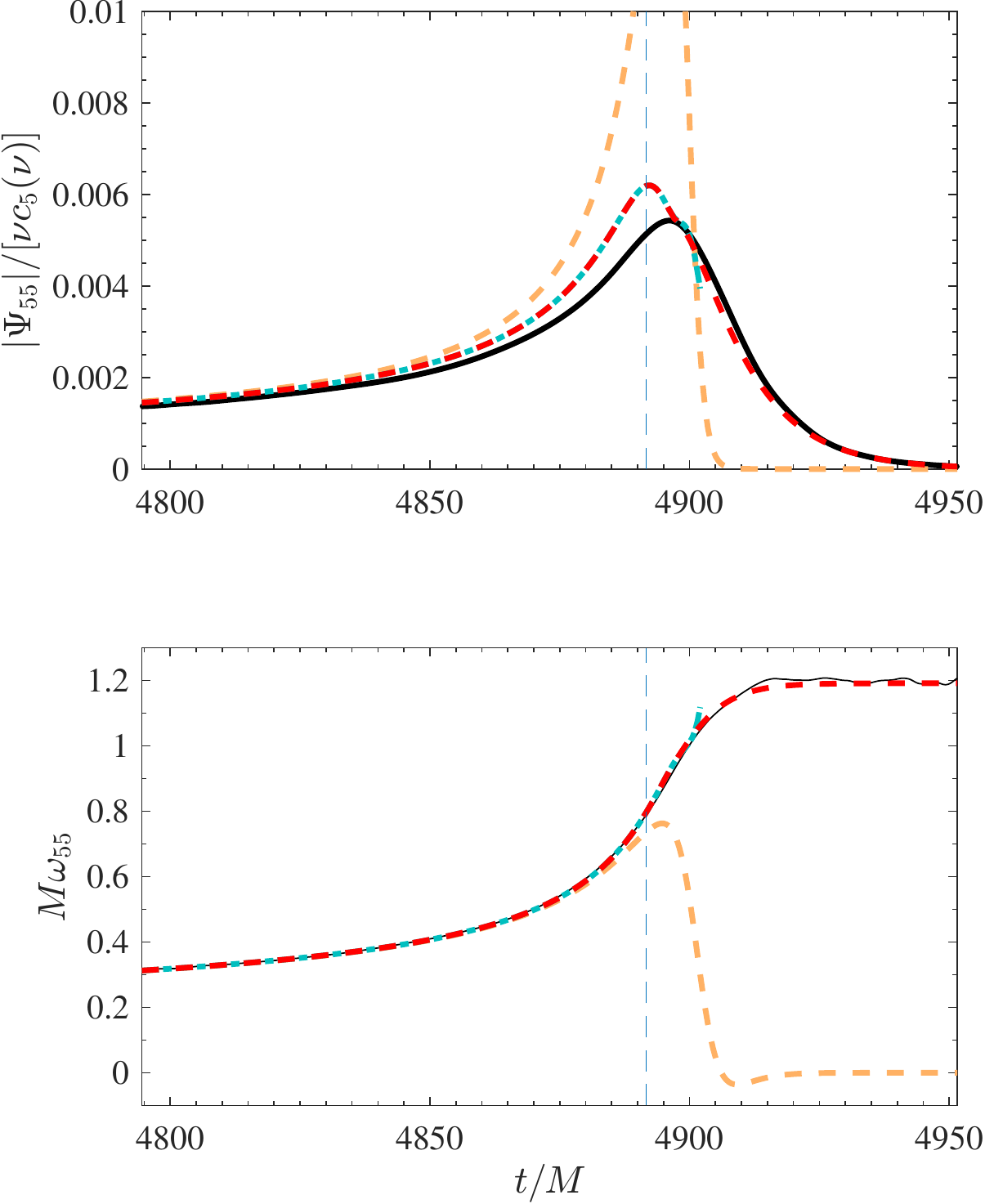} 
  \caption{  \label{fig:q6_41_55}Mass ratio $q=6$, SXS:BBH:0166. EOB/NR comparison
    for $(4,1)$ mode (left) and $(5,5)$ mode (right). Although the $(4,1)$ numerical
    frequency is rather noisy, it shows a good qualitative agreement with the
    analytical EOB prediction. In addition, the EOB and NR $(5,5)$
    frequencies are remarkably close during the full transition from
    inspiral to plunge, merger and ringdown. For both modes, the
    NQC-corrected analytical amplitude close to merger tends
    to be larger then, though consistent with, the NR one.}
\end{figure*}
\section{Conclusions}
\label{sec:conc}
We have presented \TEOBiResumSM{}, a new, NR-informed, EOB model for nonspinning black hole binaries 
that incorporates higher multipolar waveform modes. The multipoles are complete through merger and ringdown
up to $\ell=m=5$ included. The additional waveform modes up to $\ell=8$ (including $\ell=5$, $|m|<5$) are also 
included but they rely on the, purely analytical, EOB-resummed waveform. In practical terms, this means that
the corresponding EOB GW frequency $\omega_\lm^{\rm EOB}$ smoothly goes to zero after merger and does not saturate
to the plateau corresponding to the QNM excitation. Up to the merger point, it generally delivers
an excellent approximation to the multipolar NR frequency $\omega_{\lm}^{\rm NR}$.
Our main findings can be summarized as follows:
\begin{itemize}
\item[(i)] At a purely analytical level, the major novelty introduced here is that the $\nu$-dependent terms entering 
the factorized waveform amplitudes $\rho_\lm$ are hybridized with test-particle information up to (relative) 6PN 
order (i.e. each $\rho_\lm$ is given by a 6-th order polynomial in some squared velocity variable). As a second
step, such polynomials are additionally resummed using Pad\'e approximants consistent with test-particle limit
results~\cite{Messina:2018ghh}. This approach improves the robustness of the waveform amplitude across the 
parameter space, improves the stability of the NQC corrections and bridges the gap, at least for what concerns 
the waveform and radiation reaction effects, with the test-particle limit~\cite{Nagar:2016ayt,Messina:2018ghh}.
\item[(ii)] Each multipolar mode up to $\ell=m=5$ is completed through merger and ringdown by means of the 
NQC correction factor and NR-informed post-merger behavior. The transition between the inspiral-merger 
phase and the post-merger ringdown can be easily, and naturally, performed just after the peak of each 
multipole, at the NQC determination point. 
\item[(iii)] To have each separate EOB waveform multipole (both the amplitude and phase) correct around its
  own amplitude peak requires five functions of $\nu$ that  are determined using NR
  simulations: $\left\{\Delta t_\lm^{\rm NR},A_\lm^{\rm NQC},\dot{A}_\lm^{\rm NQC},\omega_\lm^{\rm NQC},\dot{\omega}_\lm^{\rm NQC}\right\}$. 
This allows us to properly determine the NQC correction factor multipole by multipole. This is a crucial piece 
of information that must be added to the purely analytical description of each waveform multipole in order to
correctly represent the very latest stages of the evolution, $\sim 50M$ before the peak. It is remarkable that
such a straightforward procedure is so efficient in improving, multipole by multipole, the circularized EOB waveform. 
This is {\it also} the case for the $m=1$ mode, where the impact of the radial-momentum dependent 
terms can be particularly large. Note, however, that this approach only works in conjunction with the 
structure of the Newtonian (multipolar) prefactors, that should be effectively modified by relaxing,
in a specific multipole dependent way, Newton's Kepler's constraint during the plunge. This allows one
to ease the action of the NQC factors.
\item[(iv)] In order to gauge some of the analytical uncertainty associated to the choices made in constructing 
a NR-informed EOB model, we have contrasted the effect of two different choices of radiation reaction. 
This corresponds to two different and independent determinations of $a_6^c(\nu)$ obtained through an 
EOB/NR phasing comparison. Eventually, we conclude that the choice made for \TEOBiResumSM{}, that
relies on Pad\'e resummed waveform amplitudes, is more accurate and robust, especially in view of
its use in a forthcoming spin-aligned multipolar waveform model.
\item[(v)] We have performed an extensive investigation of the EOB/NR unfaithfulness varying both the
  mass ratio and the viewing direction of the waveform. The global multipolar model was found to 
  yield an unfaithfulness always $<3$\% for binaries of $50\leq M \leq 200 M_{\odot}$. We could clearly probe 
  that such degradation of the EOB/NR performance, that occurs for large masses and only for some specific 
  region of the parameter space with $1\lesssim q \lesssim 2$, is mainly due to uncertainties in the NR higher 
  modes that may be amplified due to, for example, the extrapolation procedures. By contrast, 
  one has also to remark that $\bar{F}$ comfortably remains below (or around) $1\%$ up to total
  mass of the order of $100M_\odot$.
\item[(vi)] For the first time, we have provided an EOB-based description of the $(3,2)$ and $(4,3)$ 
 waveform modes through merger and ringdown, although we did so neglecting mode mixing effects. 
 We found that such approximation does not seem to especially degrade the performance of the model. 
\end{itemize}
The \TEOBiResumSM{} model presented here will be made publicly available as a stand-alone $C$-implementation
(notably complemented by the fast post-adiabatic approximation for the inspiral~\cite{Nagar:2018gnk}) as well as
within the LIGO {\tt LALSuite} library.

\acknowledgments 
We are grateful to Thibault Damour for discussions. G.~R. thanks IHES for hospitality during the final stage of development of this work. The authors thank Sascha Husa and Mark Hannam 
for use of the BAM simulation. G.~P. acknowledges support from the Spanish Ministry of Culture and Sport grant FPU15/03344, the Spanish Ministry of Economy 
and Competitiveness grants FPA2016-76821-P, the Agencia estatal de Investigaci\'on, the RED CONSOLIDER CPAN  FPA2017-90687-REDC, RED CONSOLIDER MULTIDARK: 
Multimessenger Approach for Dark Matter Detection, FPA2017-90566-REDC, Red nacional de astropart\'iculas (RENATA), FPA2015-68783-REDT, European 
Union FEDER funds, Vicepresid\`encia i Conselleria d'Innovaci\'o, Recerca i Turisme, Conselleria d'Educaci\'o, i Universitats del Govern de les Illes 
Balears i Fons Social Europeu, Gravitational waves, black holes and fundamental physics. We thank Patricia Schmidt for useful discussions 
throughout this project.


\appendix
\section{Detailed amplitude and frequency EOB/NR comparisons for higher modes}
\label{sec:AppHM}
This Appendix collects some complementary
information behind the global view plot of Fig.~\ref{fig:q6_all}.
Figures~\ref{fig:q6_l3l4} and~\ref{fig:q6_41_55} show EOB/NR
amplitude amplitude frequency comparisons for the illustrative
case SXS:BBH:0166, with $q=6$ for the $\ell=3$, $\ell=4$ and $\ell=m=5$
modes. As above, the dashed vertical line identifies the merger
point. In general, the bare EOB frequency (orange, dashed line)
gives a reliable representation of the NR one essentially up
to merger for the $\ell=m$ modes. Similarly, the NQC correction
factor is able to efficiently bridge the gap with the postpeak
(ringdown) part for all multipoles, even the $(3,2)$ and $(4,3)$,
although our model just averages on the postpeak behavior due
to mode-mixing effects. The proof of the robustness of the
ringdown-matching procedure is evident also in an (essentially)
irrelevant mode like the $(4,1)$, where the analytic model is
able to accurately interpolate in a region where NR data
are rather noisy.

\section{Interpolating fit for the  NQC extraction point}
\label{sec:nqc_fits}
This Appendix lists the fits of waveform amplitude and frequency
values extracted at the special NQC point on the NR time axis.
These fits are then used to determine the parameters $(a_1^\lm,a_2^\lm,b_1^\lm,b_2^\lm)$
entering the multipolar NQC correction factors of Eq.~\eqref{eq:hlmNQC}.
We focus here on the fits for all modes with the exception of the $(2,1)$
one that is separately treated in Sec.~\ref{sc:NQC_h21} below 
due to the special behavior in the test-particle limit.
For each mode, the NQC point is located $2M$ on the right of the peak location.
We fit the NR waveform data (amplitude, frequency and first time derivatives)
extracted there with a factorized template of the form
\begin{align}
\label{eq:NQC_hlm}
Y^{\rm NQC}_{\ell m}= Y^{0}_{\ell m}\; \hat{\hat{Y}}^{\rm NQC}_{\ell m},
\end{align}
where $Y^{0}_{\ell m}$ refers to the test-particle limit value and 
$\hat{\hat{Y}}^{\rm NQC}_{\ell m}$ captures the remaining $\nu$-dependence.
The latter is modeled with a rational function or polynomial up to second
order in $\nu$, in both denominator and numerator.
The reader should note that the amplitude is not fitted directly,
but rather we use the quantity $\hat{A}^{{\rm NQC}}_{\ell m}\equiv A^{\rm NQC}_{\ell m}/|c_{\ell + \epsilon}(\nu)|$.
The parameter of the fits are reported in the Table~\ref{tab:NQC}.  
\\
\subsection{$(\ell,m)=(2,1)$ mode}
\label{sc:NQC_h21}
\begin{figure}[t]
\begin{center}
  \includegraphics[width=0.9\columnwidth]{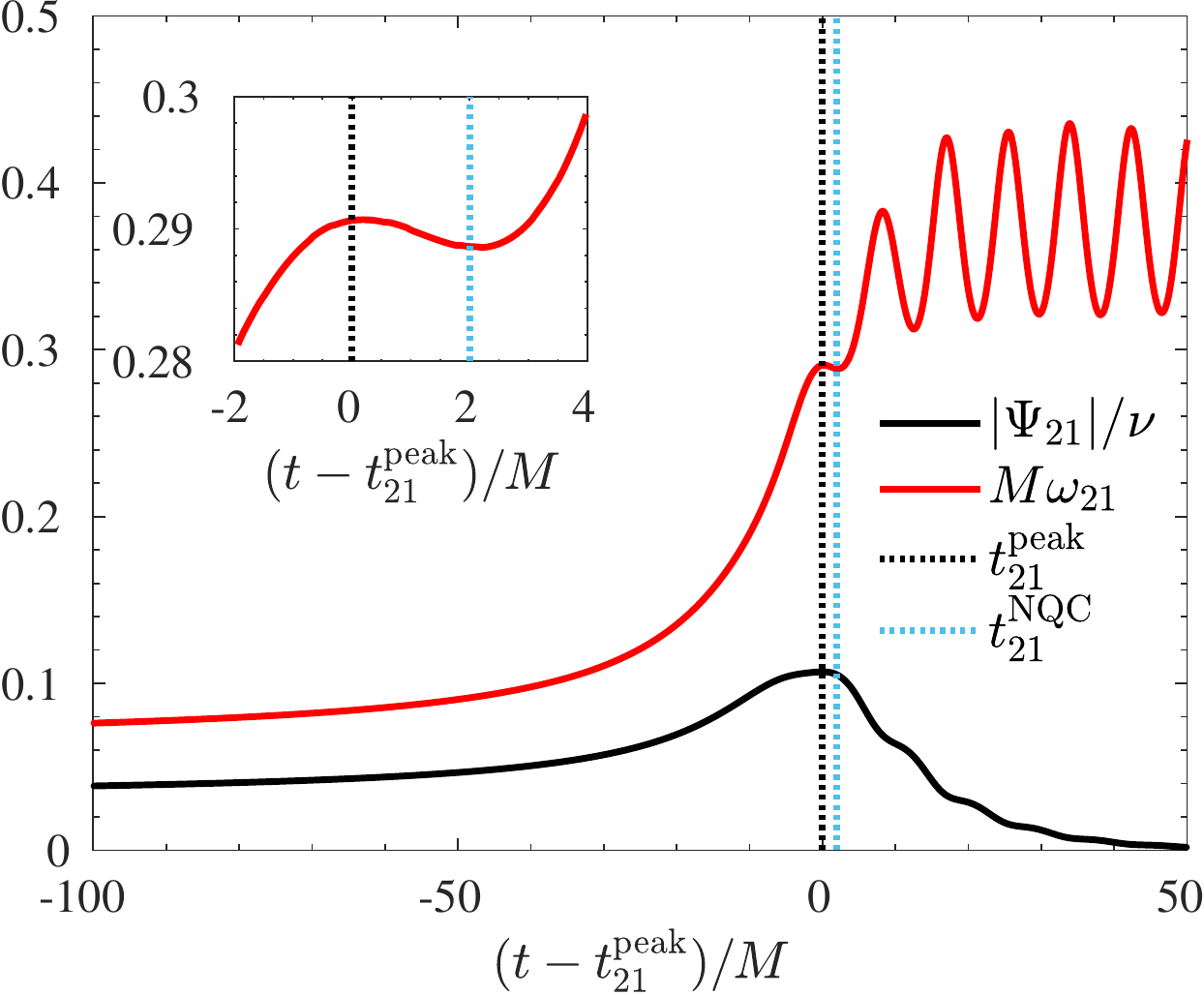}
  \caption{The amplitude $|\Psi_{21}|/\nu$ and frequency $M\omega_{21}$ of the $(2,1)$
    mode in the extreme-mass-ratio limit for a BBH coalescence in the large mass ratio limit.
    The waveform is obtained using a Regge-Wheeler-Zerilli perturbative approach, where a
    test-mass inspirals and plunges over a Schwarzschild black hole.
    The frequency $M\omega_{21}$(red) starts to oscillate after the peak because of the
    interference between negative and positive frequency QNMs~\cite{Nagar:2006xv,Damour:2007xr,Bernuzzi:2010ty}.
    The inset zooms on $M\omega_{21}$ at the peak of the $(2,1)$ mode.
}
\label{fig:21_mode_mix}
\end{center}
\end{figure}

The values of frequency and amplitude at the $(2,1)$ NQC extraction point $t_{21}^{\rm NQC}$
are directly fitted with linear and quadratic polynomials in $\nu$, and the
factorization of the test-particle values is omitted. The reason for doing
so is the peculiar (well-known) behavior of the frequency in the test-particle
limit, that is illustrated in Fig.~\ref{fig:21_mode_mix}. One sees that
the frequency $M\omega_{21}$ starts to oscillate after the peak of the
$(2,1)$ mode. These oscillations are due to the interference of negative
and positive QNMs~\cite{Nagar:2006xv,Damour:2007xr,Bernuzzi:2010ty}. 
Because this phenomenon shows up at  $t^{\rm NQC}_{21}$, 
factoring out the test-particle behavior is no longer beneficiary to the fit 
quality, at least with the sample of NR data currently at our disposal. 
The $(2,1)$ fits are listed in the second row of Table~\ref{tab:NQC}.  

\begin{table*}[t]
	\caption{\label{tab:NQC_hlm_fits} 
	The fits of the NQC functioning points $\{\hat{A}^{{\rm NQC}}_{\ell m},\dot{A}^{{\rm NQC}}_{\ell m}/\nu,\omega^{\rm NQC}_{\ell m},\dot{\omega}^{\rm NQC}_{\ell m}\}$.
	The fits are given explicitly. The fits are done after the factorization defined in eq.~\eqref{eq:NQC_hlm}.
	For all multipoles the factorization of the test-particle limit $Y^{0}_{\ell m}$ is highlighted explicitly in the third and fifth column
	of the table. The exception to this is the $(2,1)$ mode for which the test-particle behavior has not been factorized (see Sec.~\ref{sc:NQC_h21}).
	$\hat{\hat{Y}}^{\rm NQC}_{\ell m}$ is fitted for all multipoles with at most quadratic polynomials or rational functions in $\nu$.
	 The reader should note that $A^{{\rm NQC}}_{\ell m}=|c_{\ell + \epsilon}(\nu)|\hat{A}^{{\rm NQC}}_{\ell m}$.
	}
	\label{tab:NQC}
\begin{ruledtabular}
\begin{tabular}{l l | l l | l l }
 	& 		& \multicolumn{2}{c|}{$\hat{A}^{{\rm NQC}}_{\ell m}$} & \multicolumn{2}{c}{$\omega^{\rm NQC}_{\ell m}$} \\\hline
 2 	&  2 	& $0.294773$ 	& $\left(1-0.051898\nu+1.5886\nu^2\right)$ & $0.285588$ & $\left(1+0.92487\nu+1.7206\nu^2\right)$\\
 2 	&  1 	&  \multicolumn{2}{c|}{$0.097671-0.0014424\nu$} & \multicolumn{2}{c}{$0.29622+0.048182\nu+0.37472\nu^2$} \\\hline
 3 	&  3 	& $0.0512928$ 	& $\left(1+0.09537\nu+3.7217\nu^2\right)$ & $0.476647$ & $\left(1+1.1008\nu+2.84\nu^2\right)$ \\
 3 	&  2 	& $0.0178914$ 	& $\left(\frac{1-6.1472\nu+11.435\nu^2}{1-3.6362\nu}\right)$ & $0.482635$ & $\left(\frac{1-9.1403\nu+21.399\nu^2}{1-8.8647\nu+20.185\nu^2}\right)$ \\
 3 	&  1 	& $0.00520201$ 	& $\left(1-4.9441\nu+8.9339\nu^2\right)$ & $0.485186$ & $\left(1-0.4421\nu-6.8184\nu^2\right)$ \\\hline
 4 	&  4 	& $0.0144330$ 	& $\left(\frac{1-3.7335\nu-0.2895\nu^2}{1-3.7298\nu}\right)$ & $0.665507$ & $\left(1+0.95802\nu\right)$ \\
 4 	&  3 	& $0.00487784$ 	& $\left(\frac{1-5.7951\nu+12.833\nu^2}{1-3.2681\nu}\right)$ & $0.673274$ & $\left(\frac{1-9.2007\nu+22.161\nu^2}{1-9.026\nu+21.238\nu^2}\right)$ \\
 4 	&  2 	& $0.00161809$ 	& $\left(1-4.6975\nu+7.3437\nu^2\right)$ & $0.663076$ & $\left(1-0.086381\nu-8.5978\nu^2\right)$ \\
 4 	&  1 	& $0.00043987$ 	& $\left(\frac{1-8.4975\nu+27.31\nu^2}{1-1.2002\nu}\right)$ & $0.735051$ & $\left(\frac{1-8.3628\nu+20.529\nu^2}{1-7.4883\nu+18.695\nu^2}\right)$ \\\hline
 5 	&  5 	& $0.00516272$ 	& $\left(1-0.38892\nu+6.7413\nu^2\right)$ & $0.855016$ & $\left(\frac{1-2.8461\nu-3.7163\nu^2}{1-3.8378\nu}\right)$ \\\hline\hline
 	& 		& \multicolumn{2}{c|}{$\dot{A}^{{\rm NQC}}_{\ell m}/\nu$} & \multicolumn{2}{c}{$\dot{\omega}^{\rm NQC}_{\ell m}$}\\\hline
 2 	&  2 	& $-0.00119366$ & $\left(1+3.0125\nu-2.1792\nu^2\right)$ & $0.00628027$ & $\left(1+2.5374\nu+3.9341\nu^2\right)$ \\
 2 	&  1 	& \multicolumn{2}{c|}{$\left(-0.0011119+0.0042824\nu\right)/\left(1-3.0565\nu\right)$}& \multicolumn{2}{c}{$0.0020157+0.049725\nu$}\\\hline
 3 	&  3 	& $-0.00039568$ & $\left(1+1.0985\nu-13.458\nu^2\right)$ & $0.0110394$ & $\left(1+2.1358\nu +4.1544\nu^2\right)$ \\
 3 	&  2 	& $-0.00026840$ & $\left(\frac{1-8.4869\nu+18.736\nu^2}{1-5.7457\nu+7.9581\nu^2}\right)$ & $0.0141756$ & $\left(\frac{1-10.831\nu+37.969\nu^2}{1-12.954\nu+51.155\nu^2}\right)$ \\
 3 	&  1 	& $-0.00043382$ & $\left(\frac{1-9.0479\nu+23.054\nu^2}{1+88.626\nu^2}\right)$ & $0.0673118$ & $\left(\frac{1+13.318\nu}{1+70.552\nu}\right)$ \\\hline
 4 	&  4 	& $-0.00015129$ & $\left(1-2.206\nu+2.0191\nu^2\right)$ & $0.0147878$ & $\left(\frac{1-3.4516\nu +4.8703\nu^2}{1-5.7616\nu+11.286\nu^2}\right)$ \\
 4 	&  3 	& $-0.00008468$ & $\left(1-4.1848\nu+4.2192\nu^2\right)$ & $0.0172836$ & $\left(\frac{1-19.234\nu+105.04\nu^2}{1-19.837\nu+107.76\nu^2}\right)$ \\
 4 	&  2 	& $-0.00004223$ & $\left(\frac{1-5.1172\nu+5.4408\nu^2}{+6.1593\nu}\right)$ & $0.0213781$ & $\left(\frac{1-6.2629\nu +10.1\nu^2}{1-8.4232\nu+21.204\nu^2}\right)$ \\
 4 	&  1 	& $-0.00001827$ & $\left(1-2.8242\nu-3.1871\nu^2\right)$ & $0.0739078$ & $\left(1+0.99186\nu-19.435\nu^2\right)$ \\\hline
 5 	&  5 	& $-0.00006580$ & $\left(1-1.8592\nu\right)$ & $0.0178326$ & $\left(1+2.4606\nu\right)$ 
\end{tabular}
\end{ruledtabular}
\end{table*}




\bibliography{references}

\end{document}